\documentclass[11pt,tightenlines,onecolumn,preprintnumbers,superscriptaddress,nofootinbib]{revtex4}
\usepackage[a4paper,left=2.5cm,right=2.5cm, top=1.5cm,bottom=1.5cm]{geometry}

\usepackage{amsmath}
\usepackage{epsfig}
\usepackage{graphicx}
\usepackage{color}
\usepackage{hyperref}
\usepackage[utf8]{inputenc}
\hypersetup{%
	pdftitle    = {Window observable for the hadronic vacuum polarization contribution to the muon g-2 from lattice QCD},
}%

\newcommand{\fm}{{\rm{fm}}}
\newcommand{\MeV}{{\rm{MeV}}}

\newcommand{\cA}{c_{\rm{A}}}
\newcommand{\zv}{Z_{\rm V}}

\newcommand{\bv}{b_{\rm V}}
\newcommand{\bbarv}{\overline{b}_{\rm V}}
\newcommand{\cv}{c_{\rm V}}

\newcommand{\fv}{f_{\rm V}}
\newcommand{\bbA}{\overline{b}_{\rm A}}
\newcommand{\bA}{b_{\rm A}}

\newcommand{\ahvp}{a_\mu^{\rm hvp}}
\newcommand{\ahlbl}{a_\mu^{\rm hlbl}}
\newcommand{\be}{\begin{equation}}
\newcommand{\ee}{\end{equation}}
\newcommand{\bea}{\begin{eqnarray}}
\newcommand{\eea}{\end{eqnarray}}
\newcommand{\eq}[1]{Eq.\,(\ref{#1})}
\newcommand{\lesssim}{\;\raisebox{-.6ex}{$\stackrel{\textstyle{<}}{\sim}$}\;}
\newcommand{\gtrsim}{\;\raisebox{-.6ex}{$\stackrel{\textstyle{>}}{\sim}$}\;}

\newcommand{\dof}{\mathrm{d.o.f.}}

\newcommand{\psib}{\overline{\psi}}	
\newcommand{\mqav}{m_{\rm q}^{\rm av}}
\newcommand{\mql}{m_{{\rm q},l}}
\newcommand{\mqs}{m_{{\rm q},s}}

\newcommand{\stat}{\mathrm{stat}}
\newcommand{\syst}{\mathrm{syst}}

\newcommand{\amu}{a_\mu^{\mathrm{hvp}}}

\newcommand{\amud}{a_\mu^{\mathrm{hvp,disc}}}

\newcommand{\awin}{a_\mu^{\mathrm{win}}}
\newcommand{\awiniso}{a_\mu^{\mathrm{win,iso}}}
\newcommand{\awinl}{a_\mu^{\mathrm{win,ud}}}
\newcommand{\awins}{a_\mu^{\mathrm{win,s}}}
\newcommand{\awind}{a_\mu^{\mathrm{win,disc}}}
\newcommand{\awinc}{a_\mu^{\mathrm{win,c}}}
\newcommand{\awinisovec}{a_\mu^{\mathrm{win,I1}}}
\newcommand{\awinisosca}{a_\mu^{\mathrm{win,I0}}}
\newcommand{\awinany}{a_\mu^{\mathrm{win,}{\rm f}}}

\newcommand{\loc}{{\scriptsize L}}
\newcommand{\cons}{{\scriptsize C}}
\newcommand{\mloc}{ {\scriptscriptstyle\rm L} }
\newcommand{\mcons}{ {\scriptscriptstyle\rm C} }

\begin{document}
\begin{titlepage}
\begin{flushright}
MITP-22-038\\
CERN-TH-2022-098\\
DESY-22-105 
\end{flushright}

\renewcommand{\thefootnote}{\fnsymbol{footnote}}

\vskip 0.5 cm
\begin{center}
  {\Large\bf Window observable for the hadronic vacuum polarization
    contribution to the muon $g-2$ from lattice QCD
  \\[0.5ex]}
\end{center}
\vskip 1.0cm
\begin{center}
{\large M.\,C\`e$^{a,b}$,
  A.\,G\'erardin$^{c}$,
  G.\,von Hippel$^{d}$,
  R.\,J.\,Hudspith$^{e}$,
  S.\,Kuberski$^{e,f}$,
  H.\,B.\,Meyer$^{d,e}$,
  K.\,Miura$^{e,g}$,
  D.\,Mohler$^{h,f}$,
  K.\,Ottnad$^{d}$,
  S.\,Paul$^{d}$,
  A.\,Risch$^{i}$,
  T.\,San Jos\'e$^{d,e}$,
  H.\,Wittig$^{b,d,e}$
}

\vskip 1.0cm
$^{a}$\,Albert Einstein Center for Fundamental Physics (AEC) and Institut für Theoretische Physik, Universität Bern, Sidlerstrasse 5, 3012 Bern, Switzerland
\vskip 0.15cm
$^{b}$\,Department of Theoretical Physics, CERN, 1211 Geneva 23, Switzerland
\vskip 0.15cm
$^{c}$\,Aix-Marseille-Universit\'e, Universit\'e de Toulon, CNRS, CPT,
Marseille, France
\vskip 0.15cm
$^{d}$\,PRISMA$^+$ Cluster of Excellence and Institut f\"ur
Kernphysik, Johannes Gutenberg-Universit\"at Mainz, Germany
\vskip 0.15cm
$^{e}$\,Helmholtz-Institut Mainz, Johannes Gutenberg-Universit\"at
Mainz, Germany
\vskip 0.15cm
$^{f}$\,GSI Helmholtz Centre for Heavy Ion Research, Darmstadt,
Germany
\vskip 0.15cm
$^{g}$\,KEK Theory Center, High Energy Accelerator Research Organization, 1-1 Oho, Tsukuba, Ibaraki 305-0801, Japan
\vskip 0.15cm
$^{h}$\,Institut für Kernphysik, Technische Universit\"at Darmstadt,
Schlossgartenstrasse 2, D-64289 Darmstadt, Germany
\vskip 0.15cm
$^{i}$\,John von Neumann-Institut f{\"u}r Computing NIC, Deutsches Elektronen-Synchrotron DESY, Platanenallee 6, 15738 Zeuthen, Germany
\vskip 1.cm
{\bf Abstract}
\vskip 0.ex
\end{center}

\renewcommand{\thefootnote}{\arabic{footnote}}

\noindent
Euclidean time windows in the integral representation of the hadronic
vacuum polarization contribution to the muon $g-2$ serve to test the
consistency of lattice calculations and may help in tracing the
origins of a potential tension between lattice and data-driven
evaluations. In this paper, we present results for the intermediate
time window observable computed using O($a$) improved Wilson fermions
at six values of the lattice spacings below 0.1\,fm and pion masses
down to the physical value.
Using two different sets of improvement coefficients in the
definitions of the local and conserved vector currents, we perform a
detailed scaling study which results in a fully controlled
extrapolation to the continuum limit without any additional treatment
of the data, except for the inclusion of finite-volume corrections. To
determine the latter, we use a combination of the method of Hansen and
Patella and the Meyer-Lellouch-L\"uscher procedure employing the
Gounaris-Sakurai parameterization for the pion form factor.
We correct our results for isospin-breaking effects via the
perturbative expansion of QCD+QED around the isosymmetric theory. Our
result at the physical point is $\awin=(237.30\pm0.79_{\rm
stat}\pm1.22_{\rm syst})\times10^{-10}$, where the systematic error
includes an estimate of the uncertainty due to the quenched charm
quark in our calculation. Our result displays a tension of 3.9$\sigma$
with a recent evaluation of $\awin$ based on the data-driven method.

\vfill

\begin{center}
June 2022
\end{center}

\eject

\end{titlepage}

\setcounter{footnote}{0}

\tableofcontents
\newpage

\section{Introduction}

The anomalous magnetic moment of the muon, $a_\mu$, plays a central
role in precision tests of the Standard Model (SM). The recently
published result of the direct measurement of $a_\mu$ by the Muon
$g-2$ Collaboration \cite{Muong-2:2021ojo} has confirmed the earlier
determination by the E821 experiment at BNL \cite{Bennett:2006fi}.
When confronted with the theoretical estimate published in the 2020
White Paper \cite{Aoyama:2020ynm},
the combination of the two direct measurements increases the tension
with the SM to 4.2$\sigma$. The SM prediction of
Ref.~\cite{Aoyama:2020ynm} is based on the estimate of the
leading-order hadronic vacuum polarization (HVP) contribution,
$\ahvp$, evaluated from a dispersion integral involving hadronic cross
section data (``data-driven approach'') \cite{Davier:2017zfy,
Keshavarzi:2018mgv, Colangelo:2018mtw, Hoferichter:2019gzf,
Davier:2019can, Keshavarzi:2019abf}, which yields
$\ahvp=(693.1\pm4.0)\times10^{-10}$ \cite{Aoyama:2020ynm}. The quoted
error of 0.6\% is subject to experimental uncertainties associated
with measured cross section data.

Lattice QCD calculations for $\ahvp$ \cite{DellaMorte:2017dyu,
chakraborty:2017tqp, borsanyi:2017zdw, Blum:2018mom, giusti:2019xct,
shintani:2019wai, Davies:2019efs, Gerardin:2019rua, Aubin:2019usy,
giusti:2019hkz, Borsanyi:2020mff, Lehner:2020crt, Giusti:2021dvd,
Wang:2022lkq, Aubin:2022hgm} as well as for the
hadronic light-by-light scattering contribution $\ahlbl$
\cite{Blum:2014oka, Blum:2015gfa, Blum:2016lnc,
  Blum:2017cer,Blum:2019ugy, Green:2015sra, Green:2015mva,
Asmussen:2016lse,   Gerardin:2017ryf, Asmussen:2018ovy,
Asmussen:2018oip, Asmussen:2019act, Chao:2020kwq, Chao:2021tvp,
Chao:2022xzg} have become increasingly precise in recent years (see
\cite{Meyer:2018til,Gulpers:2020pnz,Gerardin:2020gpp} for recent
reviews). Although these calculations do not rely on the use of
experimental data, they face numerous technical challenges that must
be brought under control if one aims for a total error that can rival
or even surpass that of the data-driven approach.
In spite of the technical difficulties, a first
calculation of $\ahvp$ with a precision of 0.8\% has been published
recently by the BMW collaboration \cite{Borsanyi:2020mff}. Their
result of $\ahvp=(707.5\pm5.5)\times10^{-10}$ is in slight tension 
(2.1$\sigma$) with the White Paper estimate and reduces the tension
with the combined measurement from E989 and E821 to just 1.5$\sigma$.
This has triggered several investigations that study the question
whether the SM can accommodate a higher value for $\ahvp$ without
being in conflict with low-energy hadronic cross section data
\cite{Colangelo:2020lcg} or other constraints, such as global
electroweak fits \cite{Passera:2008jk, Crivellin:2020zul,
Keshavarzi:2020bfy, Malaescu:2020zuc}. At the same time, the
consistency among lattice QCD calculations is being
scrutinized with a focus on whether systematic effects such as
discretization errors or finite-volume effects are sufficiently well
controlled. Moreover, when comparing lattice results for $\ahvp$ from
different collaborations, one has to make sure that they refer to the
same hadronic renormalization scheme that expresses the bare quark
masses and the coupling in terms of measured hadronic observables.

Given the importance of the subject and in view of the enormous effort
required to produce a result for $\ahvp$ at the desired level of
precision, it has been proposed to perform consistency checks among
different lattice calculations in terms of suitable benchmark
quantities that suppress, respectively enhance individual systematic
effects. These quantities are commonly referred to as ``window
observables'', whose definition is given in section~\ref{sec:definitions}.

In this paper we report our results for the so-called ``intermediate''
window observables, for which the short-distance as well as the
long-distance contributions in the integral representation of $\ahvp$
are reduced. This allows for a straightforward and highly precise
comparison with the results from other lattice calculations and the
data-driven approach. This constitutes a first step towards a deeper
analysis of a possible deviation between lattice and phenomenology.
Indeed, our findings present further evidence for a strong tension between
lattice calculations and the data-driven method. At the physical point
we obtain $\awin=(237.30\pm1.46)\times10^{-10}$ (see \eq{eq:finalfull})
for a detailed error budget), which is $3.9\sigma$ above the recent
phenomenological evaluation of $(229.4\pm1.4)\times10^{-10}$ quoted in
Ref.~\cite{Colangelo:2022vok}.

This paper is organized as follows: We motivate and define the window
observables in Sect.~\ref{sec:definitions}, before describing the
details of our lattice calculation in Sect.~\ref{sec:lattice}. In
Sect.~\ref{sec:extraps} we discuss extensively the extrapolation to
the physical point, focussing specifically on the scaling behavior,
and present our results for different isospin components and the
quark-disconnected contribution. Sections~\ref{sec:charm}
and~\ref{sec:IB} describe our determinations of the charm quark
contribution and of isospin-breaking corrections, respectively.
Our final results are presented and compared to other determinations
in Sect.~\ref{sec:discussion}. In-depth descriptions of technical
details and procedures, as well as data tables, are relegated to
several appendices. Details on how we correct for mistunings of the
chiral trajectory are described in Appendices~\ref{sec:chirtraj}
and~\ref{sec:phenomodels}, the determination of finite-volume
corrections is discussed in Appendix~\ref{sec:fse}, while the
estimation of the systematic uncertainty related to the quenching of
the charm quark is presented in Appendix~\ref{sec:Qcharm}. Ancillary
calculations of pseudoscalar masses and decay constants that enter the
analysis are described in Appendix~\ref{app:ps}. Finally,
Appendix~\ref{sec:tables} contains extensive tables of our raw data.

\section{Window observables} \label{sec:definitions}

The most widely used approach to determine the leading HVP
contribution $\ahvp$ in lattice QCD is the ``time-momentum
representation'' (TMR) \cite{Bernecker:2011gh}, i.e.
\be\label{eq:TMRdef}
  \ahvp=\left(\frac{\alpha}{\pi}\right)^2\int_0^{\infty}dt\,\widetilde{K}(t)G(t)\,,
\ee
where $G(t)$ is the spatially summed correlation function of the
electromagnetic current
\bea
  & & G(t)=-\frac{a^3}{3}\sum_{k=1}^3\sum_{\vec{x}}\left\langle
        j_k^{\rm em}(t,\vec{x})\,j_k^{\rm em}(0) \right\rangle,
\nonumber \\
  & & j_\mu^{\rm em}=\textstyle\frac{2}{3}\bar{u}\gamma_\mu u
                    -\textstyle\frac{1}{3}\bar{d}\gamma_\mu d
                    -\textstyle\frac{1}{3}\bar{s}\gamma_\mu s
                    +\textstyle\frac{2}{3}\bar{c}\gamma_\mu c
                    +\ldots\,,
                    \label{eq:TMRcordef}
\eea
$\widetilde{K}(t)$ is a known kernel function
(see Appendix B of Ref.~\cite{DellaMorte:2017dyu}), and the integration is
performed over the Euclidean time variable~$t$. By considering the
contributions from the  light ($u,\,d$), strange and charm quarks to
$G(t)$ one can perform a decomposition of $\ahvp$ in terms of
individual quark flavors. It is also convenient to consider the
decomposition of the electromagnetic current into an isovector ($I=1$)
and an isoscalar ($I=0$) component according to
\bea
   & & j_{\mu}^{\rm em}=j_{\mu}^{I=1}+j_{\mu}^{I=0}+\ldots, \nonumber \\
   & & j_{\mu}^{I=1}={\textstyle\frac{1}{2}}(
                      \bar{u}\gamma_\mu u -\bar{d}\gamma_\mu d),\quad
       j_{\mu}^{I=0}={\textstyle\frac{1}{6}}(
       \bar{u}\gamma_\mu u +\bar{d}\gamma_\mu d -2\bar{s}\gamma_\mu s)
\eea
where the ellipsis in the first line denotes the missing charm and
bottom contributions.

One of the challenges in the evaluation of $\ahvp$ is associated with
the long-distance regime of the vector correlator $G(t)$. Owing to the
properties of the kernel $\widetilde{K}(t)$, the integrand
$\widetilde{K}(t)G(t)$ has a slowly decaying tail that makes a sizeable
contribution to $\ahvp$ in the region $t\gtrsim2$\,fm. However, the
statistical error in the calculation of $G(t)$ increases
exponentially with $t$, which makes an accurate determination a
difficult task. Furthermore, it is the long-distance regime of the
vector correlator that is mostly affected by finite-size effects.

The opposite end of the integration interval, i.e.\ the interval
$t\lesssim0.4$\,fm, is particularly sensitive to discretization effects
which must be removed through a careful extrapolation to the continuum
limit, possibly involving an {\it ansatz} that includes sub-leading
lattice artefacts, especially if one is striving for sub-percent
precision.

At this point it becomes clear that lattice results for $\ahvp$ are
least affected by systematic effects in an intermediate subinterval of
the integration in \eq{eq:TMRdef}, as already recognized in~\cite{Bernecker:2011gh}.
This led the authors of Ref.~\cite{Blum:2018mom} to introduce three ``window
observables'', each defined in terms of complementary sub-domains with
the help of smoothed step functions. To be specific, the
short-distance (SD), intermediate distance (ID) and long-distance (LD)
window observables are given by
\bea
 & & (\ahvp)^{\rm SD} \equiv \left(\frac{\alpha}{\pi}\right)^2\int_0^{\infty} dt \,
      \widetilde{K}(t)\,G(t)\,[1-\Theta(t,t_0,\Delta)] \\
 & & (\ahvp)^{\rm ID} \equiv \left(\frac{\alpha}{\pi}\right)^2\int_0^{\infty} dt \,
      \widetilde{K}(t)\,G(t)\,[\Theta(t,t_0,\Delta)-\Theta(t,t_1,\Delta)] \label{def:ID} \\
 & & (\ahvp)^{\rm LD} \equiv \left(\frac{\alpha}{\pi}\right)^2\int_0^{\infty} dt \,
      \widetilde{K}(t)\,G(t)\,\Theta(t,t_1,\Delta)\,,
\eea
where $\Delta$ denotes the width of the smoothed step function
$\Theta$ defined by
\be
   \Theta(t,t^\prime,\Delta) \equiv {\textstyle\frac{1}{2}}\left(
          1+\tanh[(t-t^\prime)/\Delta]\right).
\ee
The widely used choice of intervals and smoothing width that we will follow is
\be
t_0=0.4\,{\rm fm}, \quad t_1=1.0\,{\rm fm}\quad{\rm and}\quad \Delta=0.15\,{\rm fm}.
\ee

The original motivation for introducing the window observables in
Ref.~\cite{Blum:2018mom} was based on the observation that the
relative strengths and weaknesses of the lattice QCD and the
$R$-ratio approach complement each other when the evaluations using
either method are restricted to non-overlapping windows, thus achieving
a higher overall precision from their combination. Since then it has
been realized that the window observables serve as ideal benchmark
quantities for assessing the consistency of lattice calculations,
since the choice of sub-interval can be regarded as a filter for
different systematic effects. Furthermore, the results can be
confronted with the corresponding estimate using the data-driven
approach. This allows for high-precision consistency checks among
different lattice calculations and between lattice QCD and
phenomenology.

In this paper, we focus on the intermediate window and use the simplified notation
\be
\awin \equiv (\ahvp)^{\rm ID}.
\ee
We remark that the observable $\awin$, which accounts for about one third
of the total $\ahvp$, can be obtained from experimental data for the ratio
\be\label{eq:Rsdef}
R(s)\equiv \frac{\sigma(e^+e^-\to \,{\rm hadrons})}{\sigma(e^+e^- \to \mu^+\mu^-)}
\ee
via the dispersive representation of the correlator (\ref{eq:TMRcordef})~\cite{Bernecker:2011gh}.
How different intervals of center-of-mass energy contribute to the
different window observables in the data-driven approach is
investigated in Appendix~\ref{sec:phenomodels}; similar observations have already been made in
Refs.~\cite{Colangelo:2022xlq,DeGrand:2022lmc,Colangelo:2022vok}.
For the intermediate window $\awin$, the relative contribution of the region $\sqrt{s}<600$\,MeV 
is significantly suppressed  as compared to the quantity $\amu$.
Instead, the relative contribution of the region $\sqrt{s}>900$\,MeV, including the $\phi$ meson contribution,
is somewhat enhanced\footnote{Contributions as massive as the $J/\psi$, however, make again a smaller
relative contribution to $\awin$ than to $\amu$.}.
Interestingly, the region of the $\rho$ and $\omega$ mesons between 600 and 900\,MeV makes about the same fractional contribution 
to $\awin$ as to $\amu$, namely  55 to 60\%.
Thus if the spectral function associated with the lattice correlator $G(t)$ was for some reason enhanced by a constant factor $(1+\epsilon)$
in the interval $600<\sqrt{s}/{\rm MeV}<900$ relative to the experimentally measured spectral function $R(s)/(12\pi^2)$,
it would approximately lead to an enhancement by a factor
$(1+0.6\epsilon)$ of both $\amu$ and $\awin$.
Finally, we note that the relative contributions of the three $\sqrt{s}$ intervals are rather similar for $\awin$ as for
the running of the electromagnetic coupling from $Q^2=0$ to $Q^2=1\,{\rm GeV}^2$.

\section{Calculation of \texorpdfstring{$a_\mu^{\rm win}$}{the window quantity} on the lattice} \label{sec:lattice}

\subsection{Gauge ensembles}

Our calculation employs a set of 24
gauge ensembles generated as part of the CLS (Coordinated Lattice Simulations) initiative
using $N_f=2+1$ dynamical flavors of non-perturbatively O($a$) improved Wilson quarks and the tree-level
O($a^2$) improved L\"uscher-Weisz gauge action \cite{Bruno:2014jqa}. The gauge ensembles used in this work
were generated for constant average bare quark mass such that the
improved bare coupling $\tilde{g}_0$~\cite{Luscher:1996sc} is kept constant along the chiral
trajectory. Six of the ensembles listed in Table~\ref{tab:sim}
realize the SU(3)$_{\rm f}$-symmetric point $m_u=m_d=m_s$ corresponding to
$m_\pi=m_K\approx420$\,MeV. Pion masses lie in the range $m_\pi\approx
130-420$\,MeV. Seven of the ensembles used have periodic (anti-periodic for fermions) boundary conditions in time,
while the others admit open boundary conditions~\cite{Luscher:2012av}.
All ensembles included in the final analysis satisfy
$m_\pi L\gtrsim4$. Finite-size effects can be checked explicitly for
$m_\pi=280$ and 420\,MeV, where in each case two ensembles with different
volumes but otherwise identical parameters are available. The
ensembles with volumes deemed to be too small are marked by an asterisk in
Table~\ref{tab:sim} and are excluded from the final analysis.

The QCD expectation values are obtained from the CLS ensembles by including
appropriate reweighting factors, including a potential sign of the
latter~\cite{Mohler:2020txx}. A negative reweighting factor, which originates
from the handling of the strange quark, is found on fewer than
0.5\% of the gauge field configurations employed in this work.

For the bulk of our pion masses, down to the physical value, results
were obtained at four values of the lattice spacing in the range
$a=0.050-0.086$\,fm. At and close to the SU(3)$_{\rm f}$-symmetric point, four more ensembles
have been added that significantly extend the range of available
lattice spacings to $a=0.039-0.099$\,fm, which allows us to perform a
scaling test with unprecedented precision.

\begin{table}[t]
\caption{Parameters of the simulations: the bare coupling $\beta = 6/g_0^2$, the lattice dimensions, the lattice spacing $a$ in physical units extracted from \cite{Bruno:2016plf}, the pion and kaon masses and the physical size of the lattice, the number of gauge field configurations used for the connected light- and strange-quark contributions (penultimate column) and for the disconnected contribution (last column). Ensembles with an asterisk are not included in the final analysis but used to control finite-size effects. The ensembles A653, A654, B450, N451, D450, D452, and E250 have periodic boundary conditions in time, all others have open boundary conditions.}
\vskip 0.1in
\begin{tabular}{lcl@{\hskip 01em}c@{\hskip 02em}c@{\hskip 01em}c@{\hskip 01em}c@{\hskip 01em}c@{\hskip 01em}c@{\hskip 01em}c}
	\hline
	Id	&	$\quad\beta\phantom{\Big|}\quad$	&	$(\frac{L}{a})^3\times\frac{T}{a}$ 	&	$a\;[\fm]$	&	$m_{\pi}\;[\MeV]$	&	$m_{K}\;[\MeV]$	&	 $m_{\pi}L$	&	$L\;[\fm]$	&	\#{}confs conn	&	\#{}confs disc	\\
	\hline
A653		& 3.34	&	$24^3\times96$	& 0.0993	&	421(4) & 421(4) & 5.1 & 2.4 & 4000	& -  \\
A654		&		&	$24^3\times96$	&		&	331(3) & 451(5) & 4.0 & 2.4 & 4000 & -  \\
\hline
H101	& 3.40	&	$32^3\times96$	& 0.08636	&	416(4) & 416(4) & 5.8 & 2.8 & 2000	& -   \\
H102	&		&	$32^3\times96$	& 		&	352(4) & 437(4) & 4.9 & 2.8 & 1900  & 1900 \\
H105$^*$	&		&	$32^3\times96$	& 		&	277(3) & 462(5) & 3.9 & 2.8 & 2000  & 1000 \\
N101	&		&	$48^3\times128$	& 		&	278(3) & 461(5) & 5.8 & 4.1 & 1500  & 1300 \\
C101	&		&	$48^3\times96$	&		&	219(2) & 470(5) & 4.6 & 4.1 & 2000 & 2000 \\
\hline
B450		& 3.46	&	$32^3\times64$	& 0.07634	&	415(4) & 415(4) & 5.1 & 2.4 & 1500	& -   \\ 
S400		&		&	$32^3\times128$	& 		&	349(4) & 440(4) & 4.3 & 2.4 & 2800 & 1700 \\ 
N451	&		&	$48^3\times128$	& 		&	286(3) & 461(5) & 5.3 & 3.7 & 1000  & 1000 \\
D450	&		&	$64^3\times128$	& 		&	215(2)  & 475(5) & 5.3 & 4.9  & 500 & 500 \\
D452	&		&	$64^3\times128$	& 		&	154(2)  & 482(5) & 3.8 & 4.9 & 900 & 800 \\ 
\hline
H200$^*$	& 3.55	&	$32^3\times96$	& 0.06426	&	416(5) &  416(5) & 4.3 & 2.1 & 2000	& -   \\ 
N202	&		&	$48^3\times128$	& 		&	412(5) &  412(5) & 6.4 & 3.1 & 900 & -  \\ 
N203 	&		&	$48^3\times128$	& 		&	346(4) &  442(5) & 5.4 & 3.1 & 1500 & 1500 \\ 
N200 	&		&	$48^3\times128$	& 		&	284(3) &  463(5) & 4.4 & 3.1 & 1700  & 1700 \\
D200 	&		&	$64^3\times128$	& 		&	200(2) &  480(5) & 4.2 & 4.1 & 2000 & 1000 \\
E250 	&		&	$96^3\times192$	& 		&	128(1) & 489(5)  & 4.0 & 6.2 & 600 & 1000 \\
\hline
N300	& 3.70	&	$48^3\times128$	&0.04981	&	419(4) & 419(4) & 5.1 & 2.4 & 1700	& -   \\ 
N302	&		&	$48^3\times128$	&		&	344(4) & 450(5) & 4.2 & 2.4 & 2200 & 1000 \\ 
J303		&		&	$64^3\times192$	&		&	257(3) & 474(5) & 4.1 & 3.2  & 1000 & 500 \\ 
E300		&		&	$96^3\times192$	&		&	174(2) & 490(5) & 4.2 & 4.8 & 600 & 500 \\ 
\hline
J500		& 3.85 	&	$64^3\times192$	& 0.039	&	411(4) & 411(4) & 5.2 & 2.5  & 1200	& -  \\ 
J501		& 		&	$64^3\times192$	&		&	332(3) & 443(4)	 &4.2	& 2.5 & 400 & - \\ 
\hline
 \end{tabular} 
\label{tab:sim}
\end{table}

\subsection{Renormalization and O(\texorpdfstring{$a$}{a})-improvement}
\label{sec:improvement}

To reduce discretization effects, on-shell O($a$)-improvement has been fully implemented. CLS simulations are performed using a non-perturbatively O($a$) improved Wilson action~\cite{Bulava:2013cta}, therefore we focus here on the improvement of the vector current in the $(u,d,s)$ quark sector. To further constrain the continuum extrapolation and explicitly check our ability to remove leading lattice artefacts, two discretizations of the vector current are used, the local (\loc) and the point-split (\cons) currents
\begin{subequations}
\begin{align}
J_{\mu}^{(\mloc),a}(x) &= \psib(x) \gamma_{\mu} \frac{\lambda^a}{2} \psi(x) \,,\\
J_{\mu}^{(\mcons),a}(x) &= \frac{1}{2} \left(
\psib(x+a\hat{\mu})(1+\gamma_{\mu}) U^{\dag}_{\mu}(x)
\frac{\lambda^a}{2} \psi(x) - \psib(x) (1-\gamma_{\mu} ) U_{\mu}(x)
\frac{\lambda^a}{2} \psi(x+a\hat{\mu}) \right)  \,,
\label{eq:consvec1}
\end{align}  
\end{subequations}
where $\psi$ denotes a vector in flavor space, $\lambda$ are the
Gell-Mann matrices, and $U_{\mu}(x)$ is the gauge link in the
direction $\hat{\mu}$ associated with site $x$. With the local tensor
current defined as $\Sigma^{a}_{\mu\nu}(x) = -\frac{1}{2}\,
\overline{\psi}(x) [\gamma_{\mu}, \gamma_{\nu}] \frac{\lambda^a}{2}
\psi(x)$, the improved vector currents are given by 
\begin{equation}
\label{eq:impcv}
J^{(\alpha),a,I}_{\mu}(x) = J^{(\alpha),a}_{\mu}(x) +
a\cv^{(\alpha)}(g_0) \, \tilde{\partial}_{\nu} \Sigma^{a}_{\mu\nu}(x)
\,,\quad \alpha=\loc,\,\cons\,,
\end{equation}
where $\tilde{\partial}$ is the symmetric discrete derivative
$\tilde{\partial}_{\nu} f(x) = (1/2a) \left( f(x+a) - f(x-a) \right)$.
The coefficients $\cv^{(\alpha)}$ have been determined
non-perturbatively in Ref.~\cite{Gerardin:2018kpy} by imposing Ward
identities in large volume ensembles and independently in
\cite{Heitger:2020zaq} using the Schr\"odinger functional (SF) setup.
The availability of two independent sets allows us to perform detailed
scaling tests, which is a crucial ingredient for a fully controlled
continuum extrapolation.

The conserved vector current does not need to be further renormalized. For the local vector current, the renormalization pattern, including O$(a)$-improvement, has been derived in Ref.~\cite{Bhattacharya:2005rb}. Following the notations of Ref.~\cite{Gerardin:2018kpy}, the renormalized isovector and isoscalar parts of the electromagnetic current read
\begin{subequations}
\begin{align}
J_{\mu}^{(\mloc),3,\mathrm{R}}(x) &= Z_3 \, J_{\mu}^{(\mloc),3,\mathrm{I}}(x) \,, \\
J_{\mu}^{(\mloc),8,\mathrm{R}}(x) &= Z_8 \, J_{\mu}^{(\mloc),8,\mathrm{I}}(x) +  Z_{80} \, J_{\mu}^{(\mloc),0,\mathrm{I}}(x) \,,
\label{eq_Rimp}
\end{align}
\end{subequations}
where $J_{\mu}^{0} = \frac{1}{2} \psib \gamma_{\mu} \psi$ is the flavor-singlet current and 
\begin{subequations}
\begin{align}
Z_3 &= \zv \left[ 1 + 3\bbarv a \mqav + \bv a \mql \right] \,, \\
Z_8 &= \zv \left[ 1 + 3\bbarv a \mqav + \frac{\bv}{3} a( \mql + 2 \mqs) \right] \,, \\
Z_{80} &= \zv \left( \frac{1}{3} \bv + \fv \right) \frac{2}{\sqrt{3}} a(\mql - \mqs)  \,.
\end{align}
\end{subequations}
Here, $\mql$ and $\mqs$ are the subtracted bare quark masses of the light and strange-quarks respectively defined in Appendix~\ref{app:ps} 
and $\mqav = (2\mql+\mqs)/3$ stands for the average bare quark mass.
The renormalization constant in the chiral limit, $\zv$, and the
improvement coefficients $\bv$ and $\bbarv$, have been determined
non-perturbatively in Ref.~\cite{Gerardin:2018kpy}. Again, independent
determinations using the SF setup are available
in~\cite{Heitger:2020zaq,Fritzsch:2018zym}. The coefficient $\fv$,
which starts at order~$g_0^6$ in perturbation
theory~\cite{Gerardin:2018kpy}, is unknown but expected to be very
small and is therefore neglected in our analysis. 

Thus, in addition to having two discretizations of the vector current, we also have at our disposal two sets of improvement coefficients that can be used to benchmark our continuum extrapolation:
\begin{itemize}
\item Set 1 : using the improvement coefficients obtained in large-volume simulations in Ref.~\cite{Gerardin:2018kpy}.
\item Set 2 : using $\zv$ and $\cv$ from~\cite{Heitger:2020zaq}, $\bv$ and $\bbarv$ from~\cite{Fritzsch:2018zym}, using the SF setup.
\end{itemize}
Note, in particular, that the improvement coefficients $\cv$, $\bv$ and $\bbarv$ have an intrinsic ambiguity of order O$(a)$. Thus, for a physical observable, we expect different lattice artefacts at order O$(a^n)$ with $n \geq 2$. This will be considered in Section~\ref{sec:contextrap}.

\subsection{Correlation functions}

The vector two-point correlation function is computed with the local
vector current at the source and either the local or the point-split
vector current at the sink. The corresponding renormalized correlators are 
\begin{subequations}
\begin{align}
G^{(\mloc\mloc),R}(t) &= Z_3^2 \, G^{(\mloc\mloc),33,I}(t) + \frac{1}{3} Z_8^2 \, G^{(\mloc\mloc),88,I}(t) + \frac{1}{3} Z_8 Z_{80} \, \left( G^{(\mloc\mloc),80,I}(t) +  G^{(\mloc\mloc),08,I}(t) \right) \,, \\
G^{(\mcons\mloc),R}(t) &= Z_3 \, G^{(\mcons\mloc),33,I}(t) + \frac{1}{3} Z_8 \, G^{(\mcons\mloc),88,I}(t) + \frac{1}{3} Z_{80} \, G^{(\mcons\mloc),80,I}(t) \,, 
\end{align}
\end{subequations}
with the improved correlators
\begin{equation}
G^{(\alpha\,\mloc),ab,I}(t) = -  \frac{a^3}{3}\sum_{k=1}^3
\sum_{\vec{x}} \langle \,  J_{k}^{(\alpha),a,I}(t,\vec{x})\;
J_{k}^{(\mloc),b,I}(0) \, \rangle \,,\quad \alpha=\loc,\,\cons\,.
\end{equation}

In the absence of QED and strong isospin breaking, there are only two sets of Wick contractions, corresponding to the quark-connected part and the quark-disconnected part of the vector two-point functions. The method used to compute the connected contribution has been presented previously in \cite{Gerardin:2019rua}. In this work we have added several new ensembles and have significantly increased our statistics, especially for our most chiral ensembles. The method used to compute the disconnected contribution involving light and strange quarks is presented in detail in Ref.~\cite{Ce:2022eix}. Note that we neglect the charm quark contribution to disconnected diagrams in the present calculation. 

\subsection{Treatment of  statistical errors and autocorrelations}

Statistical errors are estimated using the Jackknife procedure with blocking to reduce the size of auto-correlations. In practice, the same number of 100 Jackknife samples is used for all ensembles to simplify the error propagation. In a fit, samples from different ensembles are then easily matched. 

Our analysis makes use of the pion and kaon masses, their decay constants, the Wilson flow observable $t_0$, as well as the Gounaris-Sakurai parameters entering the estimate of finite-size effects. These observables are always estimated on identical sets of gauge configurations and using the same blocking procedure, such that correlations are easily propagated using the Jackknife procedure. 

The light and strange-quark contributions have been computed on the
same set of gauge configurations, except for A654 where only the
connected strange-quark contribution has been calculated. The
quark-disconnected contribution is also obtained on the same set of
configurations for most ensembles (see Table~\ref{tab:sim}). When it is not, correlations are not fully propagated; this is expected to have a very small impact on the error, since the disconnected contribution has a much larger relative statistical error.

The charm quark contribution, which is at the one-percent level, is obtained using a smaller subset of gauge configurations.
Since its dependence on the ratio of pion mass to decay constant $(m_\pi/f_\pi)$ is rather flat,
the error of this ratio is neglected in the chiral extrapolation of the charm contribution.

In order to test the validity of our treatment of statistical errors, we have performed an independent check of the entire analysis using the $\Gamma$-method~\cite{Wolff:2003sm} for the estimation of autocorrelation times and statistical uncertainties. The propagation of errors is based on a first-order Taylor series expansion with derivatives obtained from automatic differentiation~\cite{Ramos:2018vgu}. Correlations of observables based on overlapping subsets of configurations are fully propagated and the results confirm the assumptions made above.

\subsection{Results for \texorpdfstring{$\awin$}{the window quantity} on individual ensembles} \label{sec:results}

\begin{figure}[t]
	\includegraphics*[width=0.49\linewidth]{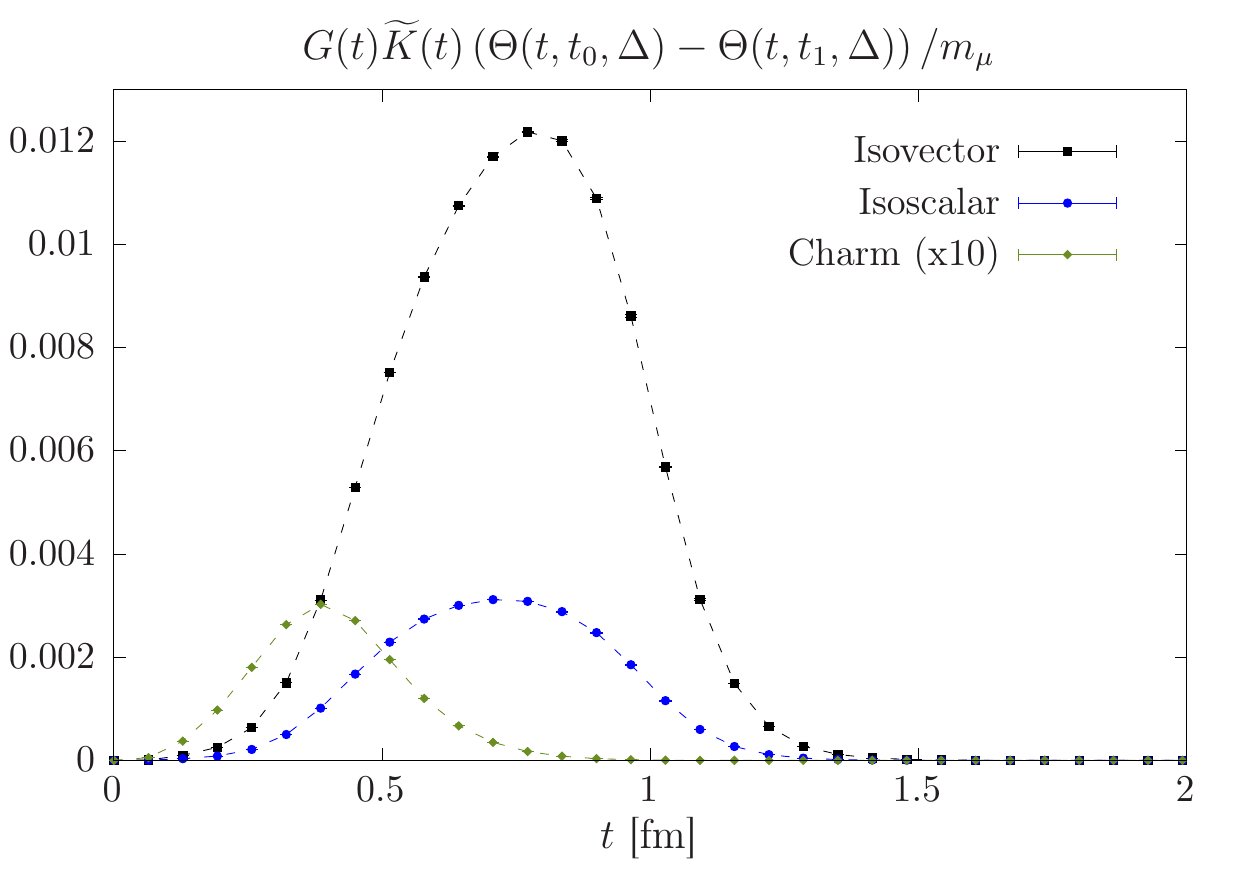} 	
	\caption{Integrands used to compute the intermediate window $a_\mu^{\rm win}$ for the isovector, isoscalar and charm quark contributions. The isoscalar contribution does not include the charm quark contribution. The data has been obtained on ensemble E250, which has close-to-physical quark masses, using two local vector currents and set~1 of renormalization and improvement coefficients.}
	\label{fig:integrand}
\end{figure}

For the intermediate window observable, the contribution from the noisy tail of the correlation function is exponentially suppressed and the lattice data are statistically very precise. Thus, on each ensemble, $\awin$ is obtained using \eq{def:ID} after replacing the integral by a discrete sum over timeslices. Since the time extent of our correlator is far longer than $t_1 = 1.0$\;fm,  we can safely replace the upper bound of \eq{def:ID} by $T/2$, with $T$ the time extent of the lattice.
The results for individual ensembles are summarized in
Tables~\ref{tab:winL1},~\ref{tab:winL2} and~\ref{tab:charm}. On
ensemble E250, corresponding to a pion mass of 130 MeV, we reach a relative
statistical precision of about two permille for both the isovector and isoscalar contributions.
The integrands used to obtain $a_\mu^{\rm win}$ are displayed in Fig.~\ref{fig:integrand}.

Our simulations are performed in boxes of finite volume $L^3$ with
$m_\pi L\gtrsim4$, and corrections due to finite-size effects (FSE)
are added to each ensemble individually prior to any continuum and
chiral extrapolation. This is the only correction applied to the raw
lattice data. FSE are dominated by the $\pi\pi$ channel and mostly
affect the isovector correlator at large Euclidean times. For the
intermediate window observable, they are highly suppressed compared to
the full hadronic vacuum polarization contribution. Despite this
suppression, FSE in the isovector channel are not negligible and
require a careful treatment. They are of the same order of magnitude
as the statistical precision for our most chiral ensemble and enhanced
at larger pion masses. In the isoscalar channel, FSE are included only
at the SU(3)$_f$ point where $m_{\pi} = m_{K}$. The methodology is
presented in Appendix~\ref{sec:fse}, and the corrections we have applied to the lattice data
are given in the last column of Tables~\ref{tab:fse2} and~\ref{tab:fse1} respectively for Strategy~1 and~2.
In our analysis, we have conservatively assigned an uncertainty of 25\% to these finite-size corrections,
in order to account for any potential effect not covered by the theoretical approaches described in Appendix~\ref{sec:fse}.
In addition to the ensembles H105 and H200 that
are only used to cross-check the FSE estimate, ensembles S400 and N302
are also affected by large finite-volume corrections. We exclude those
ensembles in the isovector channel.

\section{Extrapolation to the physical point} \label{sec:extraps}

\subsection{Definition of the physical point in iso-symmetric QCD}
\label{sec:isoQCD}

Our gauge ensembles have been generated in the isospin limit of QCD
with $m_l \equiv m_u=m_d$, neglecting strong isospin-breaking effects
and QED corrections. Naively, those effects are expected to be of
order ${\rm O}((m_d - m_u)/\Lambda_{\rm QCD}) \approx 1\%$ and ${\rm O}(\alpha) \approx 1\%$,  and are not entirely negligible at our level of precision. In Ref.~\cite{Budapest-Marseille-Wuppertal:2017okr}, although the authors used a different scheme to define their iso-symmetric setup, those corrections have been found to be of the order of 0.4\% for this window observable. A similar conclusion was reached in Ref.~\cite{Blum:2018mom} although only a subset of  the diagrams was considered. 
 This correction will be discussed in Section~\ref{sec:IB}. 
Only in full QCD+QED is the precise value of the observable unambiguously defined: the separation between its iso-symmetric value and the isospin-breaking correction is scheme dependent. In Section~\ref{sec:resIorFdecomp}, we provide the necessary information to translate our result into a different scheme. 

Throughout our calculation, we define the `physical' point in the $(m_\pi,m_K)$ plane by 
imposing the conditions~\cite{Lellouch:discussion2021workshop,Neufeld:1995mu,Urech:1994hd}
\bea
m_\pi &=& (m_{\pi^0})_{\rm phys}, \\
2 m^2_K - m^2_{\pi} &=& (m^2_{K^+} + m^2_{K^0}  - m^2_{\pi^+})_{\rm phys}.
\eea
Inserting the PDG values~\cite{Tanabashi:2018oca} on the right-hand side,
our physical iso-symmetric theory is thus defined by the values
\begin{equation}
m_\pi = 134.9768(5)~\MeV \,, \quad m_K = 495.011(10)~\MeV \,.
\end{equation}
We note that since our gauge ensembles have been generated at constant sum of the bare quark masses,
the linear combination $( m_K^2 + m_{\pi}^2 / 2 )$ is approximately constant.
Two different strategies are used to extrapolate the lattice data to the physical point.

\paragraph{Strategy 1}

We use the gradient flow observable $t_0$~\cite{Luscher:2010iy} as an intermediate scale and the dimensionless parameters
\be
\Phi_2 = 8 t_0 m_{\pi}^2,\qquad
\Phi_4 = 8 t_0 ( m_K^2 + {\textstyle\frac{1}{2}} m_{\pi}^2  )
\ee
as proxies for the light and the average quark mass as the physical point is approached.
In the expressions of $\Phi_2$ and $\Phi_4$, $t_0$ is the pion- and kaon-mass dependent flow observable;
we use the notation $t_0^{\rm sym}$ to denote its value at the SU(3)$_{\rm f}$-symmetric point.
We adopt the physical-point value $\sqrt{8t_0} = 0.4081(20)(37)$\,fm from Ref.~\cite{Strassberger:2021tsu}, obtained
by equating  the linear combination of pseudoscalar-meson decay constants 
\be
f_{K\pi} = \frac{2}{3}\Big(f_{K} + \frac{1}{2} f_{\pi} \Big) 
\ee
to its physical value, set by the PDG values of the decay constants given below.
Ref.~\cite{Strassberger:2021tsu} is an update of the work presented in~\cite{Bruno:2016plf} and includes a larger set of ensembles, including ensembles close to the physical point. We note that in Refs.~\cite{Strassberger:2021tsu,Bruno:2016plf} the absolute scale was determined assuming a slightly different definition of the physical point: the authors used the meson masses corrected for isospin-breaking effects as in~\cite{Aoki:2016frl}, $m_\pi = 134.8(3)$~MeV and $m_K = 494.2(3)$~MeV. Using the NLO $\chi$PT expressions, we have estimated the effect on $f_{K\pi}$ of these small shifts in the target pseudoscalar meson masses to be at the sub-permille level and therefore negligible for our present purposes.

\paragraph{Strategy 2}

Here we use $f_{\pi}$-rescaling, which was already presented in our
previous work~\cite{Gerardin:2019rua}, and express all dimensionful
quantities in terms of the ratio $f_{\pi}^{\rm phys} / (a f_{\pi}^{\rm
lat})$, where $a f_{\pi}^{\rm lat}$ can be computed precisely on each
ensemble.
In this case, the intermediate scale $t_0$ is not needed and we use the following dimensionless proxies
for the quark masses,
\begin{equation}
  \widetilde{y} = \frac{ m_{\pi}^2 }{ 8\pi f_{\pi}^2}, \qquad
  y_{K\pi} = \frac{m_{K}^2+ \frac{1}{2}m_\pi^2}{8\pi f_{K\pi}^2}.
\end{equation}
As $\Phi_4$, the proxy $y_{K\pi}$ is approximately constant along our
chiral trajectory. Since all relevant observables have been computed
as part of this project, this method has the advantage of being fully
self-consistent, and all correlations can be fully propagated. It will be our preferred strategy. We use the following input to set the scale in our iso-symmetric theory~\cite{Tanabashi:2018oca,Aoki:2021kgd},
\begin{equation}
  f_\pi = 130.56(14)~\MeV \,.
\end{equation}
The quantity $y_{K\pi}$ is only used to correct for a small departure of the CLS ensembles from the physical value of this quantity,
which we obtain using $f_K = 157.2(5)~\MeV$~\cite{Tanabashi:2018oca,Aoki:2021kgd}.
The latter, phenomenological value of $f_K$ implies a ratio $f_K/f_\pi$ that is consistent with the latest
lattice determinations~\cite{Bazavov:2017lyh,Miller:2020xhy,ExtendedTwistedMass:2021qui}. The impact of the uncertainty of $f_K$ on $\awin$ is
small\footnote{The sensitivity of $\awin$ to the value of $f_K$ can be derived from Table~\ref{tab:der}.},
$\delta \awin \simeq 0.10 \times 10^{-10}$, and occurs mainly through the strange contribution.
In the isosymmetric theory, we take the phenomenological values of the triplet $(m_\pi,m_K,f_\pi)$ as part of the definition 
of the target theory, and therefore only include the uncertainty from $f_K$ in our results.
By contrast, in the final result including isospin-breaking effects, which we compare to a data-driven determination of $\awin$,
we include the experimental uncertainties of all quantities used as input.

The observables $m_{\pi}$, $m_K$, $f_{\pi}$ and $f_K$, as well as
$t_0/a^2$ have been computed on all gauge ensembles and corrected for
finite-size effects~\cite{Colangelo:2005gd}. Their values for all
ensembles are listed in Table~\ref{tab:ps}.

\subsection{Fitting procedure}

We now present our strategy to extrapolate the data to the physical point in our iso-symmetric setup.  
The ensembles used in this work have been generated such that the
physical point is approached keeping
\be
X_K = \{\Phi_4,y_{K \pi}\}
\ee
approximately constant, where the two entries correspond respectively to Strategy~1 and~2.
To account for the small mistuning, only a
linear correction in $\Delta X_K = X_K^{\rm phys} - X_K$ is thus
considered. To improve the fit quality, a dedicated calculation of the
dependence of $\awin$ on $X_K$ has been performed, which is described
in Appendix~\ref{sec:chirtraj}. This analysis does not yet include all
ensembles in the final result, and hence we decided to not apply this
correction ensemble-by-ensemble prior to the global extrapolation to
the physical point. Instead, we have used $\Delta X_K$ to fix suitable
priors on the fit parameter $\gamma_0$ in \eq{eq:fit}, which
parametrizes the locally linear dependence on $X_K$. The values of these priors are given in  Appendix~\ref{sec:chirtraj}.

To describe the light quark dependence beyond the linear term in
\be
X_{\pi} = \{\Phi_2, \widetilde{y} \}\,
\ee
(respectively for Strategy~1 and~2),
we allow for different fit ansätze encoded in the function $f_{\rm ch}(X_{\pi})$. The precise
choice of $f_{\rm ch}$ is motivated on physical grounds and depends on the quark flavor. The specific forms will be discussed below. 
Since on-shell O($a$)-improvement has been fully implemented, leading discretization artefacts are expected to scale as $a^2/t_0$ up to logarithmic corrections~\cite{Symanzik:1983dc,Husung:2019ytz}. 
In the case of the vacuum polarization function, a further logarithmic
correction proportional to $a^2 \log a$ was discovered in~\cite{Ce:2021xgd}. Contrary to standard logarithmic corrections, it does not vanish as the coupling $g_0$ goes to zero due to correlators being integrated over very short distances. However, the intermediate window strongly suppresses the short-distance contribution, so that we do not expect this source of logarithmic enhancement to be relevant here. However,
in the absence of further information on the relevant exponents of $\log a$ in full QCD~\cite{Husung:2019ytz},
we still consider a possible logarithmic correction with unit exponent. 
Moreover, to check whether we are in the scaling regime, we consider
higher order terms proportional to $a^3$. 
Finally, we also allow for a term $\propto X_a^2 X_{\pi}$ that describes
pion-mass dependent discretization effects of order~$a^2$.  

Thus, for each discretization of the vector correlator, the continuum and chiral extrapolation is done independently assuming the most general functional form 
\begin{multline}
\awinany(X_a,X_{\pi}, X_K) = \awinany(0,X_{\pi}^{\exp},X_{K}^{\exp}) +
\beta_2 \, X_a^2 + \beta_3 \, X_a^3 + \delta \, X_a^2 X_{\pi}  +
\epsilon \, X_a^2 \log X_a \\ +  \gamma_0 \left( X_K - X_K^{\rm phys}
\right) + \gamma_1 \, \left( X_{\pi} - X_{\pi}^{\exp} \right)  +
\gamma_2 \left( f_{\rm ch}(X_{\pi}) -  f_{\rm ch}(X_{\pi}^{\exp})\right) \,,
\label{eq:fit}
\end{multline}
where `f' can be any flavor content and $X_a = a / \sqrt{t_0}$ parametrizes the lattice spacing. 
Despite the availability of data from six lattice spacings and more than twenty ensembles, trying to fit all parameters is not possible. Thus each analysis is duplicated by switching on/off the parameters $\beta_3$, $\delta$ and $\epsilon$ that control the continuum extrapolation.
In addition, for each functional form $f_{\rm ch}$ of the chiral
dependence, different analyses are performed by imposing cuts in the pion mass (no cut, $<400$~MeV, $<300$~MeV) and/or in the lattice spacing. 

Since several different fit ansätze can be equally well motivated, we
apply the model averaging method presented
in~\cite{Jay:2020jkz,Djukanovic:2021cgp} where the Akaike Information Criterion
(AIC) is used to weight different analyses and to estimate the
systematic error associated with the fit ansatz (see
also~\cite{1100705,Borsanyi:2020mff}).
Thus, to each analysis $(n)$ described above (defined by a specific
choice of $f_{\rm ch}$, applying cuts in the pion mass or in the
lattice spacing, and including or excluding terms proportional to
$\beta_3$, $\delta$, $\epsilon$) we associate a weight $w_n$ given
by
\begin{equation}
  \label{eq:Akweight}
w_n = N \exp\left[ - \frac{1}{2} \left( \chi^2 + 2 k - 2 n \right) \right]
\end{equation}
where $\chi^2$ is the minimum value of the chi-squared of the correlated fit, $k$ is the number of fit parameters and $n$ is the number of data points included in the fit\footnote{Different definitions of the weight factor have been proposed in the literature. In~\cite{Borsanyi:2020mff} the authors used $w_n = N \exp\left[ - \frac{1}{2} \left( \chi^2 + 2 k - n \right) \right]$ which, applied to our data for a given number of fit parameters, tends to favor fits that discard many data points. This issue will be discussed further below.}. The normalization factor $N$ is such that the sum over all the analyses' weights are equal to one. Each analysis is again duplicated by either using the local-local or the local-conserved correlators. 
For those analyses, we use a flat weight. Finally, when cuts are
performed, some fits may have very few degrees of freedom, and hence
we exclude all analyses that contain fewer than three degrees of
freedom.
The central value of an observable $\mathcal{O}$ is then obtained by a weighted average over all analyses
\begin{equation}
\bar{\mathcal{O}} = \sum_n w_n \mathcal{O}_n \,,
\label{eq:akaike_mean}
\end{equation}
and our estimate of the systematic error associated with the extrapolation to the physical point is given by
\begin{equation}
(\delta \mathcal{O})_{\rm syst}^2 = \sum_n w_n (\mathcal{O}_n - \bar{\mathcal{O}})^2 \,.
\label{eq:akaike_syst}
\end{equation}
The statistical error is obtained from the Jackknife procedure using the estimator defined by \eq{eq:akaike_mean}. 

\subsection{The continuum extrapolation at the SU(3)\texorpdfstring{$_{\rm f}$}{f}-symmetric point}
\label{sec:contextrap}

To reach sub-percent precision, a good control over the continuum limit is mandatory~\cite{Ce:2021xgd, Husung:2019ytz}. As discussed below, it is one of the largest contributions to our total error budget. Thus, before presenting our final result at the physical point, we first demonstrate our ability to perform the continuum extrapolation. We have implemented three different checks: First, two discretizations of the vector correlator are used and the extrapolations to the physical point are done independently. Both discretizations are expected to agree within errors in the continuum limit. 
Physical observables computed using Wilson-clover quarks approach the
continuum limit with a rate $\propto a^2$ once the action and all
currents are non-perturbatively O($a$)-improved~\cite{Luscher:1996sc}.
To check our ability to fully remove O($a$) lattice artefacts in the
action and the currents, two independent sets of improvement
coefficients are used: both of them should lead to an $a^2$ scaling
behavior but might differ by higher-order corrections. Finally, we
have included six lattice spacings at the SU(3)$_{\rm f}$-symmetric
point, all of them below $0.1$~fm and down to 0.039~fm, to scrutinize
the continuum extrapolation. In this section, we discuss those three
issues, with a specific focus on the ensembles with SU(3)$_{\rm f}$ symmetry.\\

Ensembles with six different lattice spacings in the range
$[0.039:0.099]$~fm are available for $m_{\pi} = m_K \approx 420~$MeV.
Since the pion masses do not match exactly, we first describe our
procedure to interpolate our SU(3)$_{\rm f}$-symmetric ensembles to a
single value of $X_{\pi} = X_{\pi}^{\star}$, to be be able
to focus solely on the continuum extrapolation. This reference point $X_{\pi}^{\star}$ is chosen to minimize the quadratic sum of the shifts $\delta X_{\pi} = X_{\pi} - X_{\pi}^{\star}$. 

We start by applying the finite-size effect correction discussed in the previous section to all ensembles. Then, a global fit over all the ensembles and simultaneously over both discretizations of the correlation function is performed using the functional form of \eq{eq:fit} without any cut in the pion mass. Thus $(\gamma_0,\gamma_1,\gamma_2)$ are fit parameters common to both discretizations, while the others are discretization-dependent. For the isovector contribution, we use the choice $f_{\rm ch}(X_{\pi})=1/X_{\pi}$ 
that leads to a reasonable $\chi^2/\dof = 1.1$. 
The good $\chi^2$, and more importantly the good description of the
light-quark mass dependence, ensures that the small interpolation to
$X_{\pi}^{\star}$ is safe and that we do not bias the result. In
practice, we have checked explicitly that using different functional
forms $f_{\rm ch}$ to interpolate the data leads to changes that are small compared to the statistical error.  Thus, for both choices of the improvement coefficients (set~1 and set~2), and for both discretizations  {\sc LL} and {\sc CL}, the data from an SU(3)$_{\rm f}$-symmetric ensemble is corrected in the pseudoscalar masses to the reference SU(3)$_{\rm f}$-symmetric point at the same lattice spacing. The correction is obtained by taking the difference of
\eq{eq:fit} evaluated with the reference-point arguments $(X_a,X^{\star}_{\pi},X_K^{\star})$ and the ensemble arguments $(X_a,X_{\pi},X_K)$, resulting in
\begin{multline}
\awinany{}^{,\alpha}(X_a,X^{\star}_{\pi},X_K^{\star}) = \awinany{}^{,\alpha}(X_a,X_{\pi}, X_K) -\delta \, X_a^2 \left( X_{\pi} -  X_{\pi}^{\star} \right) - \gamma_0 \left( X_K - X_K^{\star} \right) \\ - \gamma_1 \, \left( X_{\pi} -  X_{\pi}^{\star} \right) - \gamma_2 \left( f_{\rm ch}(X_{\pi}) -  f_{\rm ch}(X_{\pi}^{\star})\right) \,, 
\label{eq:SU3cor}
\end{multline}
where $\alpha = ({\scriptsize\rm LL}), ({\scriptsize\rm CL})$ stands for the discretization.
Note that $X_K^{\star} = X_\pi^{\star}$ and $X_K = X_\pi$ in view of the SU(3)$_{\rm f}$-symmetry.
Throughout this procedure, correlations are preserved via the Jackknife analysis.

\begin{figure}[t]
	\includegraphics*[width=0.48\linewidth]{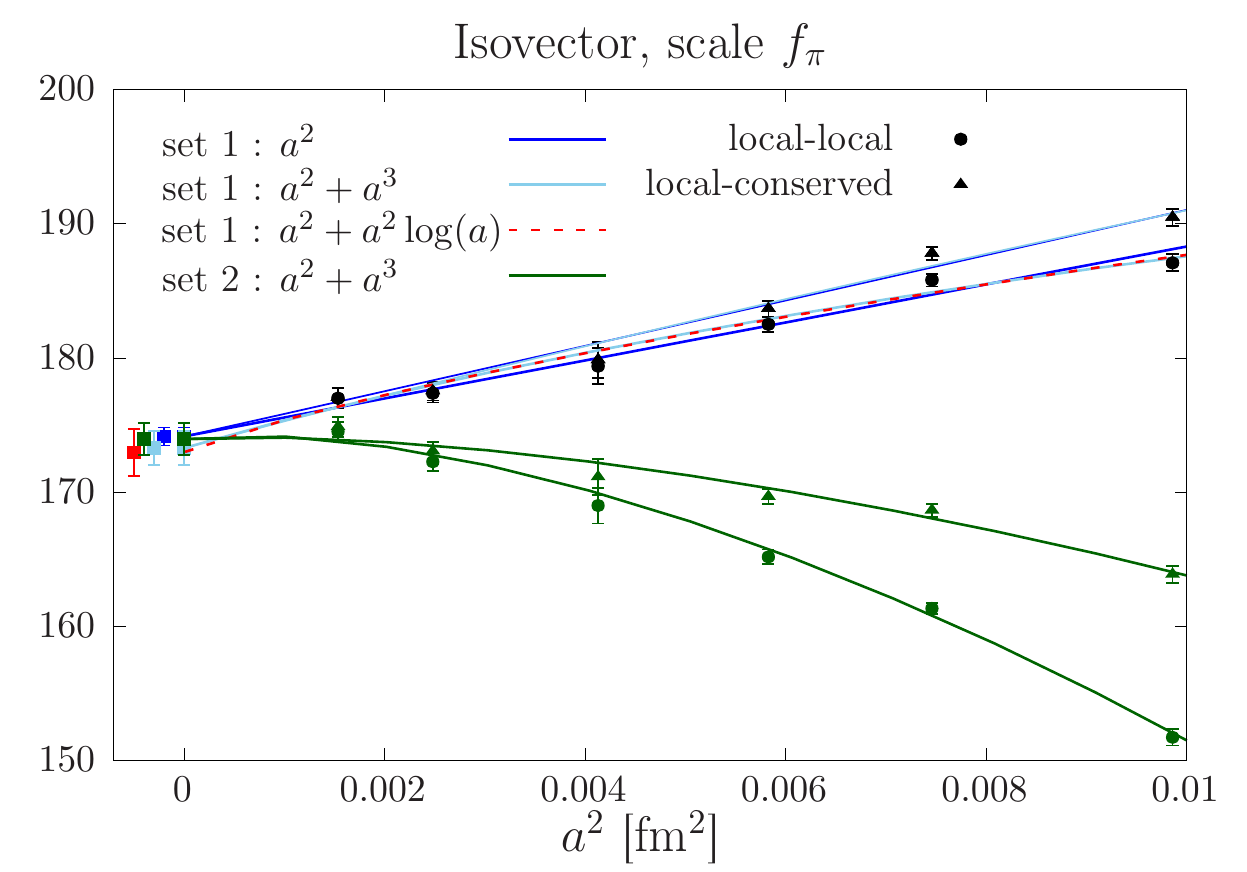} \
	\includegraphics*[width=0.48\linewidth]{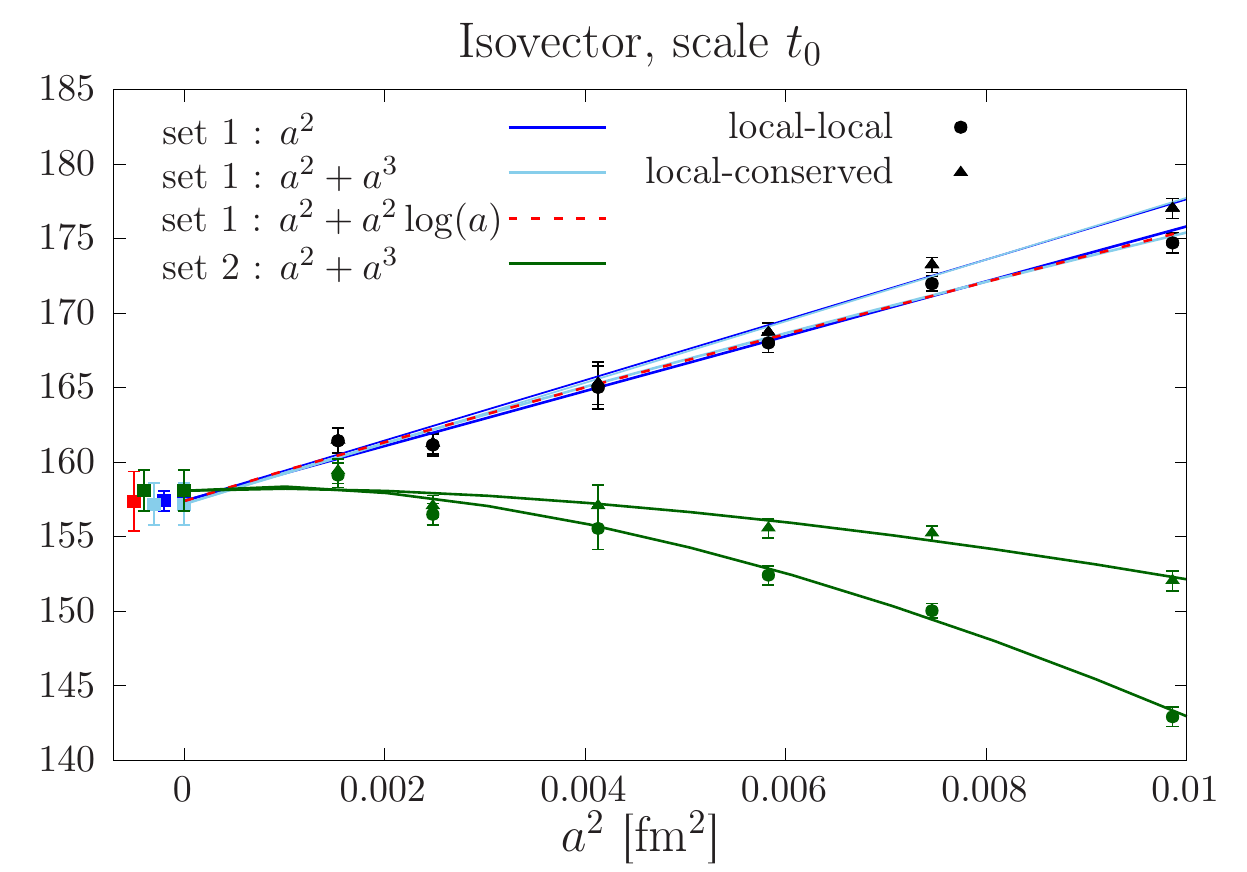} 
	
	\caption{Continuum extrapolation for the isovector quark contribution at the SU(3)$_{\rm f}$-symmetric point.
	Left: using $f_{\pi}$-rescaling. 
	Right: with $t_0$ to set the scale. The blue and green points correspond to the two different sets of improvement coefficients (see Section~\ref{sec:lattice}). For clarity, the extrapolated results have been shifted to the left.}
	\label{fig:SU3_cont}
\end{figure}

In a second step, we extrapolate both discretizations of the
correlation function to a common continuum limit, using data at all
six lattice spacings and assuming a polynomial in the lattice spacing,
\begin{equation}
\awinany{}^{,\alpha}(X_a,X_{\pi}^{\star}) = \awinany(0,X_{\pi}^{\star}) \left( 1 + \beta^{(\alpha)}_2 \, X_a^2 + \beta^{(\alpha)}_3 \, X_a^3\right) .
\label{eq:ansatzSU3}
\end{equation}
The two data sets obtained using the two different sets of improvement
coefficients are fitted independently. The results are displayed in
Fig.~\ref{fig:SU3_cont} for two cases: either applying $f_{\pi}$-rescaling (left panel) or using $t_0$ to set the scale (right panel). 
For Set~1 of improvement coefficients, we observe a remarkably linear
behavior over the whole range of lattice spacings, whether
$f_{\pi}$-rescaling is applied or not. The second set of improvement
coefficients (Set~2) leads to some visible curvature, but the continuum
limit is perfectly compatible provided that lattice artefacts of order
$a^3$ are included in the fit. 

We also tested the possibility of logarithmic corrections assuming the ansatz
\begin{equation}
  \awinany{}^{,\alpha}(X_a,X_{\pi}^{\star}) =
  \awinany(0,X_{\pi}^{\star}) \left( 1 + \beta^{(\alpha)}_2 \, X_a^2 + \epsilon^{(\alpha)} \, X_a^2 \log X_a\right) ,
\end{equation}
which is shown as the red symbol and red dashed curve in
Fig.~\ref{fig:SU3_cont}. The result is again compatible with the naive
$a^2$ scaling, albeit with larger error. We conclude that logarithmic
corrections are too small to be resolved in the data. 
We also remark that it is difficult to judge the quality of the continuum extrapolation based solely on the relative size of discretization effects between our coarsest and finest lattice spacing, as this measure strongly depends on the definition of the improvement coefficients. 

We tested the modification of the continuum extrapolation via 
$X_a^2 \rightarrow (\alpha_\mathrm{s}(1/X_a))^{\hat{\Gamma}} X_a^2$ as proposed 
in Refs.~\cite{Husung:2019ytz,Husung:2021mfl} for $\awinisovec$ and
$\awinisosca{}^{,c\!\!\!/}$ in our preferred setup, using $f_\pi$-rescaling
and set 1 of improvement coefficients. 
The strong coupling constant $\alpha_\mathrm{s}$ has been obtained 
from the three-flavor $\Lambda$ parameter of Ref.~\cite{Bruno:2017gxd}.
Several choices of $\hat{\Gamma}$ in the range from $0.76$ to $3$ were tested. 
The curvature that is introduced by this modification, especially for larger
values of $\hat{\Gamma}$, would lead to larger values of $\awin$ in the continuum
limit. However, such curvature is not supported by the data, as indicated 
by a deterioration of the fit quality when $\hat{\Gamma}$ is increased. 
Therefore, only small weights would be assigned to such fits in 
our model averaging procedure, where the modification has not been included.
\subsection{Results for the isospin and flavor decompositions \label{sec:resIorFdecomp}}

Having studied the continuum limit at the SU(3)$_{\rm f}$-symmetric point,
we are ready to present the result of the extrapolation to the physical point. The charm quark contribution is not included here and will be considered separately in Section~\ref{sec:charm}. 

\begin{figure}[t]
	\includegraphics*[width=0.46\linewidth]{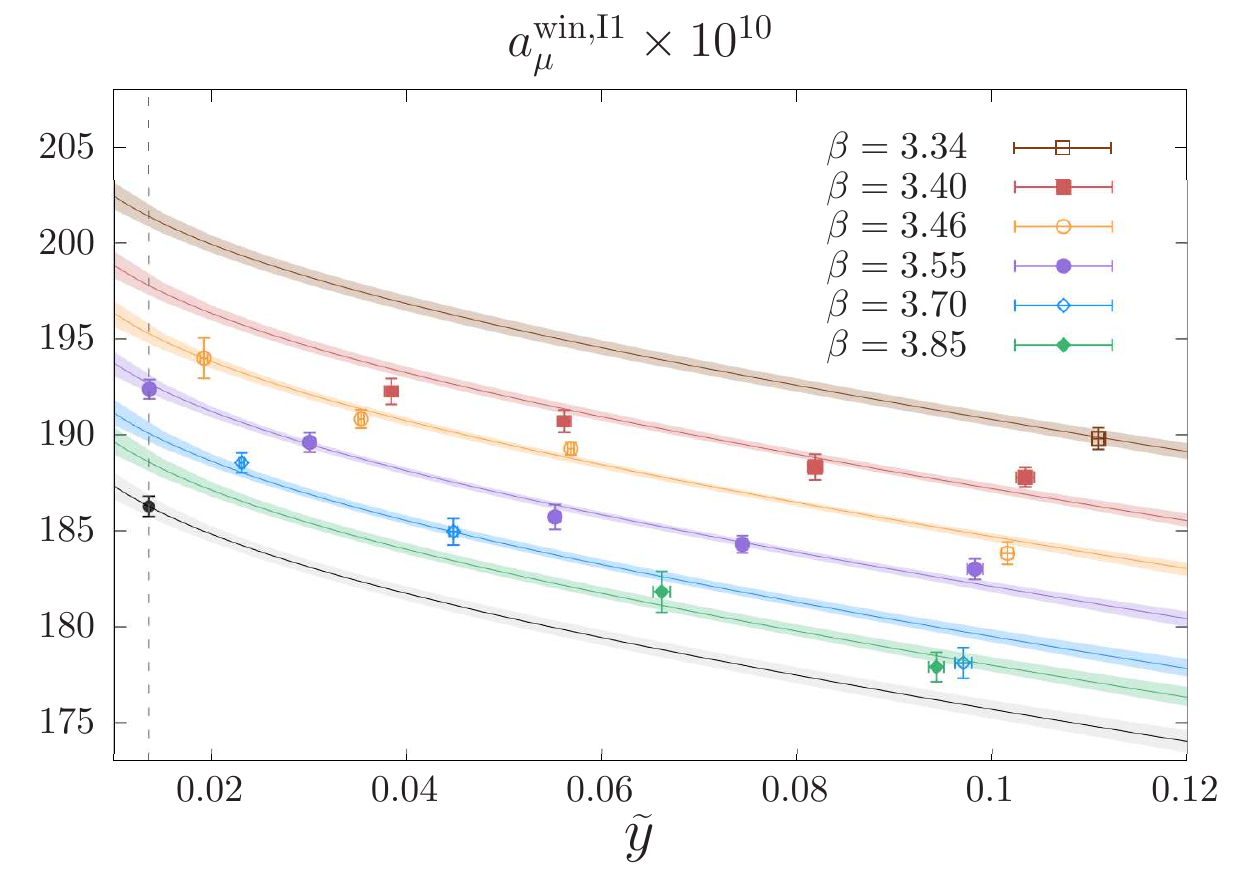} 
	\includegraphics*[width=0.46\linewidth]{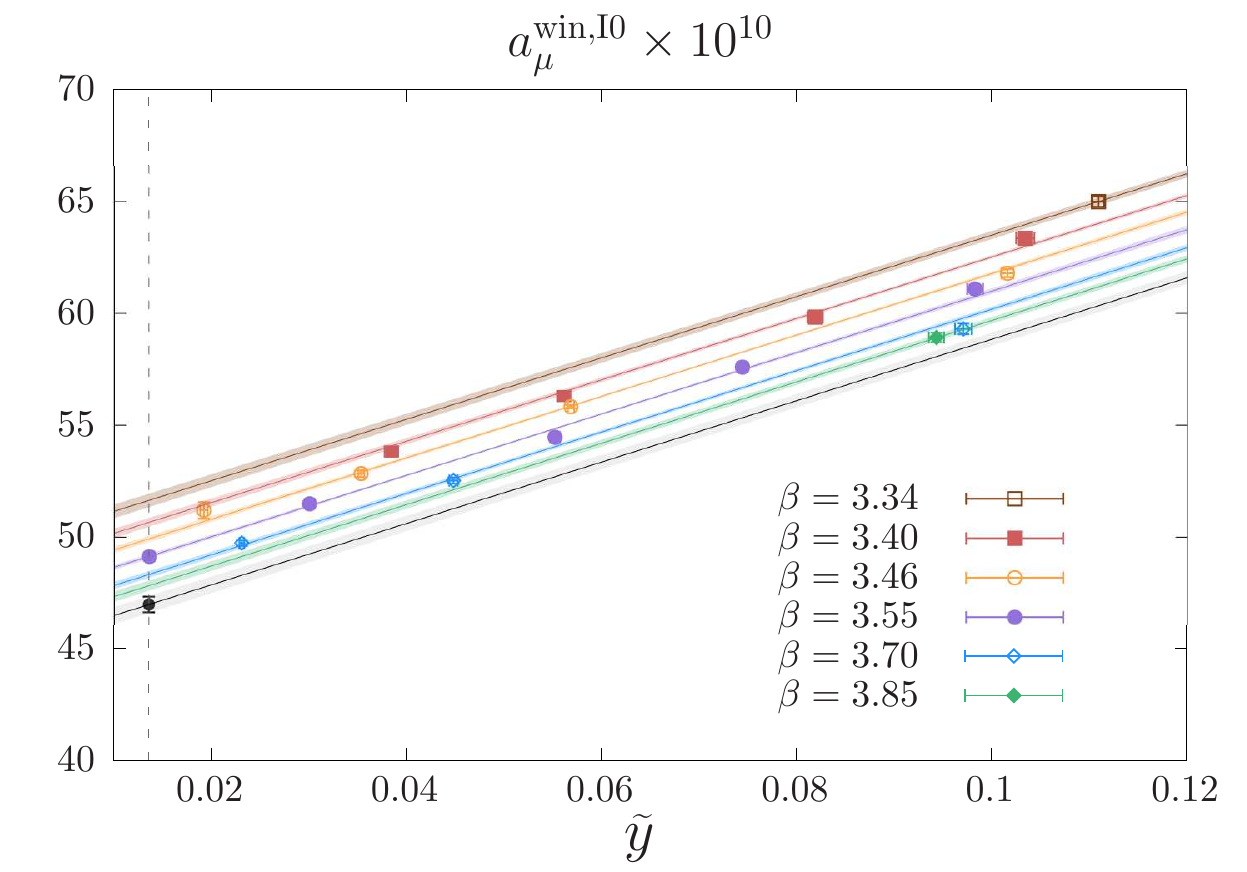} 
	\caption{Left: one typical extrapolation of the isovector contribution using $f_{\rm ch}(\widetilde{y}) = 1/\widetilde{y}$. The data corresponds to the local-conserved discretization of the correlator using the set 1 of improvement coefficients. Error bands are the results from the fit for each of the six lattice spacings. The black line is the chiral extrapolation in the continuum limit. The black point is the result at the physical point. Right: same for the isoscalar contribution but using $f_{\rm ch}(\widetilde{y}) = 0$.}
	\label{fig:winID_I1I0}
\end{figure}

For the isovector or light quark contribution we use the same set of
functional forms as in~\cite{Gerardin:2019rua} $f_{\rm ch}(X_{\pi}) = \{ \log X_{\pi}\,; X_{\pi}^2\,;1/X_{\pi}\,; X_{\pi} \log X_{\pi} \}$. 
The data shows some small curvature close to the physical pion mass.
Thus, the variation $f_{\rm ch}=0$ is excluded as it would significantly undershoot our ensemble at the physical pion mass (E250).
We use Set~1 of improvement coefficients as our preferred choice and
will use Set~2 only as a crosscheck. A typical extrapolation using
$f_{\rm ch}(\widetilde{y}) = 1/\widetilde{y}$ without any cut in the data is shown in the left panel of Fig.~\ref{fig:winID_I1I0}. 
We find that the specific functional form of $f_{\rm ch}$ has much
less impact on the extrapolation as compared to the inclusion of higher-order lattice artefacts. 
For the isoscalar and strange quark contributions, we restrict ourselves
to functions that are not singular in the chiral limit: $f_{\rm ch}(X_{\pi})
= \{ 0\,; X_{\pi}^2\,; X_{\pi} \log X_{\pi} \}$. Again, the extrapolation using $f_{\rm ch}(\widetilde{y}) = \widetilde{y} \log \widetilde{y}$ 
with $\delta \neq 0$ and without any cut in the
data is shown in the right panel of Fig.~\ref{fig:winID_I1I0}. 

Using the fit procedure described above, the AIC estimator defined in \eq{eq:akaike_mean} leads to
the following results for the isovector ($I=1$) and the isoscalar contribution, charm excluded,
\begin{align}
\awinisovec &= (186.30 \pm 0.75_{\stat} \pm 1.08_{\syst}) \times 10^{-10}	\label{res:I1} \,, \\
\awinisosca{}^{,c\!\!\!/} &= (47.41 \pm 0.23_{\stat} \pm 0.29_{\syst}) \times 10^{-10}\,, 	\label{res:I0}
\end{align}
where the first error is statistical and the second is the
systematic error from the fit form used to extrapolate our data to the
physical point. 
In Table~\ref{tab:der} , we also provide the derivatives
\begin{equation}
X \frac{\partial \awinany}{ \partial X } \,, \quad X \in \{ m_{\pi}, m_K, f_{\pi}, f_K \}  \,, \quad {\rm f} \in \{ \mathrm{I1}, \mathrm{I0} \} \,,
\label{eq:derwin}
\end{equation}
to translate our result to a different iso-symmetric scheme.

We also note that both discretizations of the vector correlator yield perfectly compatible results. For the isovector contribution, and in units of $10^{-10}$, we obtain $186.14(0.87)_{\stat}(1.29)_{\syst}$ for the local-local discretization and $186.47(0.79)_{\stat}(0.79)_{\syst}$ for the local-conserved discretization, with a correlated difference of $-0.33(0.72)$. For the isoscalar contribution, we find $47.39(0.24)_\stat(0.36)_\syst$ for the local-local discretization and $47.43(0.20)_\stat(0.19)_\syst$ for the local-conserved discretization, with a correlated difference of $-0.04(0.10)$.

As an alternative to the fit weights given by Eq.\ (\ref{eq:Akweight}), we
have tried applying the weight factors used in Ref.~\cite{Borsanyi:2020mff}; see
the footnote below Eq.\ (\ref{eq:Akweight}).  While  a major
change occurs in the subset of fits that dominate the weighted average, the
results do not change significantly. In particular, the central value
of the isovector contribution changes by no more than half a standard
deviation.

\begin{table}[b]
  \caption{Derivatives of the window quantity $a_\mu^{\rm win}$ (in units of $10^{-10}$),
    for both the isovector and isoscalar contributions, as defined by \eq{eq:derwin}.}
\vskip 0.1in
\begin{tabular}{l@{\hskip 01em}c@{\hskip 01em}c@{\hskip 01em}c@{\hskip 01em}c}
	\hline
$X$	&	$m_{\pi}$	&	$m_{\rm K}$ 	&	$f_{\pi}$	&	$f_{\rm K}$	\\
\hline
I1	&	$-7(5)$	&	$-11(7)$	&	$-66(84)$		&	7(5)			\\
I0	&	2(1)		&	$-34(2)$	&	$-29(9)$		&	25(2) 		\\
\hline
 \end{tabular} 
\label{tab:der}
\end{table} 
  
\begin{figure}[t]
	\includegraphics*[width=0.478\linewidth]{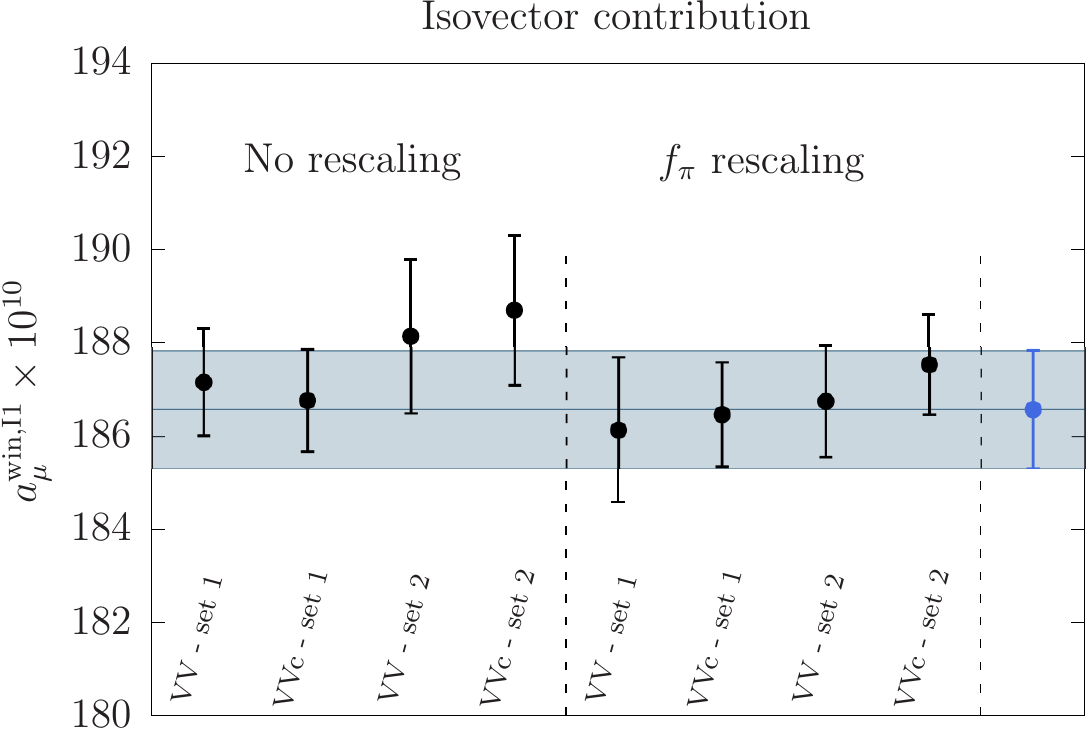}	 \quad
	\includegraphics*[width=0.47\linewidth]{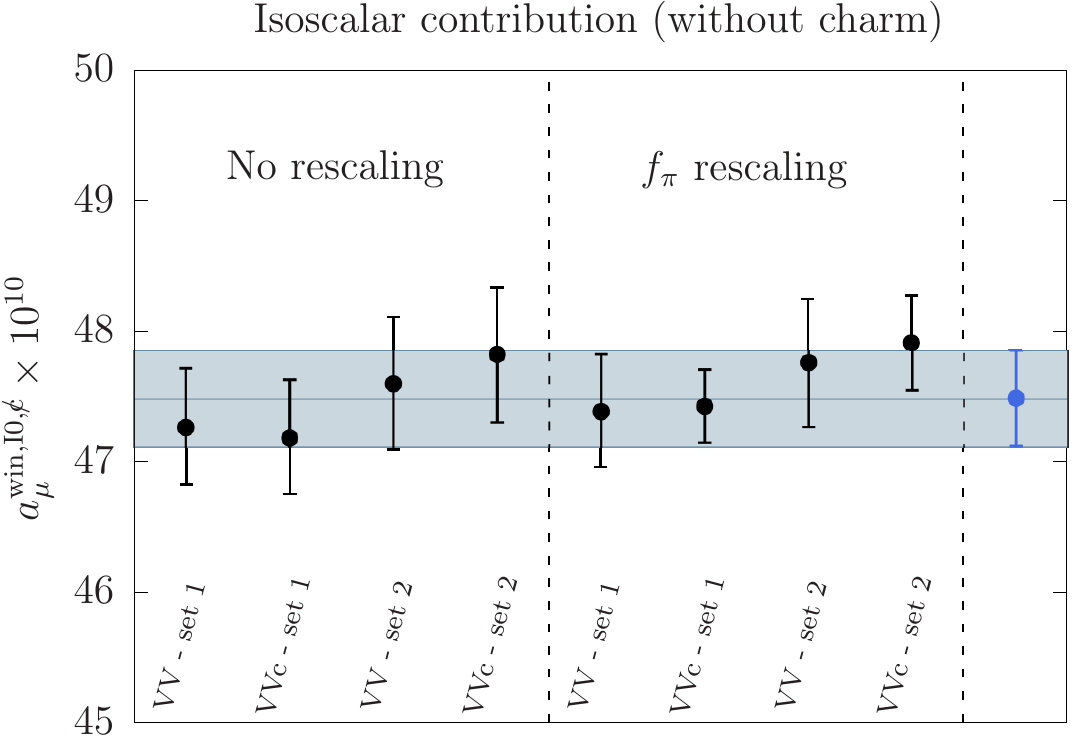}	
	\caption{Comparison of the isovector and isoscalar
contributions (without the charm) using different variations (either
using $f_\pi$ or $t_0$ to set the scale, and with both sets of
improvement coefficients). The blue point is our final estimate
obtained from the rescaling method with the set 1 of improvement
coefficients. }
	\label{fig:check}
\end{figure}

Finally, we have also performed an extrapolation to the physical point
using the second set of improvement coefficients. Since our study at
the SU(3)$_{\rm f}$-symmetric point shows curvature in the data, we
exclude those continuum extrapolations that are only quadratic in the lattice spacing. The other variations are kept identical to those used for the first set. The results are slightly larger but compatible within one standard deviation. A comparison between the two strategies to set the scale and the two sets of improvement coefficients is shown in Fig.~\ref{fig:check} for both the isovector and isoscalar contributions.

\begin{figure}[t]
	\includegraphics*[width=0.46\linewidth]{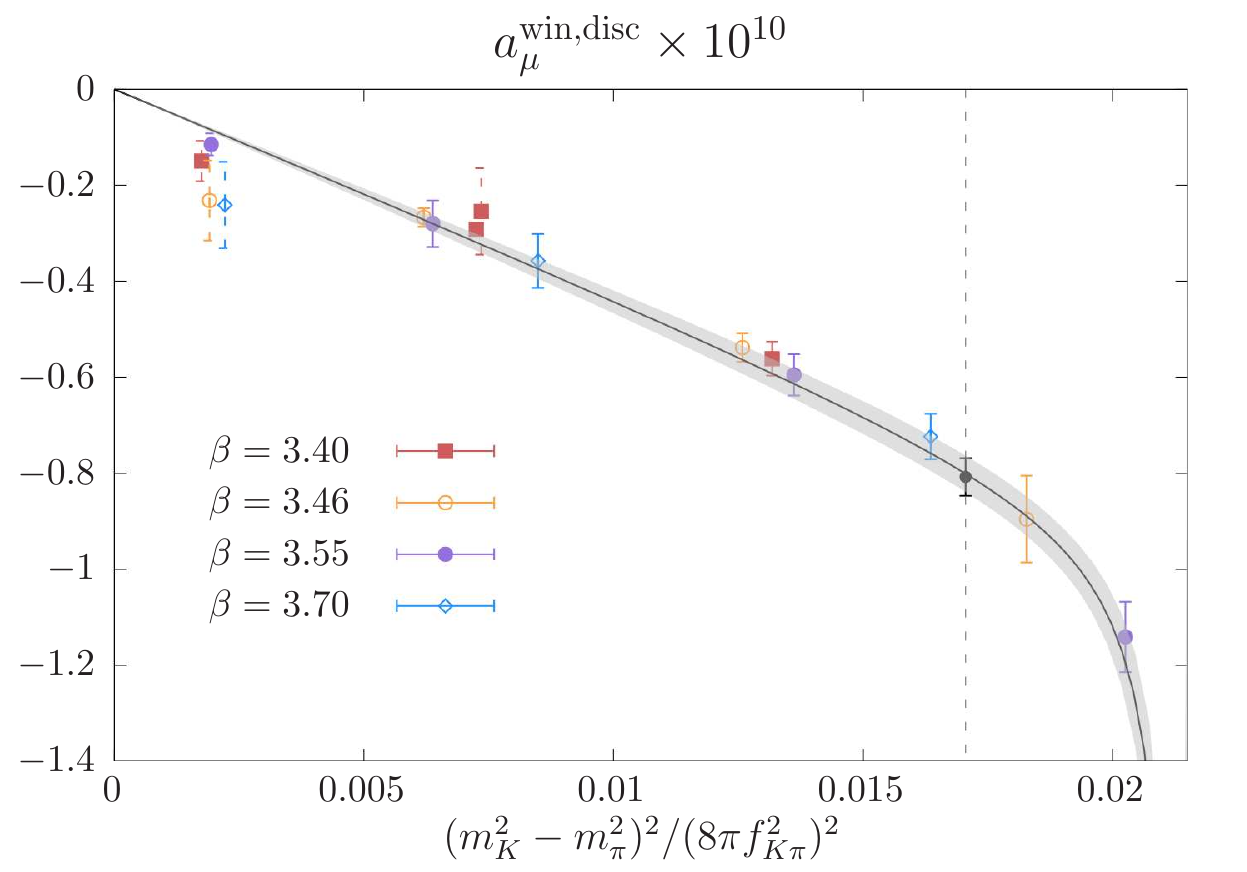} 
	\caption{Extrapolation to the physical point for the
quark-disconnected contribution using \eq{eq:extrapD}. The vertical dashed line represents
the physical point in our iso-symmetric QCD setup. The black point is
the result of the extrapolation, and the grey band represents the
extrapolation to the continuum limit with $X_K=X_K^{\star}$. Points with dashed error bars are not included in the fit.}
	\label{fig:winID_cd}
\end{figure}

In order to facilitate comparisons with other lattice collaborations, we also present
results for the light, strange and disconnected contributions separately. 
For the light and strange-quark connected contributions, we obtain 
\begin{align}
\awinl  &= (207.00 \pm 0.83_{\stat} \pm 1.20_{\syst} ) \times 10^{-10}, \label{res:ud}  \\
\awins &= (27.68 \pm 0.18_{\stat} \pm 0.22_{\syst} ) \times 10^{-10}. \label{res:s} 
\end{align}
For the disconnected contribution, the correlation function is very
precise in the time range relevant for the intermediate window, and a simple sum over lattice points is used to evaluate \eq{def:ID}. The data are corrected for finite-size effects using the method described in Section~\ref{sec:fse}.
Since our ensembles follow a chiral trajectory at fixed bare average
quark mass, we can consider $\awind$ as being, to a
good approximation, a function of the SU(3)$_{\rm f}$-breaking variable 
$\Delta_2 = \{ 8t_0(m_K^2-m_\pi^2), \; (m_K^2 - m_{\pi}^2)/(8\pi f_{K\pi}^2)\}$ (respectively for Strategy~1 and~2),
with the additional constraint that the disconnected contribution vanishes quadratically in
$\Delta_2$ for $\Delta_2\to 0$.
We apply the following ansatz  
\begin{multline}
  \label{eq:extrapD}
\awind(X_a,X_{\pi},X_{K}) = \Delta_2^2  \left( \alpha 
+ \gamma_0 \left( X_K - X_K^{\rm phys} \right)  + \beta_2 X_a^2   \right) 
\\  + \gamma_1 \left( \frac{1}{X_K^{\rm phys} - \Delta_2} - \frac{\Delta_2}{ (X_K^{\rm phys})^2 } - \frac{1}{X_K^{\rm phys}} \right)  .
\end{multline}
The ensembles close to the SU(3)$_f$ symmetric point ($m_\pi \approx 350$~MeV) are affected by significant FSE corrections and are not included in the fit. We obtain for the disconnected contribution
\begin{equation}
\awind = (-0.81 \pm 0.04_{\stat}  \pm 0.08_{\syst} ) \times 10^{-10} \label{res:d} \,,
\end{equation}
and the extrapolation is shown in Fig.~(\ref{fig:winID_cd}). The extrapolation using $t_0$ to set the scale shows less curvature close to the physical point. We use half the difference between the two extrapolations as our estimate for the systematic error. 
It is worth noting that the value for the intermediate window represents roughly 6\% of the total contribution to $\amud$.
As a crosscheck, we note that using Eqs.~(\ref{res:ud}), (\ref{res:s}) and (\ref{res:d}) we would obtain 
$\awinisosca{}^{,c\!\!\!/} = (47.57 \pm 0.20_{\stat} \pm 0.26_{\syst}) \times 10^{-10}$, in good agreement with \eq{res:I0}.

\section{The charm quark contribution}
\label{sec:charm}

In our calculation, charm quarks are introduced in the valence sector only.
A model estimate of the resulting quenching effect is provided in Appendix~\ref{sec:Qcharm}.
The method used to tune the mass of the charm quark has previously been
described in Ref.~\cite{Gerardin:2019rua} and has been applied to additional
ensembles in this work. We only sketch the general strategy here, referring the
reader to Ref.~\cite{Gerardin:2019rua} for further details. For each gauge
ensemble, the mass of the ground-state $c\bar{s}$ pseudoscalar meson is
computed at four values of the charm-quark hopping parameter. Then the value of
$\kappa_c$ is obtained by linearly interpolating the results in $1/\kappa_c$ to
the physical $D_s$ meson mass $m_{D_s} = 1968.35(0.07)$~MeV~\cite{Tanabashi:2018oca}.
We have checked that using either a quadratic fit or a linear fit in $\kappa_c$
leads to identical results at our level of precision. The results for all
ensembles are listed in the second column of Table~\ref{tab:charm}.

The renormalization factor $\hat{Z}_V^{(c)}$ of the local vector current has
been computed non-perturbatively on each individual ensemble by imposing the
vector Ward-identity using the same setup as in Ref.~{\cite{Gerardin:2018kpy}},
but with a charm spectator quark. To propagate the error from the tuning of
$\kappa_c$, both $\hat{Z}_V^{(c)}$ and $\awinc$ are computed at three values of
$\kappa$ close to $\kappa_c$. In the computation of correlation functions, the
same stochastic noises are used to preserve the full statistical correlations.
For both quantities, we observe a very linear behavior and a short
interpolation to $\kappa_c$ is performed. The systematic error introduced by
the tuning of $\kappa_c$ is propagated by computing the discrete derivatives of
both observables with respect to $\kappa_c$ (second error quoted in
Table~\ref{tab:charm}). This systematic error is considered as uncorrelated
between different ensembles.

\begin{figure}[t]
	\includegraphics*[width=0.46\linewidth]{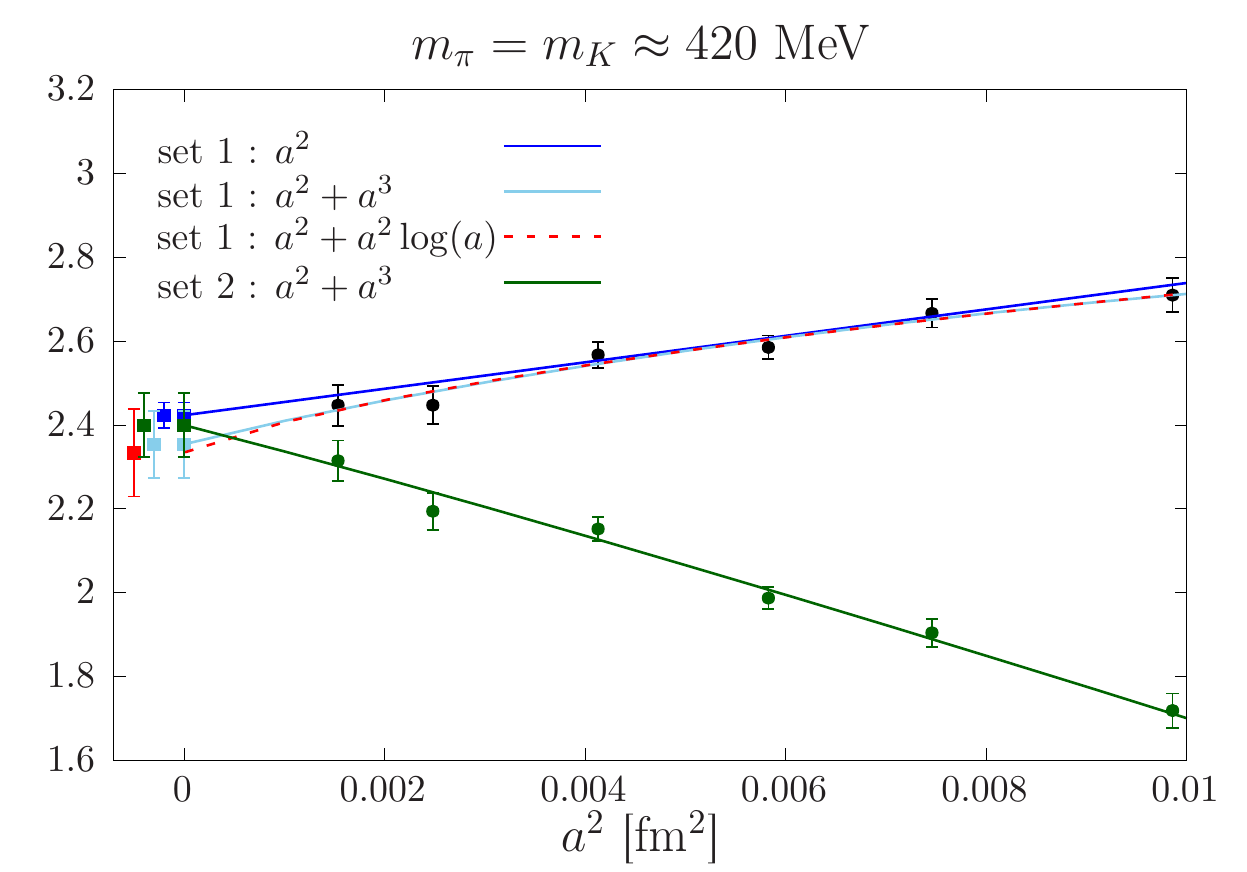}
	\includegraphics*[width=0.46\linewidth]{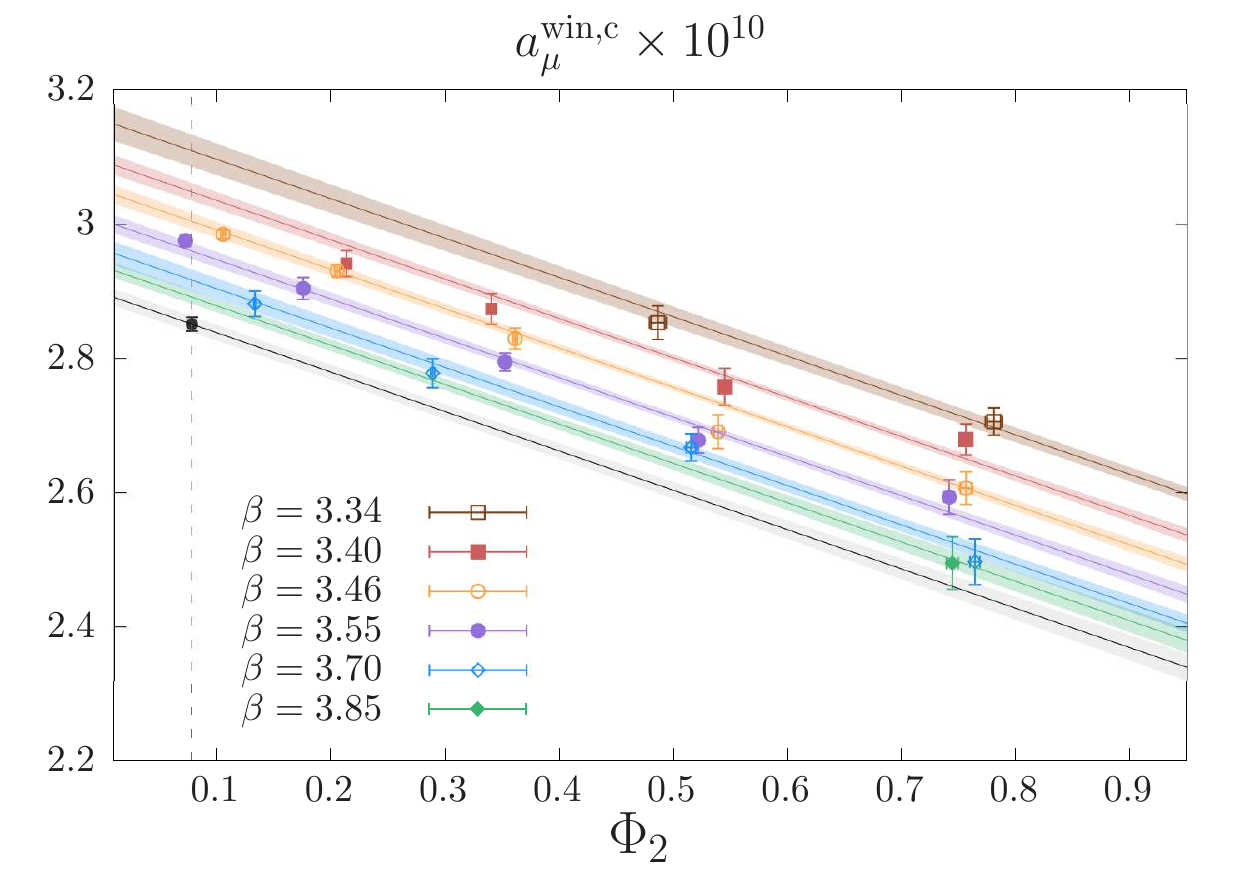}	 \qquad
	\caption{Left panel: study of the continuum extrapolation of the charm
quark contribution to $a_\mu^{\rm win}$ at the SU$(3)_{\rm f}$-symmetric point using the
local-conserved discretization of the correlation function. The black and green
points are obtained using two independent sets of improvement coefficients, as
explained in Section~\ref{sec:improvement}. Right panel: Example of a typical
extrapolation to the physical point of the charm-quark contribution. The error
from the scale setting, which is highly correlated between ensembles, is not
shown. The plain lines are obtained from the fit function (\ref{eq:extrapC})
without any cut in the pion mass.}
	\label{fig:charm}
\end{figure}

From ensembles generated with the same bare parameters but with different
spatial extents (H105/N101 or H200/N202), it is clear that FSE are negligible
in the charm-quark contribution. As in our previous
work~\cite{Gerardin:2019rua}, the local-local discretization exhibits a long
continuum extrapolation with discretization effects as large as 70\% between
our coarsest lattice spacing and the continuum limit, compared to only 12\% for
the local-conserved discretization. Thus, we discard the local-local
discretization from our extrapolation to the physical point, which assumes the
functional form
\begin{multline}
\awinc(X_a,X_{\pi}, X_K) = \awinc(0,X_{\pi}^{\exp},X_{K}^{\exp}) + \beta_2 \, X_a^2 + \beta_3 \, X_a^3 + \delta \, X_a^2 X_{\pi} + \beta_4 \, X_a^2 \log(X_a) \\ +  \gamma_0 \left( X_K - X_K^{\rm phys} \right)   + \gamma_1 \, \left( X_{\pi} - X_{\pi}^{\exp} \right)   \,.
\label{eq:extrapC}
\end{multline}
Lattice artefacts are described by a polynomial in 
$X_{a} =  a/ \sqrt{ t_0^{\rm sym} }$
and a possible logarithmic term is included;
recall that $t_0^{\rm sym}$ denotes the value of the flow observable at the SU(3)$_{\rm f}$-symmetric point.
Only the set of proxies $X_{\pi} = \phi_2$ and $X_{K} = \phi_4$ is used. The
light-quark dependence shows a very flat behavior, and a good $\chi^2/\dof =
0.9$ is obtained without any cut in the pion mass. The corresponding
extrapolation is shown on the right panel of Fig.~\ref{fig:charm}. \\

Before quoting our final result, we provide strong evidence that our continuum
extrapolation is under control by looking specifically at the SU(3)$_{\rm f}$-symmetric
point where six lattice spacings are available. As for the isovector
contribution, we use \eq{eq:extrapC} to correct for the small pion-mass
mistuning at the SU(3)$_{\rm f}$-symmetric point. The data are interpolated to
a single value of $X_{\pi}^*$ using the same strategy as in
\eq{eq:SU3cor}.
Those corrected points are finally extrapolated to the continuum limit using
the ansatz~(\ref{eq:ansatzSU3}). The result is shown in the left panel of
Fig.~\ref{fig:charm} for the two sets of improvement coefficients of the vector
current. Again, excellent agreement is observed between the two data sets.
Even for the charm-quark contribution, we observe
very little curvature when using the set~1 of improvement coefficients.

Having confirmed that our continuum extrapolation is under control, we quote
our final result for the charm contribution obtained using the ansatz~(\ref{eq:extrapC}).
Using \eq{eq:akaike_mean}, the AIC analysis described
above leads to 
\begin{equation}
\awinc = (2.89 \pm 0.03_{\stat} \pm 0.03_{\syst} \pm 0.13_{\rm scale}) \times 10^{-10} \,,
\label{res:charm}
\end{equation}
where variations include cuts in the pion masses and in the lattice spacing,
and fits where the parameters $\beta_3$, $\beta_4$ and $\delta$ have
been either switched on or off.

\section{Isospin breaking effects}
\label{sec:IB}

As discussed in the previous Sections~\ref{sec:lattice} and \ref{sec:isoQCD},
our computations are performed in an isospin-symmetric setup,
neglecting the effects due
to the non-degeneracy of the up- and down-quark masses and QED. At the
percent and sub-percent level of precision it is, however,
necessary to consider the impact of isospin-breaking effects. To estimate the
latter, we have computed $\awin$ in QCD+QED on a subset of our
isospin-symmetric ensembles using the
technique of Monte Carlo
reweighting~\cite{Ferrenberg:1988yz,Duncan:2004ys,Hasenfratz:2008fg,Finkenrath:2013soa,deDivitiis:2013xla}
combined with a leading-order perturbative expansion of QCD+QED around isosymmetric QCD
in terms of the electromagnetic coupling $e^{2}$ as well as the shifts in the
bare quark masses $\Delta m_u,\Delta m_d,\Delta m_s$~\cite{deDivitiis:2013xla,Risch:2021hty,Risch:2019xio,Risch:2018ozp,Risch:2017xxe}.
Consequently, we must evaluate additional diagrams that represent the perturbative
quark mass shifts as well as the interaction between quarks and photons.
We make use of non-compact lattice QED and regularize the manifest IR divergence with
the QED\textsubscript{L} prescription~\cite{Hayakawa:2008an}, with the boundary
conditions of the photon and QCD gauge fields chosen in
accordance~\cite{Risch:2018ozp}. We characterize the physical point of QCD+QED
by the quantities
$m_{\pi^{0}}^{2}$, $m_{K^{+}}^{2} + m_{K^{0}}^{2} - m_{\pi^{+}}^{2}$, $m_{K^{+}}^{2}-m_{K^{0}}^{2}-m_{\pi^{+}}^{2}+m_{\pi^{0}}^{2}$ and the fine-structure constant
$\alpha$~\cite{Risch:2021hty}. The first three quantities are inspired by
leading-order chiral perturbation theory including leading-order mass and
electromagnetic isospin-breaking corrections~\cite{Neufeld:1995mu}, and
correspond to proxies for the average light-quark mass, the strange-quark mass,
and the light-quark mass splitting. As we consider leading-order effects only,
the electromagnetic coupling does not renormalize~\cite{deDivitiis:2013xla},
i.e.\ we may set $e^{2}=4\pi\alpha$.
The lattice scale is also affected by isospin breaking, which we however neglect at
this stage. Making use of the isosymmetric scale~\cite{Bruno:2016plf}, we match
$m_{\pi^{0}}^{2}$ and $m_{K^{+}}^{2} + m_{K^{0}}^{2} - m_{\pi^{+}}^{2}$ in both
theories on each ensemble and set
$m_{K^{+}}^{2}-m_{K^{0}}^{2}-m_{\pi^{+}}^{2}+m_{\pi^{0}}^{2}$ to its
experimental value.

We have computed the leading-order QCD+QED quark-connected
contribution to $\awin$ as well as the pseudoscalar meson
masses $m_{\pi^{0}}$, $m_{\pi^{+}}$, $m_{K^{0}}$ and $m_{K^{+}}$ required
for the hadronic renormalization scheme on the ensembles D450, N200, N451
and H102, neglecting quark-disconnected diagrams as well as isospin-breaking
effects in sea-quark contributions. The considered quark-connected diagrams
are evaluated using stochastic U$(1)$ quark sources with support on a single
timeslice whereas the all-to-all photon propagator in Coulomb gauge is
estimated stochastically by means of $Z_{2}$ photon sources. Covariant
approximation averaging~\cite{Shintani:2014vja} in combination with the
truncated solver method~\cite{Bali:2009hu} is applied to reduce the
stochastic noise. We treat the noise problem of the vector-vector
correlation function at large time separations by means of a reconstruction
based on a single exponential function. A more detailed description of the
computation can be found in Refs.~\cite{Risch:2021nfs,Risch:2021hty,Risch:2019xio}.
The renormalization procedure of the local vector current in the QCD+QED
computation is based on a comparison of the local-local and the conserved-local
discretizations of the vector-vector correlation function and hence differs
from the purely isosymmetric QCD calculation~\cite{Gerardin:2018kpy} described in Section~\ref{sec:improvement}. We
therefore determine the relative correction by isospin breaking in the
QCD+QED setup. For $f_{\pi}$-rescaling as introduced in
Section~\ref{sec:isoQCD}, isospin-breaking effects in the determination of
$f_{\pi}$ are neglected. We observe that the size of the relative first-order
corrections for $\awin$ is compatible on each ensemble and can in total be estimated as a $(0.3\pm0.1)\%$ effect.

\section{Final result and discussion}
\label{sec:discussion}

We first quote our final result $\awiniso$ in our iso-symmetric setup as defined in Section~\ref{sec:isoQCD}. Using the isospin decomposition, and combining Eqs.~(\ref{res:I1}), (\ref{res:I0}) and (\ref{res:charm}), we find
\begin{align}
\awinisovec &= (186.30 \pm 0.75_{\stat} \pm 1.08_{\syst}) \times 10^{-10}\,,	\\
\awinisosca =  \awinisosca{}^{,c\!\!\!/} + \awinc  &= (50.30 \pm 0.23_{\stat} \pm 0.32_{\syst}) \times 10^{-10} \,,\\
\awiniso = \awinisovec + \awinisosca &= (236.60 \pm 0.79_{\stat} \pm 1.13_{\syst} \pm 0.05_{\rm Q} ) \times 10^{-10}\,,
\label{eq:final}
\end{align}
where the first error is statistical, the second is the systematic
error, and the last error of $\awiniso$
is an estimate of the quenching effect of the charm quark derived in
Appendix~\ref{sec:Qcharm}.
Overall, this uncertainty has a negligible effect on the systematic error estimate.
The small bottom quark contribution has been neglected. For $\amu$,
this contribution has been computed in~\cite{Colquhoun:2014ica} and
found to be negligible at the current level of precision. 

As stressed in Section~\ref{sec:isoQCD}, our definition of the
physical point in our iso-symmetric setup is scheme dependent. To
facilitate the comparison with other lattice collaborations, the
derivatives with respect to the quantities used to define our
iso-symmetric scheme are provided in Table~\ref{tab:der}. They can be
used to translate from one prescription to another a posteriori.

One of the main challenges for lattice calculations of both $\ahvp$
and the window observable is the continuum extrapolation of the light
quark contribution, which dominates the results by far.
To address this specific point, we have used six lattice spacings in
the range [0.039,0.0993]~fm in our calculation, along with two
different discretizations of the vector current (see the discussion in
Section~\ref{sec:contextrap}). Although this work contains many
ensembles away from the physical pion mass, we observe only a mild
dependence on the proxy used for the light-quark mass. This
observation is corroborated by the fact that, in the model averaging
analysis, most of the spread comes from fits that differ in the
description of lattice artefacts rather than on the functional form
$f_{\rm ch}$ that describes the light-quark mass dependence. 

\begin{figure}[t]
	\includegraphics*[width=0.98\linewidth]{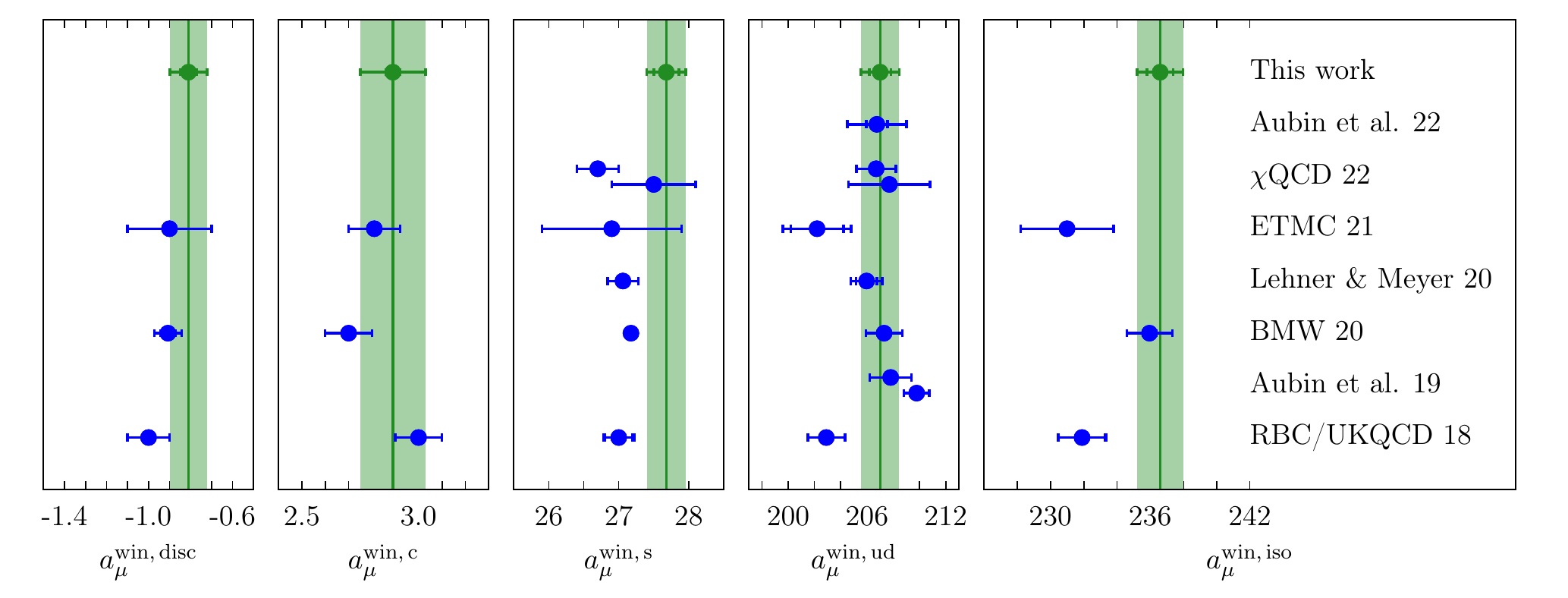}
	\caption{Comparison of our results (in units of $10^{-10}$)
with other lattice calculations \cite{Blum:2018mom, Aubin:2019usy,
Borsanyi:2020mff, Lehner:2020crt, Giusti:2021dvd, Wang:2022lkq,
Aubin:2022hgm} in isosymmetric QCD. The four panels
on the left show compilations of the individual quark-disconnected, charm,
strange and light quark contributions. The total result for $\awin$ in
the isosymmetric case is shown in the rightmost panel. Our results are
represented by green circles and vertical bands.}
	\label{fig:cmpwiniso}
\end{figure}

In Fig.~\ref{fig:cmpwiniso}, we compare our results in the isosymmetric
theory with other lattice calculations. Our estimate for $\awiniso$
agrees well with that of the BMW collaboration who quote $\awiniso =
236.3(1.4) \times 10^{-10}$ using the staggered quark
formulation~\cite{Borsanyi:2020mff}. However, our result is about
$2.3\sigma$ above the published value by the RBC/UKQCD collaboration,
$\awiniso = 232.0(1.5) \times 10^{-10}$, obtained using domain wall
fermions~\cite{Blum:2018mom}. It is also $1.7\sigma$ above the recent
estimate quoted by ETMC, based on the twisted-mass formalism
\cite{Giusti:2021dvd}, which reads $\awiniso = 231.0(2.8) \times
10^{-10}$. The difference with the latter two calculations can be
traced to the light-quark contribution $\awinl$, which is shown in the
second panel from the right. 
In this context, it is interesting to note that, apart from BMW, two
independent calculations using staggered quarks (albeit with a different
action as compared to the BMW collaboration) have quoted results for
$\awinl$~\cite{Aubin:2019usy,Aubin:2022hgm,Lehner:2020crt} that are
in good agreement with our estimate, as can be seen in
Fig.~\ref{fig:cmpwiniso}. 
The middle panel of the figure shows that our
estimate for the strange quark contribution is slighly higher compared
to other groups, but due to the relative smallness of $\awins$ this
cannot account for the difference between our result for $\awiniso$
and Refs.~\cite{Giusti:2021dvd} and~\cite{Blum:2018mom}.
Good agreement with the BMW, ETMC and RBC/UKQCD collaborations is
found for both the charm and quark-disconnected contributions.

If one accepts that most lattice estimates for the light-quark
connected contribution $\awinl$ have stabilized around
$\approx207\times10^{-10}$, one may search for an explanation why the
results by RBC/UKQCD \cite{Blum:2018mom} and ETMC
\cite{Giusti:2021dvd} are smaller by about 2\%. This is particularly
important since $\awinl$ contributes about 87\% to the entire
intermediate window observable.
One possibility is that the extrapolations to the physical point in
Refs. \cite{Blum:2018mom} and \cite{Giusti:2021dvd} are both quite
long. For instance, the minimum pion mass among the set of ensembles
used by ETMC is only about 220~MeV, while the result by RBC/UKQCD has
been obtained from two lattice spacings, i.e.\ 0.084\,fm and 0.114\,fm.
Further studies using additional ensembles at smaller pion mass and
lattice spacings are highly desirable to clarify this important issue.

\begin{figure}[t]
	\includegraphics*[width=0.55\linewidth]{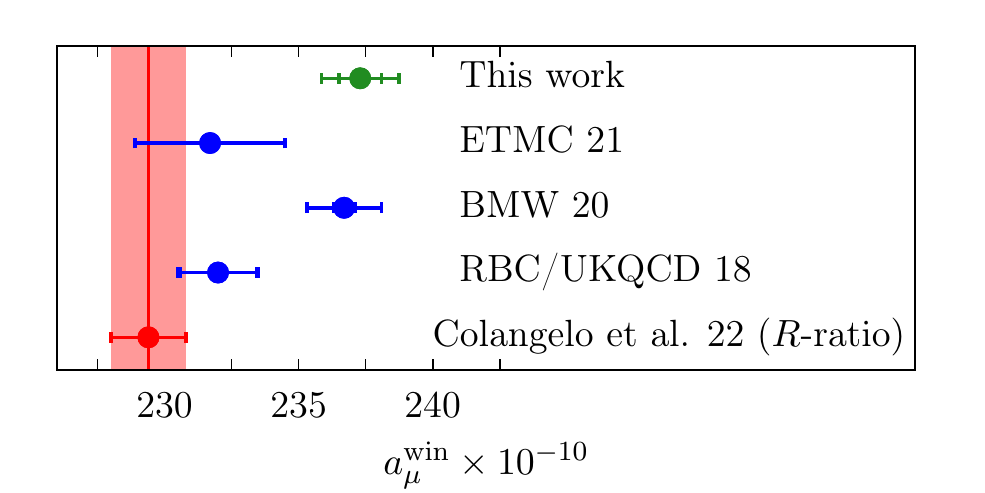}
	\caption{Comparison of our result for $\awin$ including
isospin-breaking corrections with the estimates by ETMC
\cite{Giusti:2021dvd}, BMW \cite{Borsanyi:2020mff} and RBC/UKQCD
\cite{Blum:2018mom}. The estimate based on the data-driven method of
Ref.~\cite{Colangelo:2022vok} is shown in red.}
	\label{fig:cmpwinfull}
\end{figure}

In order to compare our result with phenomenological determinations of
the intermediate window observable, we must correct for the effects of
isospin-breaking. Our calculation of isospin-breaking corrections,
described in Section \ref{sec:IB}, has been performed on a subset of
our ensembles and is, at this stage, lacking a systematic assessment
of discretization and finite-volume errors. Furthermore, only
quark-connected diagrams have been considered so far. To account for
this source of uncertainty, we double the error and thereby apply a relative
isospin-breaking correction of $(0.3\pm0.2)\%$ to $\awiniso$, which
amounts to a shift of $+(0.70\pm0.47)\times10^{-10}$. Thus, our final
result including isospin-breaking corrections is
\begin{equation}
   \awin= (237.30 \pm 0.79_{\stat} \pm 1.13_{\syst} \pm 0.05_{\rm Q} \pm0.47_{\rm IB} ) \times 10^{-10}\,.
\label{eq:finalfull}
\end{equation}
Adding all errors in quadrature yields $237.30(1.46)\times10^{-10}$
which corresponds to a precision of 0.6\%. A comparison with other
lattice calculations is shown in Fig.~\ref{fig:cmpwinfull}. Since
corrections due to isospin breaking are small, the same features are
observed as in the isosymmetric theory: while our result
agrees well with the published estimate from BMW
\cite{Borsanyi:2020mff}, it is larger than the values quoted by ETMC
\cite{Giusti:2021dvd} and RBC/UKQCD \cite{Blum:2018mom}. Our result
lies $3.9\sigma$ above the recent evaluation using the data-driven
method~\cite{Colangelo:2022vok}, which yields
$\awin=229.4(1.4)\times 10^{-10}$ and is shown in red in
Fig.~\ref{fig:cmpwinfull}.
Our result for $\awin$ is also consistent with the observation
that the central value of our 2019 result for the complete hadronic
vacuum polarization contribution~\cite{Gerardin:2019rua} lies higher
than the phenomenology estimate, albeit with much larger
uncertainties. In Ref. \cite{Ce:2022eix} we observed a similar, but statistically
much more significant enhancement in the hadronic running of the electromagnetic coupling,
$\Delta\alpha_{\mathrm{had}}(-Q^2)$ relative to the data-driven evaluation,
especially for $Q^2\lesssim3\,{\rm GeV}^2$. As pointed out at the end of
Section~\ref{sec:definitions}, the relative contributions from 
the three intervals of center-of-mass energy separated by 
$\sqrt{s}=600\,$MeV and $\sqrt{s}=900\,$MeV
are similar for $\awin$ and $\Delta\alpha_{\mathrm{had}}(-1{\rm GeV}^2)$, even though the respective weight functions
in the time-momentum representation are rather different.
The fact that the lattice determination is larger by more than three percent for both quantities,
in each case with a combined error of less than one percent,
suggests that a genuine difference exists at the level of the underlying spectral function, $R(s)/(12\pi^2)$,
between lattice QCD and phenomenology.

If one were to subtract the data-driven evaluation of $\awin$ from the
White Paper estimate \cite{Aoyama:2020ynm} and replace it by our
result in \eq{eq:finalfull}, the tension between the SM
prediction for $a_\mu$ and experiment would be reduced to
$2.9\sigma$. This observation illustrates the relevance of the window
observable for precision tests of the SM. Our findings also strengthen
the evidence supporting a tension between data-driven and lattice
determinations of $\ahvp$.

In our future work we will extend the calculation to other windows and
focus on the determination of the full hadronic vacuum polarization
contribution, $\ahvp$.

\section*{Acknowledgements}
Calculations for this project have been performed on the HPC clusters
Clover and HIMster-II at Helmholtz Institute Mainz and Mogon-II at
Johannes Gutenberg-Universität (JGU) Mainz, on the HPC systems
JUQUEEN, JUWELS and JUWELS Booster at J\"ulich Supercomputing Centre
(JSC), and on the GCS Supercomputers HAZEL HEN and HAWK at
H\"ochstleistungsrechenzentrum Stuttgart (HLRS).
The authors gratefully acknowledge the support of the Gauss Centre for
Supercomputing (GCS) and the John von Neumann-Institut für Computing
(NIC) for project HMZ21,  HMZ23 and HINTSPEC at JSC and project GCS-HQCD at HLRS.
This work has been supported by Deutsche Forschungsgemeinschaft
(German Research Foundation, DFG) through project HI 2048/1-2 (project
No.\ 399400745) and through the Cluster of Excellence ``Precision Physics,
Fundamental Interactions and Structure of Matter'' (PRISMA+ EXC
2118/1), funded within the German Excellence strategy (Project ID 39083149).
D.M.\ acknowledges funding by the
  Heisenberg Programme of the Deutsche Forschungsgemeinschaft (DFG, German
  Research Foundation) – project number 454605793.
The work of M.C.\ has been supported by the European Union's Horizon 2020
research and innovation program under the Marie Skłodowska-Curie Grant
Agreement No.\ 843134.
A.G. received funding from the Excellence Initiative of Aix-Marseille
University - A*MIDEX, a French \emph{Investissements d'Avenir}
programme, AMX-18-ACE-005 and from the French National Research Agency
under the contract ANR-20-CE31-0016. We are grateful to our colleagues
in the CLS initiative for sharing ensembles.

\newpage
\appendix
\section{Mistuning of the chiral trajectory } \label{sec:chirtraj}

The ensembles used in our work have been generated with a constant
bare average sea quark mass which differs from a constant renormalized
mass by $\mathrm{O}(a)$ cutoff effects. When the sum of the
renormalized quark masses is kept constant, the dimensionless
parameters $\phi_4$ and $y_{K\pi}$, which have been
introduced in Section \ref{sec:isoQCD} to define the chiral
trajectories towards the physical point, are constant to leading order
in chiral perturbation theory ($\chi$PT). Therefore, $\phi_4$ and
$y_{K\pi}$ cannot be constant across our set of ensembles
due to cutoff effects and higher-order effects from $\chi$PT.

We have to correct for the sources of mistuning of our ensembles with respect to the chiral trajectories of strategies 1 and 2. This can be done by parameterizing the dependence of our observables on $X_{K} \in \{y_{K\pi}, \phi_4\}$ in the combined chiral-continuum extrapolation. However, since the pion and kaon masses are not varied independently within our set of ensembles, the dependence on $\Delta X_K = X_K^{\rm phys} - X_K$ cannot be resolved reliably in our fits. A different strategy has to be employed to stabilize our extrapolation to the physical point.

Explicit corrections of the mistuning prior to the chiral extrapolation have been used in \cite{Bruno:2016plf} to approach the physical point at constant $\phi_4 = \phi_4^\mathrm{phys}$. These corrections are based on small shifts defined from the first order Taylor expansion of the quark mass dependence of lattice observables. The expectation value of a shifted observable is given by
\begin{align}
\langle \mathcal{O} \rangle \rightarrow \langle \mathcal{O}\rangle +\sum_{i=1}^{N_\mathrm{f}}\Delta m_{\mathrm{q},i}\frac{\mathrm{d} \langle\mathcal{O}\rangle}{\mathrm{d} m_{\mathrm{q}, i}}\,, \label{eq:shifted_observable}
\end{align}
with the $N_\mathrm{f}=3$ sea quark mass shifts $\Delta m_{\mathrm{q},i}$. 
Within this appendix, we work with observables and expectation values that
are defined after integration over the fermion fields, i.e.~the expectation 
values are taken with respect to the gauge configurations.
The total derivative of an observable with respect to the quark masses is decomposed via
\begin{align}
\frac{\mathrm{d} \langle\mathcal{O}\rangle}{\mathrm{d} m_{\mathrm{q}, i}} =
\left\langle \frac{\partial \mathcal{O}}{\partial m_{\mathrm{q}, i}} \right \rangle
- \left\langle \mathcal{O} \frac{\partial S}{\partial m_{\mathrm{q}, i}} \right\rangle
+  \left\langle\mathcal{O}\right\rangle \left\langle \frac{\partial S}{\partial 
	m_{\mathrm{q}, i}} \right\rangle\,. \label{eq:mass_shifts}
\end{align}
The partial derivative of an observable with respect to a quark mass of flavor $i$ captures the effect of shifts of valence quark masses. The second and third terms that contain the derivative of the action $S$ with respect to the quark masses account for sea quark effects. The chain rule is used to compute the derivatives of derived observables.

The chain rule relating the derivatives with respect to the quark masses to those with respect to the variables $X_j=X_\pi,X_K$ can be written
\begin{align}
\sum_{i=1}^{N_\mathrm{f}} n_i \; \frac{\mathrm{d} \langle{\cal O}\rangle}{\mathrm{d} m_{\mathrm{q}, i}}  & =
\sum_{j=\pi,K} \Delta_j(\vec n) \, \frac{\mathrm{d}\langle\mathcal{O}\rangle}{\mathrm{d}X_{j}},
\qquad \Delta_j(\vec n)  \equiv \sum_{i=1}^{N_\mathrm{f}} n_i\; \frac{\mathrm{d} X_{j}}{\mathrm{d} m_{\mathrm{q}, i}}
\end{align}
$ \forall\,\vec n = (n_1,n_1,n_3)$, the condition $n_1=n_2$ being imposed to remain in the isosymmetric theory.
In particular, if the direction of the vector $\vec n$ in the space of quark masses
is chosen such that $\Delta_\pi(\vec n)$ vanishes,
the following expression~\cite{Strassberger:2021tsu} for the derivative of an observable with respect to $X_K$ is obtained,
\begin{align}
	\frac{\mathrm{d}\langle\mathcal{O}\rangle}{\mathrm{d}X_{K}} =
\frac{1}{\Delta_K(\vec n)}	\sum_{i=1}^{N_\mathrm{f}} n_i \frac{\mathrm{d}
\langle\mathcal{O}\rangle}{\mathrm{d} m_{\mathrm{q}, i}}. 
\label{eq:obs_deriv_XK}
\end{align}
In \cite{Bruno:2016plf} the shifts $n_i$ have been chosen to be degenerate for all three sea quarks. In \cite{Strassberger:2021tsu} the same approach is taken at the $\mathrm{SU}(3)_{\rm f}$-symmetric point and $\vec{n} = (0, 0, 1)$ is used when $a\mql \neq a\mqs$. To stabilize the predictions for the derivatives, they are modeled as functions of lattice spacing and quark mass.

To improve the reliability of our chiral extrapolation, we have
determined the derivatives of $\awinl$ and $\awins$ with respect to
light and strange quark masses on a large subset of the ensembles in
Table~\ref{tab:sim}. Whereas the computation of the first term in
\eq{eq:mass_shifts} shows a good signal for the vector-vector
correlation function, the second and third term carry significant
uncertainties. In the case of $f_\pi$-rescaling, a non-negligible
statistical error that has its origin in ${\mathrm{d} f_\pi}/{\mathrm{d}
m_{\mathrm{q}, i}}$ enters the derivative of $\awin$. 

Our computation does not yet cover all ensembles in this work and has
significant uncertainties on some of the included ensembles. Moreover,
we have not computed the mass-derivative of $\awind$ that enters
$\awinisosca$. Therefore, we have decided not to correct our
observables prior to the global extrapolation but to determine the
coefficient $\gamma_0$ in \eq{eq:fit} instead. We do not aim
for a precise determination here but focus instead on the
determination of a sufficiently narrow prior width, in order to
stabilize the chiral-continuum extrapolation.

We compute the derivatives with respect to $X_{K}$ as specified in
\eq{eq:obs_deriv_XK} with the shift vector $\vec{n}$ chosen
such that $\Delta_\pi(\vec n)$ vanishes ensemble by ensemble, i.e.\
the shift is taken in a direction in the quark mass plane where
$X_\pi$ remains constant. The derivatives are therefore sensitive to
shifts in the kaon mass. A residual shift of $X_a$ is present at
the permille level. 

\begin{figure}[h!]
	\includegraphics*[width=0.46\linewidth]{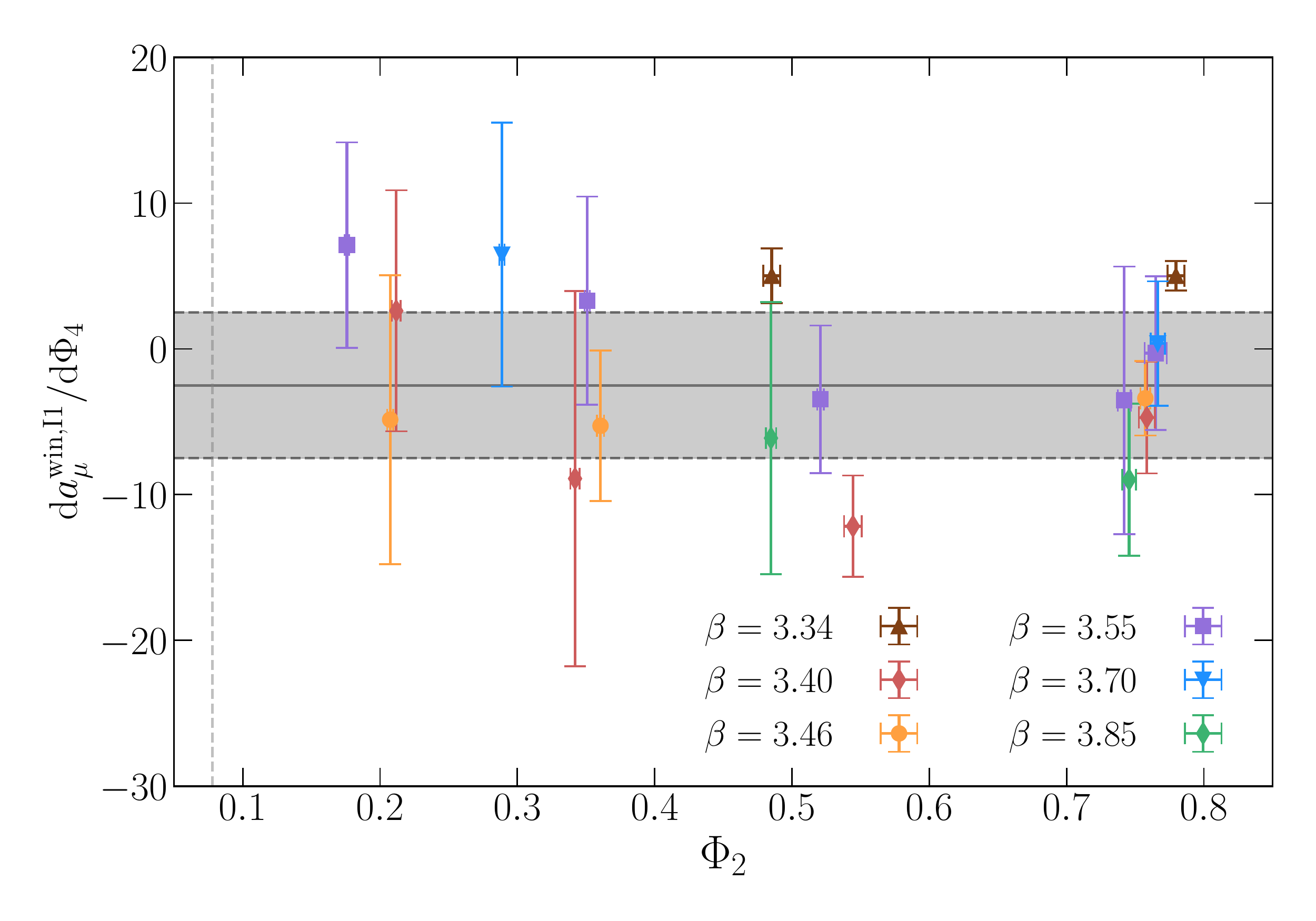}
	\includegraphics*[width=0.46\linewidth]{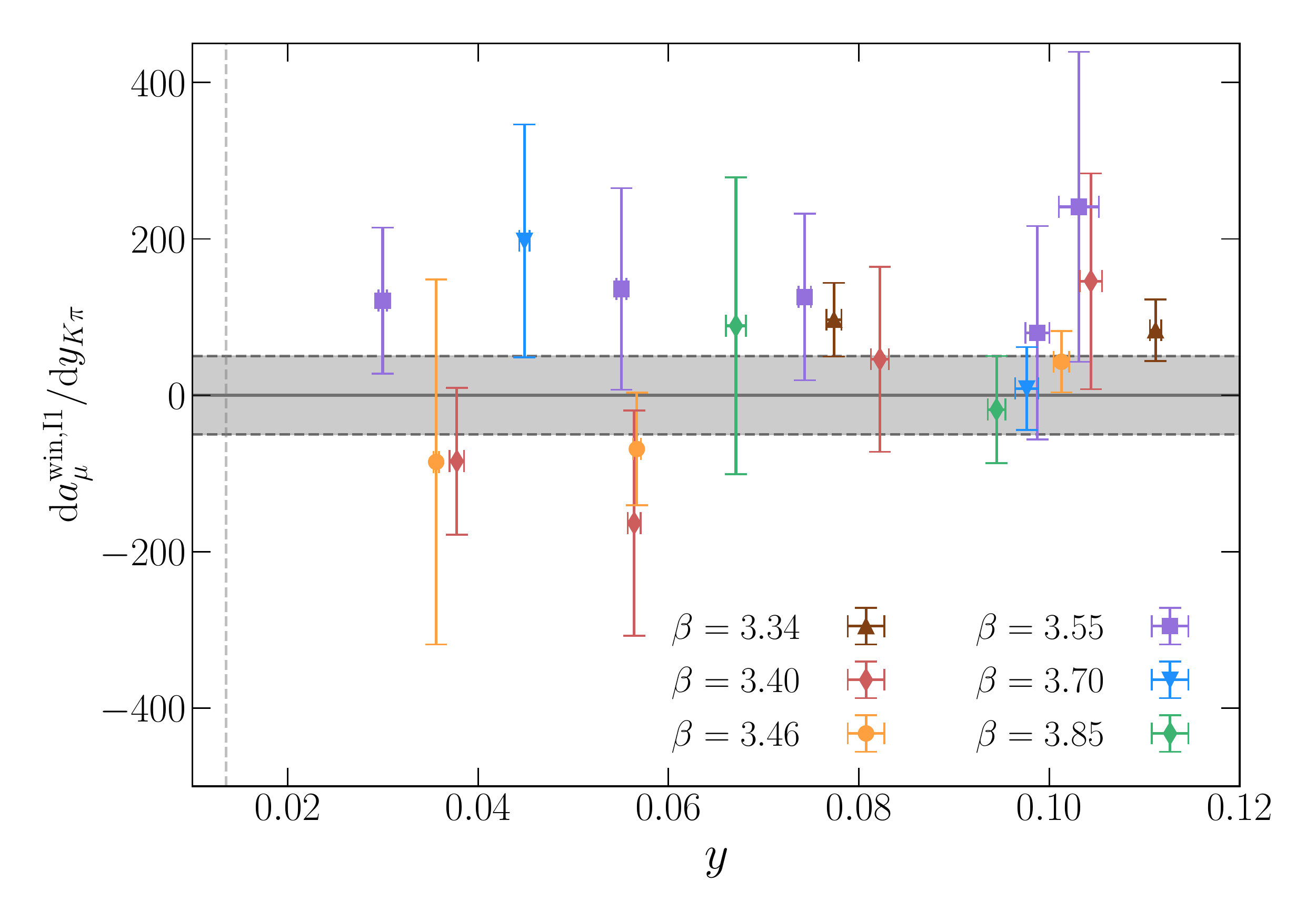} \\[4mm]
	
	\includegraphics*[width=0.46\linewidth]{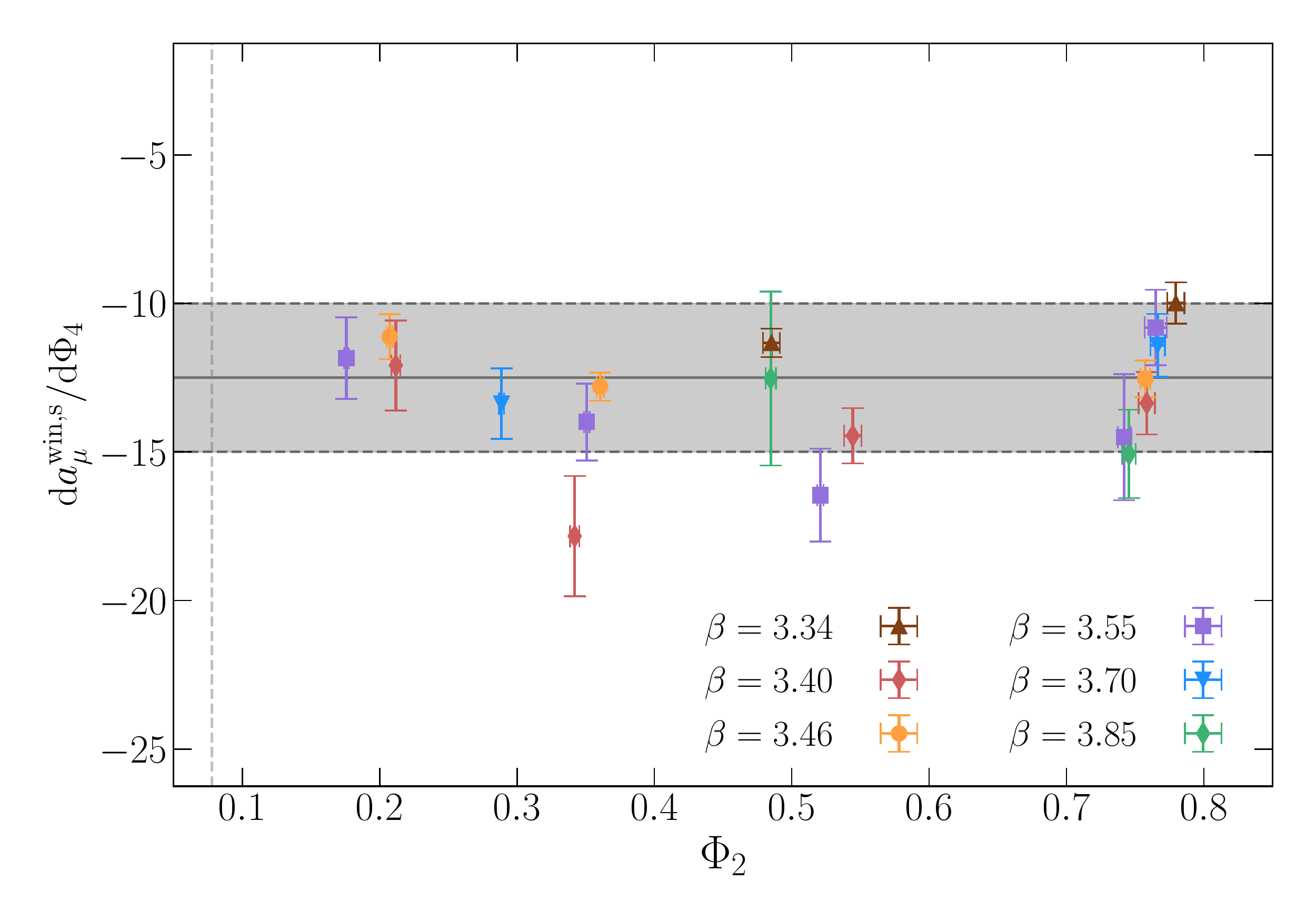}
	\includegraphics*[width=0.46\linewidth]{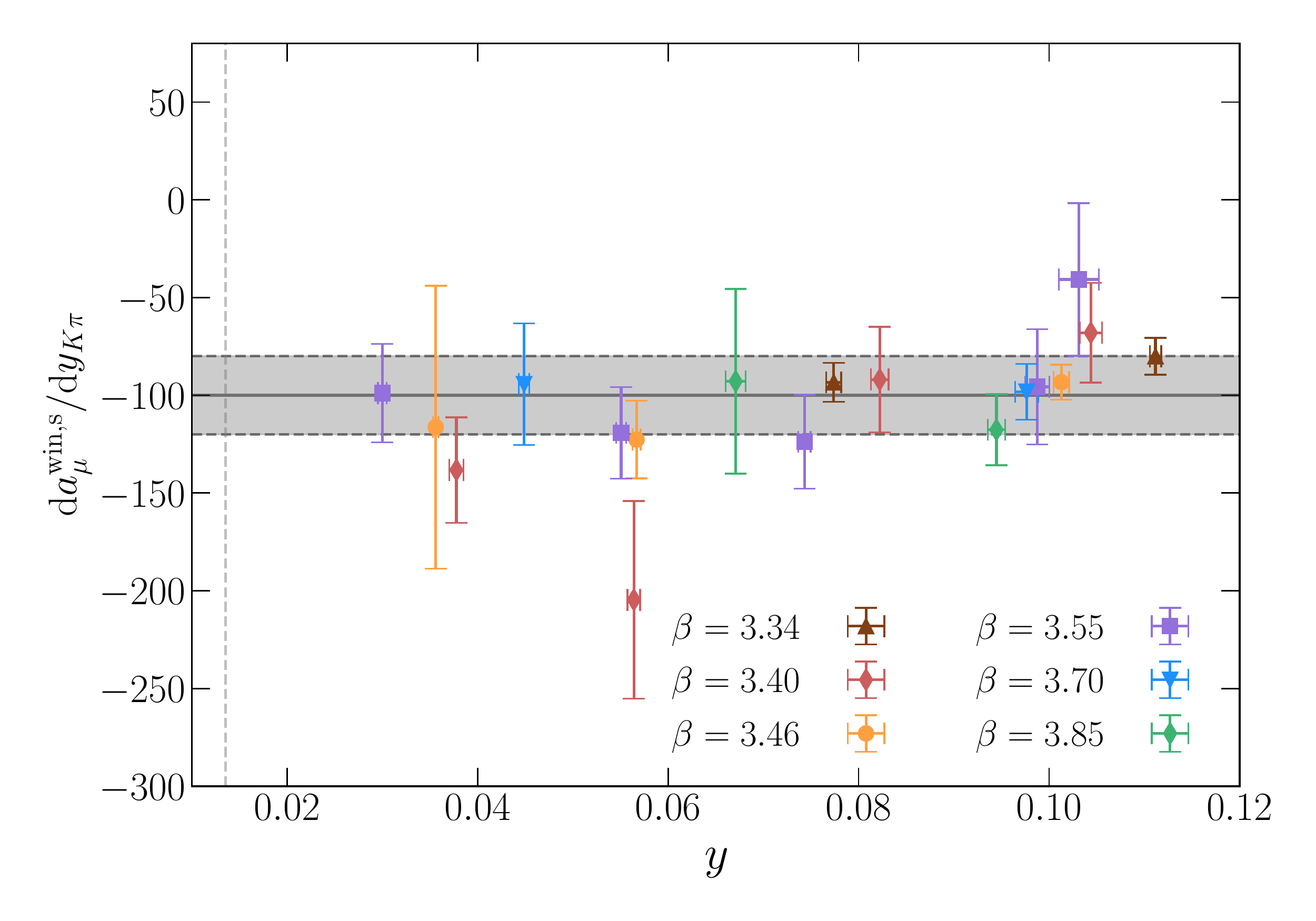} \\[4mm]
	
	\caption{Derivatives of the isovector and the strange-connected contributions to the window observable with respect to $X_\pi$. The gray areas illustrate the priors that are used in the global extrapolation.
	\label{fig:mass_shift}}
\end{figure}

We collect our results for the derivatives with respect to $\phi_4$ and $y_{K\pi}$ in Table~\ref{tab:mass_shifts}.
Throughout this appendix, we use units of $10^{-10}$ for $\awin$, as well as for coefficient $\gamma_0$.
The results are based on the local-local discretization of the correlation functions and the improvement coefficients and renormalization constants of set 1. As can be seen, the derivative of the isovector contribution to the window observable vanishes within error on most of the ensembles. This is expected from the order-of-magnitude
estimate in \eq{eq:sQEDdamuwin1_phi4}. No clear trend regarding
a dependence on $X_{\pi}$, $X_{K}$ or $X_{a}$ can be resolved. We show
the derivative of $\awinisovec$ with respect to $X_\pi$ in the upper
panels of Fig.~\ref{fig:mass_shift}. For the corresponding priors for
the chiral-continuum extrapolation we choose
\begin{align}
	\gamma_0^{{\mathrm{win,I1}, y_{K\pi}}} = 0(50)\qquad\gamma_0^{{\mathrm{win,I1}, \phi_4}} = -2.5(5.0)\,.
\end{align}

The derivative of the strange-connected contribution of the window observable with respect to $X_{K}$ is negative and can be determined to good precision. Our results are shown in the lower panels of Fig.~\ref{fig:mass_shift}. We choose our priors such that their width encompasses the spread of the data. For the strange-connected and the isoscalar contribution, we choose
\begin{align}
\gamma_0^{{\mathrm{win,s}, y_{K\pi}}} = -100(20)\qquad\gamma_0^{{\mathrm{win,s}, \phi_4}} = -12.5(2.5)\,.
\end{align}
These values are compatible with the estimate in \eq{eq:dawindmK2model}.

Discretization effects in the data may be inspected by comparing the derivatives based on the two sets of improvement coefficients. Such effects are largest for the two ensembles at $\beta=3.34$, but are still smaller than the spread in the data and therefore not significant with respect to our prior widths. In our global extrapolations, we use a single set of priors irrespective of the improvement procedure.

\begin{table}[h!]
	\caption{Derivatives of the isovector and the strange-connected contributions to the window observable with respect to $X_K$ in units of $10^{-10}$. The data is based on the the local-local discretization of the vector-vector correlation function and the improvement coefficients of set 1.}
	\vskip 0.1in
	\renewcommand{\arraystretch}{1.1}  
	\begin{tabular}{l@{\hskip 01em}|@{\hskip 01em}c@{\hskip 02em}c@{\hskip 02em}|@{\hskip 02em}c@{\hskip 02em}r@{\hskip 02em}}
\hline
 id   & $\frac{\mathrm{d}a_\mu^{\mathrm{win,I1}}}{\mathrm{d}\Phi_4}$   & $\frac{\mathrm{d}a_\mu^{\mathrm{win,I1}}}{\mathrm{d}y_{K \pi}}$   & $\frac{\mathrm{d}a_\mu^{\mathrm{win,s}}}{\mathrm{d}\Phi_4}$   & $\frac{\mathrm{d}a_\mu^{\mathrm{win,s}}}{\mathrm{d}y_{K \pi}}$   \\
\hline
 A653 & $\phantom{-}5.0(1.1)$                                                    & $\phantom{-}83(39)$                                                         & $-10.0(0.7)$                                                  & $-80(10)$                                                        \\
 A654 & $\phantom{-}5.0(1.9)$                                                    & $\phantom{-}96(47)$                                                         & $-11.3(0.5)$                                                  & $-93(10)$                                                        \\
 \hline
H101 & $-4.7(3.9)$                                                    & $\phantom{-}145(137)$                                                       & $-13.4(1.1)$                                                  & $-68(26)$                                                        \\
 H102 & $-12.2(3.5)$                                                   & $\phantom{-}46(118)$                                                        & $-14.5(1.0)$                                                  & $-91(27)$                                                        \\
 N101 & $-8.9(12.9)$                                                   & $-163(143)$                                                       & $-17.8(2.1)$                                                  & $-204(51)$                                                       \\
 C101 & $\phantom{-}2.6(8.3)$                                                    & $-84(93)$                                                         & $-12.1(1.6)$                                                  & $-138(27)$                                                       \\
 \hline
B450 & $-3.4(2.6)$                                                    & $\phantom{-}42(39)$                                                         & $-12.5(0.7)$                                                  & $-93(9)$                                                         \\
 N451 & $-5.3(5.2)$                                                    & $-68(71)$                                                         & $-12.8(0.5)$                                                  & $-122(20)$                                                       \\
 D450 & $-4.9(10.0)$                                                   & $-85(233)$                                                        & $-11.1(0.8)$                                                  & $-116(73)$                                                       \\
 \hline
H200 & $-0.3(5.3)$                                                    & $\phantom{-}241(198)$                                                       & $-10.8(1.3)$                                                  & $-40(40)$                                                        \\
 N202 & $-3.5(9.2)$                                                    & $\phantom{-}79(136)$                                                        & $-14.5(2.2)$                                                  & $-95(30)$                                                        \\
 N203 & $-3.5(5.1)$                                                    & $\phantom{-}125(106)$                                                       & $-16.5(1.6)$                                                  & $-123(25)$                                                       \\
 N200 & $\phantom{-}3.3(7.2)$                                                    & $\phantom{-}136(128)$                                                       & $-14.0(1.3)$                                                  & $-119(24)$                                                       \\
 D200 & $\phantom{-}7.1(7.1)$                                                    & $\phantom{-}121(93)$                                                        & $-11.8(1.4)$                                                  & $-98(26)$                                                        \\
 \hline
N300 & $\phantom{-}0.4(4.3)$                                                    & $ \phantom{-}8(53)$                                                         & $-11.4(1.1)$                                                  & $-98(15)$                                                        \\
 J303 & $\phantom{-}6.5(9.1)$                                                    & $\phantom{-}197(148)$                                                       & $-13.4(1.2)$                                                  & $-94(32)$                                                        \\
 \hline
J500 & $-9.0(5.3)$                                                    & $-18(68)$                                                         & $-15.1(1.5)$                                                  & $-117(19)$                                                       \\
 J501 & $-6.1(9.4)$                                                    & $\phantom{-}88(189)$                                                        & $-12.5(3.0)$                                                  & $-92(48)$                                                        \\
\hline
\end{tabular}
	\label{tab:mass_shifts}
\end{table}

 \newpage

\section{Phenomenological models \label{sec:phenomodels}}

In the first subsection of this appendix, we collect estimates of the sensitivity
of the window observables to various intervals in $\sqrt{s}$ in the dispersive approach.
The observable $\awin$ can indeed be obtained from experimental data for the ratio $R(s)$ defined in \eq{eq:Rsdef} via
\bea
\awin  &=&  \int_0^\infty ds \, f_{\rm win}(s) \,R(s),
\eea
where the weight function is given by
\bea
f_{\rm win}(s) &=& \frac{\alpha^2\,\sqrt{s}}{24\pi^4}  \int_0^\infty dt \; e^{-t\sqrt{s}} \widetilde{K}(t)\,
[\Theta(t,t_0,\Delta)-\Theta(t,t_1,\Delta)]\ .\label{eq:fwin}
\eea
In practice, since the integrand is very strongly suppressed beyond 1.5\,fm,
we have used the short-distance expansion of $\widetilde{K}(t)$ given by Eq.~(B16) of Ref.~\cite{DellaMorte:2017dyu},
which is very accurate up to 2\,fm.

The second and the third subsection contain phenomenological estimates of the derivatives
of the strangeness and the isovector contributions to $a_\mu^{\rm win}$ with respect
to the kaon mass at fixed pion mass, as a cross-check 
of the lattice results presented in Appendix~\ref{sec:chirtraj}.

\subsection{Sensitivity of the window quantity}

In~\cite{Bernecker:2011gh}, a semi-realistic model for the $R$-ratio was used
for the sake of comparisons with lattice data generated in the $(u,d,s)$ quark sector with exact isospin symmetry.
In particular, the model does not include the charm contribution, nor final states containing a photon, such as $\pi^0\gamma$.
It leads to the following values for the window observables and their sum, the full $a_\mu^{\rm hvp}$,
\bea
(a_\mu^{\rm hvp})^{\rm SD}|_{\rm model} &=& ~\;56.0 \times 10^{-10},
\\
a_\mu^{\rm win}|_{\rm model} = (a_\mu^{\rm hvp})^{\rm ID}|_{\rm model}  &=& 231.9 \times 10^{-10},
\\
(a_\mu^{\rm hvp})^{\rm LD}|_{\rm model} &=& 384.8 \times 10^{-10},
\\
a_\mu^{\rm hvp}|_{\rm model} &=& 672.7 \times 10^{-10}.
\label{eq:amuhvpmod}
\eea
Given the omission of the aforementioned channels, these values are quite realistic.\footnote{
For orientation, the charm contribution to $a_\mu^{\rm hvp}$ is $14.66(45)\times 10^{-10}$~\cite{Gerardin:2019rua},
and the $\pi^0\gamma$ channel contributes $4.5(1)\times 10^{-10}$~\cite{Aoyama:2020ynm}.
Adding these to \eq{eq:amuhvpmod}, the total is $691.9\times 10^{-10}$, consistent within errors with the White Paper
evaluation of $693.1(4.0)\times 10^{-10}$.}
Here we only use the model to provide the partition of the quantities above
into three commonly used intervals of $\sqrt{s}$, in order to illustrate
what the relative sensitivities of these quantities are to different energy intervals.
These percentage contributions are given in Table~\ref{tab:Rmodelperc}, along with the corresponding
figures for the subtracted vacuum polarization,
\be
\overline\Pi(Q^2) \equiv \Pi(Q^2) -\Pi(0) = \frac{Q^2}{12\pi^2} \int_0^\infty ds \, \frac{R(s)}{s(s+Q^2)} .
\ee
The model yields for this quantity the value $385.5\times 10^{-4}$ at $Q^2=1\,{\rm GeV}^2$.
We expect the fractions in the table to be reliable with an uncertainty at the five to seven percent level.

\begin{table}[ht]
\centerline{  \begin{tabular}{l@{\quad}r@{\qquad}r@{~~~}r@{~~~}r@{\qquad}r}
\hline\hline
$\sqrt{s}$ interval & $a_\mu^{\rm hvp}$ & $(a_\mu^{\rm hvp})^{\rm SD}$  &   $(a_\mu^{\rm hvp})^{\rm ID}$ &   $(a_\mu^{\rm hvp})^{\rm LD}$
 & $ \overline\Pi(1{\rm GeV}^2)$\\
\hline
below 0.6\,GeV  &  15.5   &  1.5  &  5.5 & 23.5  &  8.2\\
0.6 to 0.9\,GeV &     58.3   & 23.1 &   54.9 & 65.4 & 52.6  \\
above 0.9\,GeV  & 26.2    & 75.4   & 39.6  & 11.1 & 39.2  \\
\hline
Total   &      100.0   & 100.0 &    100.0 & 100.0  & 100.0\\
\hline\hline
  \end{tabular}}
\caption{\label{tab:Rmodelperc}Fractional contributions in percent from different regions in $\sqrt{s}$
  to $a_\mu^{\rm hvp}$ and the partial quantities $(a_\mu^{\rm hvp})^{\rm SD,ID,LD}$, as well as the subtracted vacuum polarization
  at scale $Q^2=1\,{\rm GeV}^2$, according to the
  $R$-ratio model given in~\cite{Bernecker:2011gh}. Note that this model includes neither the charm nor
final states containing a photon, such as $\pi^0\gamma$.}
  \end{table}

The model value for the intermediate window is best compared to the sum of Eqs.\ (\ref{res:I1},\ref{res:I0}).
The difference is $(1.8\pm 1.4)\times 10^{-10}$, which represents agreement at the $1.3\sigma$ level.
The main reason the $R$-ratio model agrees better with the lattice result than a state-of-the-art analysis~\cite{Colangelo:2022vok}
is that the model  does not account for the strong suppression of the experimentally measured
$R$-ratio in the region $1.0<\sqrt{s}/{\rm GeV} <1.5$ relative to the parton-model prediction.
This observation suggests a possible scenario where the higher lattice value of $\awin$ as compared to its
data-driven evaluation is explained by a too pronounced dip of the $R$-ratio just above the $\phi$ meson mass.
In such a scenario, the relative deviation between the central values  of $\ahvp$
obtained on the lattice and using $e^+e^-$ data would be smaller than for $\awin$ by a factor of about 1.5,
given the entries in Table \ref{tab:Rmodelperc}.
Indeed, it has been shown~\cite{Colangelo:2022xlq} that the central values 
of the BMW collaboration~\cite{Borsanyi:2020mff} cannot be explained  by a
modification of the experimental $R(s)$ ratio below $s=1\,{\rm GeV}^2$ alone.

\subsection{Model estimate of \texorpdfstring{$(\partial/\partial m_K^2) a_\mu^{{\rm win},s}(m_\pi^2,m_K^2)$}{d/dmK2 awin,s}}

In~\cite{Ce:2022eix}, we have used two closely related $R$-ratio models for the strangeness correlator and the light-quark
contribution to the isoscalar correlator,
\bea\label{eq:Rsmodel_l}
R_{I=0}^\ell(s) &=& \frac{A_\omega}{18} m_\omega^2 \delta(s-m_\omega^2) + \frac{N_c}{18}  \theta(s-s_0)\Big(1+ \frac{\alpha_s}{\pi}\Big),
\\
R^s(s) &=& \frac{A_\phi}{9} m_\phi^2 \delta(s-m_\phi^2) + \frac{N_c}{9}  \theta(s-s_1)\Big(1+ \frac{\alpha_s}{\pi}\Big),
\label{eq:Rsmodel_s}
\eea
with
\be
\sqrt{s_0}=1.02\,{\rm GeV}, \qquad \sqrt{s_1}=1.24\,{\rm GeV},
\ee
$m_\omega=0.78265\,{\rm GeV}$, $m_\phi=1.01946\,{\rm GeV}$ and~\cite{ParticleDataGroup:2020ssz}
\bea
\frac{A_\omega}{18} =   \frac{9\pi}{\alpha^2} \frac{\Gamma_{ee}(\omega)}{m_\omega} = \frac{7.33(24)}{18},
\\
\frac{A_\phi}{9} =   \frac{9\pi}{\alpha^2} \frac{\Gamma_{ee}(\phi)}{m_\phi} = \frac{5.86(10)}{9}.
\eea
The threshold values $s_0$ and $s_1$ have been adjusted to reproduce the corresponding lattice results
for $a_\mu^{\rm hvp}$. The model $R$-ratios of Eqs.~(\ref{eq:Rsmodel_l}--\ref{eq:Rsmodel_s}) were used~\cite{Ce:2022eix}
in the linear combination $(18 R_{I=0}^\ell - 9 R^s)$ in order to model the SU(3)$_{\rm f}$ breaking
contribution $\Pi^{08}$, which enters the running of the electroweak mixing angle.
Our model for this linear combination also obeys an exact sum rule, $\int_0^\infty ds\,(18 R_{I=0}^\ell - 9 R^s)=0$,
within the statistical uncertainties.
We now evaluate the window quantity for the models of Eqs.~(\ref{eq:Rsmodel_l}--\ref{eq:Rsmodel_s}).
For the strangeness contribution, we have
\be
a_\mu^{\rm win,s} = (27.6\pm 0.3_{\rm stat} )\times 10^{-10},
\ee
and for the full isoscalar contribution, the model predicts
\be
a_\mu^{\rm win,I0}= (47.4\pm 0.5_{\rm stat}) \times 10^{-10}.
\ee
Given the modelling uncertainties,
these values are in excellent agreement with the lattice results presented in the main part of the text,
respectively Eqs.~(\ref{res:s}) and (\ref{res:I0}).
We also record some useful values of the kernel,
\bea
&& f_{\rm win}(m_\phi^2) = 29.5\times 10^{-10}\,{\rm GeV}^{-2},\qquad
f_{\rm win}(s_1)= 16.1\times 10^{-10}\,{\rm GeV}^{-2}, \qquad 
\\
&& \frac{d}{ds}(s f_{\rm win}(s))_{s=m_\phi^2} = -11.3\times 10^{-10}\,{\rm GeV}^{-2}.
\eea
In the following, we evaluate the strange-quark mass dependence of $a_\mu^{\rm win,s}$,
based on the idea that the parameters $A_\phi$, $m_\phi$ and $s_1$ only depend
on the mass of the valence (strange) quark. This general assumption is reflected
in Eqs.~(\ref{eq:Aphimphideriv},~\ref{eq:mphimKderiv},~\ref{eq:s1mKderiv}) below.

It was noted a long time ago~\cite{Sakurai:1978xb}
that the electronic decay widths of vector mesons, normalized
by the relevant charge factor, is only very weakly dependent on their mass:
\bea
18\cdot \Gamma_{ee}(\omega)  &=& 10.8(4)\,{\rm keV},
\\
9\cdot \Gamma_{ee}(\phi) &=& 11.4(4)\,{\rm keV},
\\
{\textstyle\frac{9}{4}}\cdot \Gamma_{ee}(J/\psi) &=& 12.4(2)\,{\rm keV}.
\eea
This suggests that, unlike in QED,  $(A_V\cdot m_V)$ depends less strongly on $m_V$ than $A_V$ itself
for QCD vector mesons.
Therefore it is best to estimate the derivative of interest as follows,
\bea
\frac{\partial a_\mu^{{\rm win},\phi}}{\partial m_K^2}\Big|_{m_\pi^2}
&\simeq & \frac{\partial}{\partial m_K^2} \Big(\frac{A_\phi m_\phi}{9}\Big) \,m_\phi\,f_{\rm win}(m_\phi^2)
+ \Big(\frac{A_\phi m_\phi}{9}\Big) \frac{\partial m_\phi^2}{\partial m_K^2}\, \frac{\partial }{\partial m_\phi^2} (m_\phi f_{\rm win}(m_\phi^2)).
\eea
We estimate the following derivatives by taking a finite difference between the $\omega$ and the $\phi$ meson properties,
\be\label{eq:Aphimphideriv}
\frac{\partial}{\partial m_K^2} \Big(\frac{A_\phi m_\phi}{9}\Big) \simeq
\frac{1}{9} \frac{A_\phi m_\phi - A_\omega m_\omega}{m_K^2-m_\pi^2} = 0.12(10)\,{\rm GeV}^{-1}.
\ee
and
\be\label{eq:mphimKderiv}
\frac{\partial m_\phi^2}{\partial m_K^2} = 2m_\phi \frac{\partial m_\phi}{\partial m_K^2}
\simeq 2m_\phi \frac{m_\phi-m_\omega}{m_K^2-m_\pi^2} = 2.13.
\ee
Thus
\be
\frac{\partial a_\mu^{{\rm win},\phi}}{\partial m_K^2}\Big|_{m_\pi^2} \simeq
((3.5\pm3.1)  -36.1)\times 10^{-10}\,{\rm GeV}^{-2} = (-32.6\pm3.1)\times 10^{-10}\,{\rm GeV}^{-2}.
\label{eq:damudmK2Sak}
\ee


Next, we estimate the dependence originating from the valence-mass dependence of $s_1$,
\be\label{eq:s1mKderiv}
\frac{\partial s_1}{\partial m_K^2} \simeq 2\sqrt{s_1}\frac{\sqrt{s_1}-\sqrt{s_0}}{m_K^2-m_\pi^2}
= 2.4.
\ee
Thus the derivative of the perturbative continuum $a_\mu^{{\rm win},s,{\rm cont}}$ with respect to the squared kaon mass yields
\be
\frac{\partial a_\mu^{{\rm win},s,{\rm cont}}}{\partial m_K^2}
= -\frac{N_c}{9} (1+\alpha_s/\pi) f_{\rm win}(s_1) \frac{\partial s_1}{\partial m_K^2}
= -14.1\times 10^{-10}.
\ee
Adding this contribution to \eq{eq:damudmK2Sak}, we get in total
\be\label{eq:dawindmK2model}
\frac{\partial a_\mu^{{\rm win},s}}{\partial m_K^2} = (-46.6\pm3.1_{\rm stat}\pm 7.0_{\rm model})\times 10^{-10}\,{\rm GeV}^{-2}.
\ee
To the statistical error from the electronic widths of the $\omega$ and $\phi$ mesons, we have added a modelling error of 15\%.
Using $t_0$, the value above translates into 
\be
\frac{\partial a_\mu^{{\rm win},s}}{\partial \phi_4}\Big|_{\phi_2} \simeq (-10.9\pm 0.7_{\rm stat}\pm 1.6_{\rm model})\times 10^{-10},
\ee
which can directly be compared to the values from lattice QCD listed in Table~\ref{tab:mass_shifts}. The agreement is excellent.

In \eq{eq:Rsmodel_s}, we have written the perturbative contribution above the threshold $s_1$ in the massless limit.
We now verify that the  mass dependence of the perturbative contribution  is negligible for fixed $s_1$.
The leading mass-dependent perturbative contribution to the $R$-ratio well above threshold
is (see e.g.~\cite{Chetyrkin:1997qi}, Eqs.\ 11 and 12)
\be
R^s_{{\rm pert}}(m_s^2,\,s) - R^s_{{\rm pert}}(0,\,s)
= \frac{N_c}{9} \Big( -6 \Big(\frac{m_s^2}{s}\Big)^2 + 12 \frac{\alpha_s}{\pi} \frac{m_s^2}{s} + \dots\Big).
\ee
From here we have estimated $\frac{\partial}{\partial m_K^2} a_\mu^{\rm win,s,{\rm pert}} \approx 0.5\times 10^{-10}\, {\rm GeV}^{-2}$.
Since this contribution to $\frac{\partial}{\partial m_K^2} a_\mu^{\rm win,s} $
is about one sixth the statistical uncertainty from the vector meson electronic decay widths,
we neglect the perturbative mass dependence of $a_\mu^{\rm win,s}$.

For future reference, we evaluate in the same way as in \eq{eq:dawindmK2model} the derivative of $a_\mu^{{\rm hvp},s}$
and find
\be\label{eq:dahvpdmK2}
\frac{\partial a_\mu^{{\rm hvp},s}}{\partial m_K^2}\Big|_{m_\pi^2} = (-129\pm 6_{\rm stat} \pm 19_{\rm model})\times 10^{-10}\,{\rm GeV}^{-2}.  
\ee
Here, the dependence on $s_1$ only contributes 18\% of the total. We have again assigned a 15\% modelling uncertainty to the prediction.
Since we expect valence-quark effects to dominate, the prediction (\ref{eq:dahvpdmK2}) can also be applied
to the full isoscalar prediction.

\subsection{Model estimate of \texorpdfstring{$(\partial/\partial m_K^2) a_\mu^{\rm win, I1}(m_\pi^2,m_K^2)$}{d/dmK2 awin,I1}}

The influence of the strange quark mass on the isovector channel is a pure sea quark effect,
and is as such harder to estimate. Based on the OZI rule, one would also expect a smaller relative
sensitivity than in the strangeness channel addressed in the previous subsection.

One effect of the presence of strange quarks on the isovector channel
is that kaon loops can contribute. No isovector vector resonances with a strong coupling
to $\bar KK$ are known, therefore we attempt to use scalar QED (sQED) to evaluate the effect of the kaon loops.
Note that at the SU(3)$_{\rm f}$ symmetric point, the sum of the $\bar K^0 K^0$ and $K^+K^-$ contributions
to the isovector channel amounts to half as much as that of the pions.
We find, integrating in $s$ from threshold up to $4\,{\rm GeV}^2$ with $m_K=0.495$\,GeV,
\bea
a_\mu^{{\rm win},I=1} &=& 0.99 \times 10^{-10}, \qquad \textrm{kaon loops in sQED}
\\
\frac{\partial}{\partial m_K^2} a_\mu^{{\rm win},I=1} &=& -7.0 \times 10^{-10}\, {\rm GeV}^{-2}.
\label{eq:sQEDdamuwin1}
\eea

A further, more indirect effect of two-kaon intermediate states is that they can
affect the properties of the $\rho$ meson. On general grounds, one expects the two-kaon 
states to reduce the $\rho$ mass, since energy levels repel each other.
However, for the window quantity it so happens that $s f_{\rm win}(s)$ has a maximum practically at the $\rho$ mass,
therefore the derivative of this function is extremely small, 
\be\label{eq:dws}
\frac{2}{f_{\rm win}(s)}\frac{d}{ds} (s f_{\rm win}(s)) \Big|_{s=m_\rho^2}  = -0.043.
\ee
The effect of a shift in the $\rho$ meson mass is therefore heavily suppressed.\footnote{But note that this
effect must be revisited when addressing the strange-quark mass dependence of the isovector contribution to the
full $a_\mu^{\rm hvp}$.}
Reasonable estimates of the order-of-magnitude of the derivative $\partial m_\rho/\partial m_K^2|_{m_\pi^2}$
lead to a contribution to $\frac{\partial}{\partial m_K^2} a_\mu^{{\rm win},I=1}$ which is smaller
than the sQED estimate. These estimates are based on the observation that the ratio $m_\rho/f_\pi$
is about 5\% higher at a pion mass of 311\,MeV in the $N_{\rm f}=2$ QCD calculation~\cite{Erben:2019nmx}
than if one interpolates the corresponding $N_{\rm f}=2+1$ QCD results~\cite{Andersen:2018mau,Gerardin:2019rua}
to the same pion mass, though a caveat is that neither result is continuum-extrapolated.
The effect of the kaon intermediate states on the $\pi\pi$ line-shape is even
harder to estimate, but we note that even in $N_{\rm f}=2$ QCD calculations~\cite{Erben:2019nmx},
i.e.\ in the absence of kaons,
the obtained $g_{\rho\pi\pi}$  coupling is consistent with $N_{\rm f}=2+1$ QCD
calculations~\cite{Andersen:2018mau,Gerardin:2019rua} carried out at comparable pion masses.

In summary, we use the sQED evaluation of \eq{eq:sQEDdamuwin1} to provide the order-of-magnitude estimate
\be\label{eq:sQEDdamuwin1_phi4}
\frac{\partial}{\partial \phi_4}\Big|_{\phi_2} a_\mu^{{\rm win},I=1} \approx -1.6 \times 10^{-10}\,. 
\ee
We note that the statistical precision of our lattice-QCD results for this derivative in Table~\ref{tab:mass_shifts}
is not sufficient to resolve the small effect estimated here.

 \newpage
\section{Finite-volume correction} \label{sec:fse}

Corrections for finite-size effects (FSE) have been estimated using a 
similar strategy to the one presented in our previous publication on the hadronic contributions to the muon $g-2$~\cite{Gerardin:2019rua}. The main difference lies in the treatment of small Euclidean times, where we have replaced NLO $\chi$PT by the Hansen-Patella method as described below.
We have also investigated finite-size corrections in $\chi$PT at NNLO~\cite{Bijnens:2017esv,Borsanyi:2020mff}. Overall, we found it to be comparable in size to the values found in Tables~\ref{tab:fse1}--\ref{tab:fse2}, the level of agreement improving for increasing volumes and decreasing pion masses. Given that the NNLO $\chi$PT correction term is in many cases not small compared to the NLO term, we refrain from using $\chi$PT to compute finite-size effects in our analysis of $\awin$ (see~\cite{Aubin:2022hgm} for a more detailed discussion of the issue).

\subsection{The Hansen-Patella method}
\label{sec:HP}

In~\cite{Hansen:2019rbh,Hansen:2020whp}, finite-size effects for the hadronic contribution to the muon $(g-2)$ are expressed in terms of the forward Compton amplitude of the pion as an expansion in $\exp\left( -|\vec{n}| m_\pi L \right)$ for $|\vec{n}|^2 = 1,2,3,6,\dots$. Here, $n_k$ schematically represents the number of times the pion propagates around the $k^{\rm th}$ spatial direction of the lattice. Corrections that start at order $\exp\left( -n_{\rm eff} m_\pi L \right)$ with $n_{\rm eff} = \sqrt{2 + \sqrt{3}}  \approx 1.93$ are neglected: they appear when at least two pions propagate around the torus. 
The results for the first three leading contributions
($|\vec{n}|^2\leq 3$) can thus be used consistently to correct the
lattice data on each timeslice separately. We decided to use the size
of the $|\vec{n}|^2=3$ term, i.e. the last one that is parametrically
larger than the neglected $n_{\rm eff}\approx 1.93$ contribution, as
an estimate of the inherent systematic error. 

In this work we follow the method presented in~\cite{Hansen:2020whp}, where the forward Compton amplitude is approximated by the pion pole term, which is determined by the electromagnetic form factor of the pion in the space-like region. Since the form-factor is only used to evaluate the small finite-volume correction, a simple but realistic model is sufficient. Here we use a monopole parametrization obtained from $N_f=2$ lattice QCD simulations~\cite{QCDSFUKQCD:2006gmg},
\begin{equation}
F(q^2) = \frac{1}{1 + q^2/M^2} \,, \quad M^2(m_{\pi}^2) = 0.517(23) \mathrm{GeV}^2 +0.647(30)m_{\pi}^2 \,.
\label{eq:FF}
\end{equation}
The statistical error on the finite-size correction is obtained by propagating the jackknife error on the pion and monopole masses. The results obtained using this method are summarized in the third and fourth columns of Table~\ref{tab:fse1} and Table~\ref{tab:fse2}.

\subsection{The Meyer-Lellouch-Lüscher formalism with Gounaris-Sakurai parametrization}

As an alternative, we also consider the Meyer-Lellouch-L\"uscher (MLL) formalism. The isovector correlator in both finite and infinite volume is written in terms of spectral decompositions
\begin{align}
G^{I=1}(t, \infty) &= \frac{1}{48\pi^2} \int_{2m_{\pi}}^{\infty} \mathrm{d}\omega \, \omega^2 \, \left(1 - \frac{4 m_{\pi}^2}{\omega^2} \right)^{3/2} |F_{\pi}(\omega)|^2 \, e^{-\omega t}  \,, \\
G^{I=1}(t, L) &= \sum_{i} |A_i|^2 \, e^{-E_i t} \,, \quad  E_i = 2 \sqrt{m_{\pi}^2 + k_i^2} \,,
\label{eq:MLL}
\end{align}
where $F_{\pi}(\omega)$ is the time-like pion form factor. Following the L\"uscher formalism, the discrete energy levels $E_{i} = 2 \sqrt{m_{\pi}^2 + k_i^2}$ in finite volume are obtained by solving the equation
\begin{equation}
\delta_1(k_i) + \phi(q) = n\pi \,, \quad q = \frac{k_i L}{2\pi} \,,
\label{eq:luescher}
\end{equation}
where $\phi(q)$ is a known function~\cite{Luscher:1991cf,Luscher:1990ux}, $n$ a strictly positive integer and $\delta_1$ is the scattering phase shift in the isospin $I=1$, p-wave channel. Strictly speaking, this relation holds exactly only below the four-particle threshold that starts at $4m_{\pi}$. This is only a restriction at light pion mass where many states are needed to saturate the spectral decomposition in finite volume. We will see below how to circumvent this difficulty. In~\cite{Meyer:2011um}, the overlap factors $A_i$ that enter the spectral decomposition in finite volume were shown to be related to the form factor in infinite volume through the relation
\begin{equation}
| F_{\pi}(E_i) |^2 = \left( q\phi^{\prime}(q) + k \frac{\partial \delta_1}{\partial k} \right) \frac{3\pi E_i^2}{2 k_i^5} |A_i|^2 \,.
\label{eq:Meyer}
\end{equation}
The time-like pion form factor has been computed on a subset of our
lattice simulations~\cite{Andersen:2018mau,Gerardin:2019rua}. Since
the form factor is only needed to estimate the small finite-volume
correction, an approximate model can be used. Here, we assume a
Gounaris-Sakurai (GS) parametrization that contains two parameters:
the $g_{\rho\pi\pi}$ coupling and the vector meson mass
$m_{\rho}$~\cite{Gounaris:1968mw}. A given choice of those parameters
allows us to compute both the finite-volume and infinite-volume
correlation function in the isovector channel at large Euclidean times
using Eq.~(\ref{eq:MLL}). The difference $G^{I=1}(t, \infty) -
G^{I=1}(t, L)$, when inserted into \eq{def:ID}, yields our estimate of the FSE. 
In practice, the GS parameters are obtained from a fit to the isovector correlation function $G^{I=1}(t,L)$ at large Euclidean times, using Eqs.~(\ref{eq:MLL}), (\ref{eq:luescher}) and (\ref{eq:Meyer}). Statistical errors on the GS parameters can easily be propagated using the Jackknife procedure. 

Since this method is expected to give a good description only up to the inelastic threshold, Eq.~(\ref{eq:luescher}) being formally valid below $4m_{\pi}$, we opt to use the MLL formalism only above a certain cut in Euclidean time, given by $t^{*} = (m_{\pi} L/4)^2 / m_{\pi}$. Below the cut, we always use the HP method described above. Above the cut, the lightest few finite-volume states in the spectral decomposition saturate the integrand. The results using the MLL formalism are summarized in the fifth column of Table~\ref{tab:fse1} and Table~\ref{tab:fse2}. 

\subsection{Corrections applied to lattice data}

In Tables~\ref{tab:fse1} and \ref{tab:fse2} we summarize the FSE
correction applied to the raw lattice data. We find that finite-size
corrections computed using either the HP or the MLL method for
($t>t^\star$) show good agreement within their respective
uncertainties. Our final estimates, shown in the rightmost column, are
obtained by adding the result from the HP method at short times
($t<t^\star$) to that of the MLL method above $t^\star$ and the
kaon loop contribution. The latter has been computed in 
$\chi$PT at NLO (see for instance \cite{Francis:2013fzp})
on ensembles without SU(3) flavor symmetry.
At the SU(3) symmetric point, the kaon loop contribution has been 
accounted for by scaling the HP and MLL corrections by a factor of $3/2$.
We have included the scale factor in the respective entries in
Tables~\ref{tab:fse1} and \ref{tab:fse2}.

The uncertainty quoted in the rightmost column is given by the
statistical error computed as described in the two previous
sections. It includes the statistical error on the GS parameters and
on the monopole mass that appears in the parametrization of the form
factor in \eq{eq:FF}. The systematic error on the HP contribution is
estimated as described in Section~\ref{sec:HP}.



For our final estimates of finite-volume corrections, we adopt a more
conservative approach regarding the overall uncertainty. As in our
earlier paper \cite{DellaMorte:2017dyu}, we base our uncertainty estimate on the
comparison to the NLO $\chi$PT correction, which leads us to assign an error of 25\% of
the estimated correction for each ensemble, which replaces the
uncertainties quoted in the last column of Tables~\ref{tab:fse1} and
\ref{tab:fse2}.
For example, the finite-size correction applied to $a_\mu^{\rm win,I1}$ in the
case of ensemble J303 with $f_\pi$-rescaling is 
$(1.62\pm 0.405)\times 10^{-10}$.

\begin{table}[h!]
  \caption{Finite-size effects in the isovector channel with
    $f_{\pi}$-rescaling, in units of $10^{-10}$, for our ensembles
    described in Table~\ref{tab:sim}. The correction obtained using
    the HP method is given in the third and fourth columns. The MLL
    estimate in the long-distance region is listed in the fifth
    column. The contribution of the kaon is given in column six, where
    dashes for ensembles at the SU(3) symmetric point indicate that this
    contribution is contained in the HP and MLL estimates.
    Our final estimate is given in the last column. 
    Only statistical errors are shown. We assign an
    uncertainty of 25\% of the FSE on each ensemble (see text).}
\vskip 0.1in
\renewcommand{\arraystretch}{1.1}    
\begin{tabular}{l@{\hskip 02em}c@{\hskip 02em}c@{\hskip 02em}c@{\hskip 01em}c@{\hskip 01em}c@{\hskip 02em}c}
\hline
id				&	$t^\star$~[fm]	&	HP$(t<t^\star)$	&	HP$(t>t^\star)$	&	MLL$(t>t^\star)$	&	Kaon loop	&	Final Estimate \\
\hline
   A653 & 0.79 & 0.98(0.01) & 0.81(0.03) & 0.78(0.01) & - & 1.75(0.03) \\
\hline
H101 & 1.04 & 0.71(0.01) & 0.03(0.00) & 0.03(0.00) & - & 0.74(0.01) \\
H102 & 0.86 & 0.70(0.01) & 0.40(0.02) & 0.36(0.01) & 0.19 & 1.25(0.10) \\
H105 & 0.69 & 0.58(0.03) & 1.95(0.11) & 1.87(0.08) & 0.14 & 2.59(0.15) \\
N101 & 1.47 & 0.28(0.01) & 0.00(0.00) & 0.00(0.00) & 0.01 & 0.29(0.01) \\
C101 & 1.21 & 0.75(0.02) & 0.03(0.00) & 0.03(0.00) & 0.01 & 0.78(0.02) \\
\hline
B450 & 0.76 & 0.83(0.01) & 0.77(0.02) & 0.74(0.56) & - & 1.57(0.37) \\
S400 & 0.69 & 0.61(0.01) & 1.55(0.05) & 1.54(0.03) & 0.34 & 2.50(0.18) \\
N451 & 1.22 & 0.51(0.01) & 0.01(0.00) & 0.01(0.00) & 0.02 & 0.53(0.01) \\
D450 & 1.60 & 0.32(0.01) & 0.00(0.00) & 0.00(0.00) & 0.00 & 0.32(0.01) \\
D452 & 1.15 & 0.89(0.02) & 0.10(0.01) & 0.10(0.01) & 0.00 & 1.00(0.03) \\
\hline
H200 & 0.58 & 0.68(0.02) & 3.35(0.07) & 3.17(0.09) & - & 3.84(0.16) \\
N202 & 1.22 & 0.38(0.01) & 0.00(0.00) & 0.00(0.00) & - & 0.38(0.00) \\
N203 & 1.03 & 0.56(0.01) & 0.04(0.00) & 0.03(0.00) & 0.09 & 0.69(0.05) \\
N200 & 0.84 & 0.73(0.01) & 0.64(0.02) & 0.61(0.01) & 0.07 & 1.41(0.05) \\
D200 & 1.09 & 0.95(0.01) & 0.11(0.00) & 0.10(0.00) & 0.01 & 1.06(0.02) \\
E250 & 1.54 & 0.57(0.02) & 0.00(0.00) & 0.00(0.00) & 0.00 & 0.57(0.02) \\
\hline
N300 & 0.75 & 0.89(0.01) & 0.79(0.02) & 0.75(0.01) & - & 1.64(0.03) \\
N302 & 0.65 & 0.61(0.01) & 1.79(0.03) & 1.73(0.02) & 0.34 & 2.68(0.20) \\
J303 & 0.85 & 0.90(0.01) & 0.71(0.02) & 0.67(0.01) & 0.05 & 1.62(0.06) \\
E300 & 1.25 & 0.76(0.01) & 0.02(0.00) & 0.02(0.00) & 0.00 & 0.78(0.01) \\
\hline
J500 & 0.82 & 0.98(0.01) & 0.41(0.01) & 0.40(0.01) & - & 1.37(0.01) \\
J501 & 0.67 & 0.60(0.01) & 1.52(0.04) & 1.55(0.02) & 0.29 & 2.44(0.16) \\
\hline
\end{tabular} 
\label{tab:fse1}
\end{table}

\begin{table}[h!]
  \caption{Same as Table~\ref{tab:fse1} using $t_0$ to set the scale.}
\vskip 0.1in
\renewcommand{\arraystretch}{1.1}    
\begin{tabular}{l@{\hskip 02em}c@{\hskip 02em}c@{\hskip 02em}c@{\hskip 01em}c@{\hskip 01em}c@{\hskip 02em}c}
\hline
id				&	$t^\star$~[fm]	&	HP$(t<t^\star)$	&	HP$(t>t^\star)$	&	MLL$(t>t^\star)$	&	Kaon loop	&	Final Estimate \\
\hline
   A653 & 0.79 & 0.80(0.01) & 1.19(0.04) & 1.11(0.01) & - & 1.90(0.06) \\
\hline
   H101 & 1.04 & 0.73(0.02) & 0.13(0.00) & 0.12(0.00) & - & 0.85(0.01) \\
   H102 & 0.86 & 0.62(0.01) & 0.57(0.02) & 0.52(0.01) & 0.19 & 1.33(0.11) \\
   H105 & 0.69 & 0.54(0.01) & 2.10(0.06) & 2.01(0.02) & 0.14 & 2.68(0.15) \\
   N101 & 1.47 & 0.29(0.01) & 0.00(0.00) & 0.00(0.00) & 0.01 & 0.30(0.01) \\
   C101 & 1.21 & 0.73(0.02) & 0.03(0.00) & 0.03(0.00) & 0.01 & 0.76(0.02) \\
\hline
   B450 & 0.76 & 0.63(0.01) & 1.21(0.03) & 1.12(0.79) & - & 1.75(0.53) \\
   S400 & 0.69 & 0.50(0.01) & 1.87(0.04) & 1.82(0.02) & 0.34 & 2.65(0.19) \\
   N451 & 1.22 & 0.54(0.01) & 0.01(0.00) & 0.01(0.00) & 0.02 & 0.57(0.01) \\
   D450 & 1.60 & 0.32(0.01) & 0.00(0.00) & 0.00(0.00) & 0.00 & 0.32(0.01) \\
   D452 & 1.15 & 0.88(0.02) & 0.07(0.00) & 0.08(0.00) & 0.00 & 0.95(0.02) \\
\hline
   H200 & 0.58 & 0.45(0.01) & 4.14(0.09) & 3.77(0.12) & - & 4.22(0.28) \\
   N202 & 1.22 & 0.44(0.01) & 0.01(0.00) & 0.01(0.00) & - & 0.45(0.01) \\
   N203 & 1.03 & 0.57(0.01) & 0.11(0.00) & 0.10(0.00) & 0.09 & 0.76(0.05) \\
   N200 & 0.84 & 0.66(0.01) & 0.81(0.02) & 0.76(0.01) & 0.07 & 1.49(0.06) \\
   D200 & 1.09 & 0.96(0.01) & 0.12(0.00) & 0.11(0.00) & 0.01 & 1.07(0.02) \\
   E250 & 1.54 & 0.53(0.01) & 0.00(0.00) & 0.00(0.00) & 0.00 & 0.53(0.01) \\
\hline
   N300 & 0.75 & 0.63(0.01) & 1.37(0.03) & 1.24(0.02) & - & 1.87(0.09) \\
   N302 & 0.65 & 0.45(0.01) & 2.29(0.05) & 2.13(0.03) & 0.33 & 2.91(0.25) \\
   J303 & 0.85 & 0.81(0.01) & 0.93(0.02) & 0.87(0.01) & 0.05 & 1.73(0.07) \\
   E300 & 1.25 & 0.76(0.01) & 0.02(0.00) & 0.02(0.00) & 0.00 & 0.78(0.01) \\
\hline
   J500 & 0.82 & 0.74(0.01) & 0.92(0.03) & 0.85(0.01) & - & 1.60(0.05) \\
   J501 & 0.67 & 0.43(0.01) & 2.02(0.05) & 1.97(0.01) & 0.29 & 2.69(0.17) \\
\hline
\end{tabular} 
\label{tab:fse2}
\end{table}

 \newpage

\section{Quenching of the charm quark} \label{sec:Qcharm}

The gauge configurations used in this work contain the dynamical effects of up, down and strange quarks.
As for the charm quarks, we have only taken into account the connected valence contributions.
In this appendix, we estimate the systematic error from the missing effect of charm sea-quark contributions.
The question we are after can be formulated as, ``What is the charm-quark effect on the $R$-ratio in a world
in which the charm quark is electrically neutral?''.

As in~\cite{Ce:2022eix}, we adopt a phenomenological approach.
There, we evaluated the perturbative prediction for the charm sea quark effect
and found it to be small for the running of the electromagnetic coupling from $Q^2=1$\,GeV$^2$ to 5\,GeV$^2$.
Alternatively,
we considered $D$-meson pair creation in the electromagnetic-current correlator
of the $(u,d,s)$ quark sector.
The contribution of the $D^+ D^-$ channel to the $R$-ratio reads
\begin{align}
R_{D^+D^-}(s) = \frac{1}{4} \left(1-\frac{4m_{D^+}^2}{s}\right)^{3/2}\ |F_{D^+}(s)|^2\ ,
\end{align}
and similar expressions hold for the $D^0 \bar D^0$ and $D_s^+D_s^-$ channels.
Since the form factor $F_{D^+}$ is not known precisely and
 our goal is only to estimate the order of magnitude of the effect, we will approximate it by its value at $s=0$,
which amounts to treating $D$-mesons in the scalar QED framework and replacing their form factors by the
relevant electromagnetic charges:
$\{ F_{D^0}(s), F_{D^+}(s), F_{D_s^+} \} \to \{ 2/3, -1/3, -1/3 \}$.
Note that up-, down-, or strange-quarks play the role of the valence quarks giving the mesons their respective charges.

The corresponding contributions to $a_{\mu}^{\text{hvp}}$ are evaluated using the expression
\begin{align}
\Delta^{c\text{-sea}}a_{\mu}^{\text{hvp}} &=
\int_0^{\infty}ds\ f_{\rm hvp}(s)\bigl(R_{D^0D^0} + R_{D^+D^-} + R_{D_s^+D_s^-}\bigr)(s)\ ,\label{eq:app_amu_csea}\\
f_{\rm hvp}(s)&:=
\Bigl(\frac{\alpha^2\sqrt{s}}{24\pi^4}\Bigr)\int_0^{\infty}dt\, e^{-t\sqrt{s}}\,\tilde{K}(t)
= \left(\frac{\alpha m_\mu}{3\pi}\right)^2 \,\frac{\hat K(s)}{s^2}\,,
\end{align}
where $m_\mu$ is the muon mass and the analytic form of $\hat K(s)$ can be found e.g.\ in~\cite{Jegerlehner:2009ry}, section 4.1. 
Similarly, the counterpart for the intermediate window reads
\begin{align}
\Delta^{c\text{-sea}}a_{\mu}^{\text{win}} &=
\int_0^{\infty}ds\ f_{\text{win}}(s)\bigl(R_{D^0D^0} + R_{D^+D^-} + R_{D_s^+D_s^-}\bigr)(s)\ .\label{eq:app_amu_csea_win}
\end{align}
where $f_{\text{win}}(s)$ is defined in Eq.~(\ref{eq:fwin}).

For the $D$-meson masses, we  use the values provided by the Particle Data Group 2020~\cite{ParticleDataGroup:2020ssz}.
Our results are
\begin{align}
\frac{\Delta^{c\text{-sea}}a_{\mu}^{\text{hvp}}}{a_{\mu}^{\text{hvp}}} &=
\frac{0.314}{720.0}\quad (\sim 0.04\%)\ ,\\
\frac{\Delta^{c\text{-sea}}a_{\mu}^{\text{win}}}{a_{\mu}^{\text{win}}} &=
\frac{0.015}{236.60}\quad (\sim 0.006\%)\ ,
\label{eq:damuwinch}
\end{align}
where we have inserted the $a_{\mu}^{\text{hvp}}=720.0$ value from Ref.~\cite{Gerardin:2019rua}.
The charm sea-quark contributions are thus negligible at the current level of precision.

We notice that
${\Delta^{c\text{-sea}}a_{\mu}^{\text{hvp}}} / {a_{\mu}^{\text{hvp}}}$
is much smaller than the effects found in the HVP contributions to the QED running coupling, namely $\sim$ 0.4\%~\cite{Ce:2022eix}.
We interpret the difference as follows:
the typical scale in $a_{\mu}^{\text{hvp}}$ is given by the muon mass, which is well separated from the $D$-meson masses.
Therefore the $D$-meson effects are strongly suppressed.
In comparison, the running coupling was investigated at the GeV scale and the suppression is less strong.

In the intermediate window, the charm sea-quarks are even more suppressed,
as seen in the tiny value of ${\Delta^{c\text{-sea}}a_{\mu}^{\text{win}}} / {a_{\mu}^{\text{win}}}$.
This results from the following fact:
creating $D$-meson pairs requires a center-of-mass energy of $\sim 4$ GeV, corresponding to $t\sim 0.05$ fm,
which is much smaller than the lower edge of the intermediate window, $t_0 = 0.4$ fm.
Therefore, the $D$-meson pair creation contributes mostly to the short-distance window $(a_\mu^{\rm hvp})^{\rm SD}$.
In fact, the effect in the intermediate window $\Delta^{c\text{-sea}}a_{\mu}^{\text{win}}$
amounts to at most 5\% of the total $\Delta^{c\text{-sea}}a_{\mu}^{\text{hvp}}$.

Charm sea quarks lead not only to on-shell $D$ mesons in the $R(s)$
ratio, but also to virtual effects below the threshold for charm
production. This is seen explicitly in the perturbative
calculation~\cite{Hoang:1994it}, where the two effects are of the same
order.  At present, we do not have a means to estimate these virtual
effects on the quantity $\awin$, in which they are less kinematically
suppressed.  Therefore, we will conservatively amplify the uncertainty
that we assign to the neglect of sea charm quarks by a factor of three
relative to the prediction of Eq.~\ref{eq:damuwinch}.  This estimate
also generously covers the effect on $\awin$ which follows from
adopting the perturbative charm-loop effect on $R(s)$ down to
$s=1\dots1.5\,{\rm GeV}^2$. Thus, rounding the uncertainty to one significant
digit, we quote
\be
\Delta^{c\text{-sea}}\awin = 0.05\times 10^{-10}
\ee
as the uncertainty on $\awin$ due to the quenching of the charm in the final result Eq.\ (\ref{eq:final})
for the isosymmetric theory.

 \newpage
\newpage
\section{Light pseudoscalar quantities}  \label{app:ps}

In Table~\ref{tab:ps},
we provide our results for the light pseudoscalar masses and decay constants, in lattice units, for all our lattice ensembles. 

The pseudoscalar decay constant on ensembles with open boundary conditions 
is computed using the same procedure as in~\cite{Bruno:2016plf}.
We construct the ratio
\begin{equation}
R(x_0,y_0) = \sqrt{ \frac{2}{m_P} } \left[ \frac{C_A(x_0,y_0) C_A(x_0,T-y_0) }{ C_P(T-y_0,y_0) } \right]^{1/2}
\end{equation}
as an estimator for the (improved, but unrenormalized) decay constant, 
with $m_P$ the pseudoscalar mass. The two-point correlation functions are 
\begin{align}
C_P(x_0,y_0) &= -\frac{a^6}{L^3} \sum_{\vec{x},\vec{y}} \langle P(x_0,\vec{x}) P(y_0,\vec{y}) \rangle \,, \\
C_A(x_0,y_0) &= -\frac{a^6}{L^3} \sum_{\vec{x},\vec{y}} \langle A_0(x_0,\vec{x}) P(y_0,\vec{y}) \rangle \,,
\end{align}
with $P = \overline{\psi}_r \gamma_5 \psi_{r'}$ and
$A_\mu = \overline{\psi}_{r} \gamma_0 \gamma_5 \psi_{r'} + ac_A \partial_{\mu} (\overline{\psi}_r \gamma_5 \psi_{r'})$
the local O$(a)$-improved interpolating operators for the pseudoscalar and axial densities respectively.
The coefficient $\cA$
has been determined non-perturbatively in Ref.~\cite{Bulava:2015bxa} and
the valence flavors are denoted by $r$ and $r'$, with $r\neq r'$.
In practice we average the results between the two source positions $y_0=2a$ and $y_0=T-2a$, close to the temporal boundaries. As shown in~\cite{Bruno:2016plf}, a plateau $R_{\rm avg}$ is obtained at large $x_0$ where excited state contributions are small.
On ensembles with periodic boundary conditions, we use the estimator
\begin{equation}
R_{\rm avg} = \frac{2 Z_{P} }{ m_{P}^2 } \times m^{_{\rm PCAC}}_{rr'} \,,
\end{equation}
where $m^{_{\rm PCAC}}_{rr'}$ is the average PCAC quark mass of flavors $r$ and $r'$,
and $Z_{P}$ the overlap factor of the pseudoscalar meson.
The average PCAC mass is defined from an average in the interval $[t_{\rm i}, t_{\rm f}]$ via
\begin{equation}
	m^{_{\rm PCAC}}_{rr'} = \frac{a}{t_{\rm f} - t_{\rm i} + a} \sum_{x_0=t_{\rm i}}^{t_{\rm f}}\frac{\tilde{\partial}_0C_A(x_0, y_0)}{2 C_P(x_0, y_0)}\,,
\end{equation}
where the source position $y_0$ is fixed as specified above for open boundary
conditions and randomly chosen for periodic boundary conditions.
The interval is chosen such that deviations from a plateau 
which occur at short 
source-sink separations and close to the time boundaries 
are excluded from the average.

From the bare matrix element $R_{\rm avg}$, the renormalized and $\mathcal{O}(a)$-improved pseudoscalar decay constant is given by
\begin{equation}
f_{P}(X_a, X_\pi) = Z_{\rm A}(\widetilde{g}_0) \left( 1 + 3 \bbA a \mqav + \bA am_{\rm q,rr'}  \right) \, R_{\rm avg}  \,.
\end{equation}
In this equation,  $Z_{\rm A}$ is the renormalization factor in the chiral limit and $b_{\rm A}$, $\overline{b}_{\rm A}$ are improvement coefficients
of the axial current. These quantities are known from
Refs.~\cite{DallaBrida:2018tpn,Korcyl:2016ugy,Bali:2021qem}. 
The average valence quark mass 
$m_{\rm q,rr'} = (m_{{\rm q},r} + m_{{\rm q},r'})/2$  
and the average sea quark mass
$\mqav = ( 2m_{{\rm q},l} + m_{{\rm q},s})/3$ are defined in terms of the
bare subtracted quark masses $m_{{\rm q},r}\equiv (2\kappa_r)^{-1} - (2\kappa_{\rm crit})^{-1}$,
with $\kappa_{\rm crit}$ the critical value of the hopping parameter 
at which all three PCAC masses vanish.
In practice, we use the relation \cite{Bhattacharya:2005rb}
\begin{equation}
	m_{\rm q,rr'} = \frac{m^{_{\rm PCAC}}_{rr'}}{Z} - \frac{(r_{\rm m} - 1)}{Z r_{\rm m}} m^{_{\rm PCAC}}_{\rm av} + \mathrm{O}(am^{_{\rm PCAC}}_{rr'}, am^{_{\rm PCAC}}_{\rm av}) \label{eq:msub_from_mPCAC}
\end{equation}
where $m^{_{\rm PCAC}}_{\rm av} = (m^{_{\rm PCAC}}_{ll'} + 2m^{_{\rm PCAC}}_{ ls})/3$ 
is the average sea PCAC quark mass and the coefficients 
$Z(\widetilde{g}_0)=Z_{\rm m}Z_{\rm P}/Z_{\rm A}$ and 
$r_{\rm m}(\widetilde{g}_0)$ have been determined
non-perturbatively in \cite{deDivitiis:2019xla,Heitger:2021bmg}.

The lattice data for the light pseudoscalar masses and decay constants
are corrected for finite-size effects using chiral perturbation theory ($\chi$PT) as described in Ref.~\cite{Colangelo:2005gd}. Those corrections are small (the negative shift is at most $1.3~\sigma$) and we find that they correctly account for FSE on the ensembles H105/N101, which are generated using the same action parameters but different lattice volumes.

\newpage
\section{Tables} \label{sec:tables}

\subsection{Pseudoscalar observables}

\begin{table}[h!]
\caption{Pseudoscalar masses and decay constants in lattice units, including finite-size corrections. Value of the gluonic observable $t_0/a^2$ and the two dimensionless variables $\widetilde{y}$ and $\phi_2$ used in the extrapolation to the physical point.}
\vskip 0.1in
\renewcommand{\arraystretch}{1.1}  
\begin{tabular}{l|@{\hskip 01em}c@{\hskip 01em}c@{\hskip 01em}c@{\hskip 01em}c@{\hskip 01em}|@{\hskip 01em}c@{\hskip 01em}c@{\hskip 0.5em}c}
\hline
id	&	 $a m_{\pi}$	&	$a m_{K}$	 	&	$a f_{\pi}$		&	$a f_{K}$	&	 $t_0/a^2$	&	$\widetilde{y}$	&	$\phi_2$ \\
\hline
A653 &  0.21193(91)  &   0.21193(91)   &   0.07164(23)    &   0.07164(23)   &  2.171(08)  &  0.1108(06)  &  0.7803(70)\\
A654 &  0.16647(121)  &   0.22712(89)   &   0.06723(33)   &   0.07206(23)   &  2.192(11)  &  0.0777(08)  &  0.4860(77)\\
\hline
H101 &  0.18217(62)  &   0.18217(62)   &   0.06377(26)    &   0.06377(26)   &  2.846(08)  &  0.1034(09)  &  0.7557(56)\\
H102 &  0.15395(71)  &   0.19144(57)   &   0.06057(30)    &   0.06365(23)   &  2.872(13)  &  0.0818(08)  &  0.5445(54)\\
H105 &  0.12136(124)&   0.20230(61)   &   0.05800(110)  &   0.06431(29)   &  2.890(08)  &  0.0555(26)  &  0.3405(70)\\
N101 &  0.12150(55)  &   0.20158(31)   &   0.05772(31)    &   0.06418(20)   &  2.881(03)  &  0.0561(07)  &  0.3403(32)\\
C101 &  0.09569(73)  &   0.20579(34)   &   0.05496(31)    &   0.06330(15)   &  2.912(05)  &  0.0384(07)  &  0.2133(33)\\
\hline
B450 &  0.16063(45)  &   0.16063(45)   &   0.05674(15)    &   0.05674(15)   &  3.662(13)  &  0.1015(06)  &  0.7559(48)\\
S400 &  0.13506(44)  &   0.17022(39)   &   0.05394(38)    &   0.05675(32)   &  3.691(08)  &  0.0794(10)  &  0.5387(37)\\
N451 &  0.11072(29)  &   0.17824(18)   &   0.05228(13)    &   0.05789(08)   &  3.681(07)  &  0.0568(03)  &  0.3610(19)\\
D450 &  0.08329(43)  &   0.18384(18)   &   0.04989(21)    &   0.05766(12)   &  3.698(06)  &  0.0353(03)  &  0.2052(21)\\
D452 &  0.05941(55)  &   0.18651(15)   &   0.04827(49)    &   0.05704(08)   &  3.725(01)  &  0.0192(04)  &  0.1052(19)\\
\hline
H200 &  0.13535(60)  &   0.13535(60)   &   0.04799(27)    &   0.04799(27)   &  5.151(33)  &  0.1008(15)  &  0.7549(86)\\
N202 &  0.13424(31)  &   0.13424(31)   &   0.04821(17)    &   0.04821(17)   &  5.140(26)  &  0.0982(08)  &  0.7410(53)\\
N203 &  0.11254(24)  &   0.14402(20)   &   0.04645(14)    &   0.04907(12)   &  5.146(08)  &  0.0744(05)  &  0.5214(24)\\
N200 &  0.09234(31)  &   0.15071(23)   &   0.04424(16)    &   0.04901(16)   &  5.163(07)  &  0.0552(05)  &  0.3522(25)\\
D200 &  0.06507(28)  &   0.15630(15)   &   0.04226(13)    &   0.04910(11)   &  5.181(11)  &  0.0300(04)  &  0.1755(16)\\
E250 &  0.04170(41)  &   0.15924(09)   &   0.04026(19)    &   0.04864(06)   &  5.204(04)  &  0.0136(03)  &  0.0724(14)\\
\hline
N300 &  0.10569(23)  &   0.10569(23)   &   0.03819(14)    &   0.03819(14)   &  8.545(33)  &  0.0970(09)  &  0.7636(38)\\
N302 &  0.08690(34)  &   0.11358(28)   &   0.03663(15)    &   0.03860(15)   &  8.524(25)  &  0.0713(09)  &  0.5150(43)\\
J303 &  0.06475(18)  &   0.11963(16)   &   0.03444(12)    &   0.03872(16)   &  8.612(23)  &  0.0448(04)  &  0.2888(18)\\
E300 &  0.04393(16)  &   0.12372(10)   &   0.03255(09)    &   0.03832(17)   &  8.622(06)  &  0.0231(02)  &  0.1331(10)\\
\hline
J500 &  0.08153(19)  &   0.08153(19)   &   0.02989(10)    &   0.02989(10)   &  13.990(69)  &  0.0942(08)  &  0.7439(51)\\
J501 &  0.06582(23)  &   0.08794(22)   &   0.02882(15)    &   0.03059(15)   &  13.992(67)  &  0.0661(09)  &  0.4850(41)\\
\hline
\end{tabular} 
\label{tab:ps}
\end{table}

\newpage
\subsection{Isovector contribution}

\begin{table}[h!]
\caption{Values of the isovector contributions, with and without $f_{\pi}$-rescaling, in units of $10^{-10}$, for the local-local (${\scriptstyle\rm LL}$) and for the local-conserved (${\scriptstyle\rm CL}$) discretizations of the correlation function, as described in the main text.  
The finite-size correction has been applied. }
\vskip 0.1in
\renewcommand{\arraystretch}{1.1}  
\begin{tabular}{l|@{\hskip 0.5em}c@{\hskip 0.5em}c@{\hskip 0.5em}|@{\hskip 0.5em}c@{\hskip 0.5em}c@{\hskip 0.5em}|@{\hskip 0.5em}c@{\hskip 0.5em}c@{\hskip 0.5em}|@{\hskip 0.5em}c@{\hskip 0.5em}c}
\hline
	&	\multicolumn{2}{c|@{\hskip 0.5em}}{Scale $t_0$ - Set 1} &	\multicolumn{2}{c|@{\hskip 0.5em}}{Scale $f_\pi$ - Set 1}	& \multicolumn{2}{c|@{\hskip 0.5em}}{Scale $t_0$ - Set 2} &	\multicolumn{2}{c}{Scale $f_\pi$ - Set 2} \\
\hline
id	&	 ${\scriptstyle(LL)}$	&	${\scriptstyle(CL)}$	 &
		 ${\scriptstyle(LL)}$	&	${\scriptstyle(CL)}$	 &	
		 ${\scriptstyle(LL)}$	&	${\scriptstyle(CL)}$	&	
		 ${\scriptstyle(LL)}$	&	${\scriptstyle(CL)}$	\\
\hline
A653  &  173.94(36)  &  176.25(37)  &  185.71(28)  &  189.09(32)     &  142.15(35)  &  151.27(37)  &  150.38(21)  &  162.53(26) \\
\hline
H101  &  172.10(39)  &  173.35(39)  &  185.49(47)  &  187.48(49)     &  150.16(39)  &  155.36(39)  &  161.03(40)  &  168.34(46) \\
H102  &  178.54(52)  &  179.75(52)  &  186.34(56)  &  187.95(58)     &  157.27(53)  &  162.26(53)  &  163.73(51)  &  169.87(56) \\
H105$^{*}$  &  184.82(50)  &  186.01(49)  &   188.15(189)  &  189.51(199)  &  164.28(53)  &  169.09(51)  &   167.07(159)  &  172.35(187) \\
N101  &  186.31(43)  &  187.56(42)  &  188.94(60)  &  190.28(61)     &  165.61(44)  &  170.48(43)  &  167.80(54)  &  173.07(58) \\
C101  &  192.19(41)  &  193.40(41)  &  190.56(62)  &  191.69(64)     &  172.25(43)  &  176.94(42)  &  170.87(57)  &  175.33(62) \\
\hline
B450  &  168.12(38)  &  168.82(38)  &  182.47(35)  &  183.62(36)     &  152.53(38)  &  155.68(38)  &  165.14(33)  &  169.63(34) \\
N451  &  183.40(28)  &  184.05(28)  &  188.25(29)  &  189.04(29)     &  168.49(27)  &  171.40(27)  &  172.83(27)  &  176.17(28) \\
D450  &  189.36(26)  &  190.03(27)  &  189.80(46)  &  190.49(47)     &  174.95(26)  &  177.79(26)  &  175.35(43)  &  178.28(45) \\
D452  &  194.96(33)  &  195.61(33)  &  192.97(101)  &  193.58(104)     &  181.21(34)  &  183.97(34)  &  179.42(93)  &  182.00(101) \\
\hline
H200$^{*}$  &  165.17(91)  &  165.44(91)  &   179.46(90)  &  179.92(90)  &  155.70(89)  &  157.21(89)  &   169.07(86)  &  171.19(87) \\
N202  &  168.14(68)  &  168.45(69)  &  182.46(52)  &  182.97(53)     &  158.36(67)  &  159.92(68)  &  171.77(50)  &  173.98(52) \\
N203  &  173.75(43)  &  174.11(43)  &  183.80(44)  &  184.25(44)     &  164.22(43)  &  165.77(43)  &  173.65(43)  &  175.60(44) \\
N200  &  180.17(43)  &  180.43(42)  &  185.21(50)  &  185.53(50)     &  171.02(44)  &  172.41(43)  &  175.77(49)  &  177.37(50) \\
D200  &  188.37(38)  &  188.69(37)  &  189.03(38)  &  189.36(38)     &  179.52(39)  &  180.91(38)  &  180.14(37)  &  181.56(38) \\
E250  &  194.75(26)  &  194.96(26)  &  191.77(45)  &  191.96(46)     &  186.36(27)  &  187.61(26)  &  183.54(44)  &  184.66(45) \\
\hline
N300  &  160.99(59)  &  161.08(59)  &  177.99(61)  &  178.15(60)     &  156.34(59)  &  156.89(59)  &  172.86(60)  &  173.65(60) \\
J303  &  179.51(54)  &  179.57(55)  &  184.77(56)  &  184.84(56)     &  175.24(55)  &  175.67(55)  &  180.39(56)  &  180.88(56) \\
E300  &  188.05(49)  &  188.13(49)  &  188.14(47)  &  188.21(47)     &  183.96(49)  &  184.38(50)  &  184.05(47)  &  184.47(47) \\
\hline
J500  &  162.00(72)  &  162.04(72)  &  178.03(65)  &  178.07(65)     &  159.69(72)  &  159.97(72)  &  175.52(65)  &  175.86(65) \\
J501  &  170.16(98)  &  170.15(98)  &  182.04(83)  &  182.07(83)     &  167.92(98)  &  168.13(98)  &  179.68(83)  &  179.98(83) \\
\hline
\end{tabular} 
\label{tab:winL1}
\end{table}

\newpage
\subsection{Isoscalar contribution}

\begin{table}[h!]
\caption{Values of the isoscalar contributions, with and without $f_{\pi}$-rescaling, in units of $10^{-10}$, for the local-local (${\scriptstyle\rm LL}$) and for the local-conserved (${\scriptstyle\rm CL}$) discretizations of the correlation function, as described in the main text.  
The finite-size correction has been applied. }
\vskip 0.1in
\renewcommand{\arraystretch}{1.1}  
\begin{tabular}{l|@{\hskip 0.5em}c@{\hskip 0.5em}c@{\hskip 0.5em}|@{\hskip 0.5em}c@{\hskip 0.5em}c@{\hskip 0.5em}|@{\hskip 0.5em}c@{\hskip 0.5em}c@{\hskip 0.5em}|@{\hskip 0.5em}c@{\hskip 0.5em}c}
\hline
	&	\multicolumn{2}{c|@{\hskip 0.5em}}{Scale $t_0$ - Set 1} &	\multicolumn{2}{c|@{\hskip 0.5em}}{Scale $f_\pi$ - Set 1}	& \multicolumn{2}{c|@{\hskip 0.5em}}{Scale $t_0$ - Set 2} &	\multicolumn{2}{c}{Scale $f_\pi$ - Set 2} \\
\hline
id	&	 ${\scriptstyle(LL)}$	&	${\scriptstyle(CL)}$	 &
		 ${\scriptstyle(LL)}$	&	${\scriptstyle(CL)}$	 &	
		 ${\scriptstyle(LL)}$	&	${\scriptstyle(CL)}$	&	
		 ${\scriptstyle(LL)}$	&	${\scriptstyle(CL)}$	\\
\hline
A653  &  57.98(12)  &  58.75(12)  &  61.90(9)  &  63.03(11)     &  47.38(12)  &  50.42(12)  &  50.13(7)  &  54.18(9) \\
\hline
H101  &  57.36(13)  &  57.78(13)  &  61.83(16)  &  62.49(16)     &  50.05(13)  &  51.78(13)  &  53.68(13)  &  56.11(15) \\
H102  &  55.30(16)  &  55.71(16)  &  58.28(19)  &  58.82(20)     &  47.94(16)  &  49.70(16)  &  50.38(17)  &  52.53(18) \\
H105$^{*}$  &  53.16(16)  &  53.57(15)  &   54.65(81)  &  55.12(84)  &  45.83(15)  &  47.61(15)  &   47.05(66)  &  49.01(76) \\
N101  &  53.55(11)  &  53.97(11)  &  54.70(25)  &  55.16(26)     &  46.18(11)  &  47.99(11)  &  47.12(21)  &  49.06(24) \\
C101  &  52.67(11)  &  53.08(11)  &  51.89(26)  &  52.27(27)     &  45.39(11)  &  47.18(11)  &  44.74(22)  &  46.46(24) \\
\hline
B450  &  56.04(13)  &  56.27(13)  &  60.82(12)  &  61.21(12)     &  50.84(13)  &  51.89(13)  &  55.05(11)  &  56.54(11) \\
N451  &  52.80(06)  &  53.01(06)  &  54.92(10)  &  55.18(10)     &  47.51(06)  &  48.60(06)  &  49.37(09)  &  50.61(09) \\
D450  &  51.47(06)  &  51.70(07)  &  51.70(19)  &  51.93(19)     &  46.23(06)  &  47.33(06)  &  46.43(17)  &  47.55(18) \\
D452  &  50.90(10)  &  51.12(10)  &  49.82(46)  &  50.06(46)     &  45.74(10)  &  46.84(10)  &  44.79(41)  &  45.82(43) \\
\hline
H200$^{*}$  &  55.06(30)  &  55.15(30)  &   59.82(30)  &  59.97(30)  &  51.90(30)  &  52.40(30)  &   56.36(29)  &  57.06(29) \\
N202  &  56.05(23)  &  56.15(23)  &  60.82(17)  &  60.99(18)     &  52.79(22)  &  53.31(23)  &  57.26(17)  &  57.99(17) \\
N203  &  53.41(13)  &  53.50(13)  &  57.33(15)  &  57.46(15)     &  50.12(13)  &  50.65(13)  &  53.77(14)  &  54.45(15) \\
N200  &  51.61(11)  &  51.70(10)  &  53.81(15)  &  53.92(15)     &  48.35(11)  &  48.88(10)  &  50.39(14)  &  51.00(15) \\
D200  &  50.36(10)  &  50.46(10)  &  50.70(13)  &  50.80(13)     &  47.11(10)  &  47.67(09)  &  47.42(12)  &  47.99(12) \\
E250  &  49.65(09)  &  49.76(09)  &  47.90(24)  &  48.00(24)     &  46.45(09)  &  47.01(09)  &  44.82(22)  &  45.34(22) \\
\hline
N300  &  53.66(20)  &  53.69(20)  &  59.33(20)  &  59.38(20)     &  52.11(20)  &  52.30(20)  &  57.62(20)  &  57.88(20) \\
J303  &  49.80(12)  &  49.82(12)  &  52.21(16)  &  52.23(16)     &  48.25(12)  &  48.43(12)  &  50.59(16)  &  50.80(16) \\
E300  &  48.77(08)  &  48.80(08)  &  48.82(12)  &  48.84(12)     &  47.24(08)  &  47.44(08)  &  47.29(11)  &  47.48(11) \\
\hline
J500  &  54.00(24)  &  54.01(24)  &  59.34(22)  &  59.36(22)     &  53.23(24)  &  53.32(24)  &  58.51(22)  &  58.62(22) \\
\hline
\end{tabular} 
\label{tab:winL2}
\end{table}

\newpage
\subsection{Charm quark contribution}

\begin{table}[h!]
\caption{Charm hopping parameter $\kappa_c$, renormalization factor of the local vector current $Z_V^{(c)}$, value of $\awinc$ for the two discretizations of the correlator: local-local (LL) and local-conserved (LC). The first error is statistical and the second is the systematic error arising from the tuning of the charm quark hopping parameter as explained in the main text.}
\vskip 0.1in
\renewcommand{\arraystretch}{1.2}    
\begin{tabular}{l@{\hskip 02em}c@{\hskip 02em}c@{\hskip 02em}c@{\hskip 02em}c}
\hline
id				&	$\kappa_c$	&	$Z_V^{(c)}$	&	$(\awinc)_{({\scriptscriptstyle\rm LL})}$	&	$(\awinc)_{({\scriptscriptstyle\rm LC})}$\\
\hline
A653				&	0.119743(17)	&	1.32284(15)(71)	& 5.338(03)(23)  &  2.729(02)(12)	\\
A654				&	0.120079(25)	&	1.30495(11)(105)	& 5.523(06)(46)  &  2.870(04)(25)	\\
\hline
H101			&	0.122897(18)	&	1.20324(11)(70)  	&  4.546(09)(27)  &  2.692(06)(16) \\
H102			&	0.123041(26)	&	1.19743(08)(99)	&  4.641(09)(39)  &  2.765(06)(24) \\
H105 			&	0.123244(19)	&	1.18964(08)(74)	&  4.795(13)(30)  &  2.878(09)(19) \\
N101			&	0.123244(19) 	&	1.18964(08)(74) 	&  4.794(17)(30)  &  2.879(11)(19) \\
C101			&	0.123361(12)	&	1.18500(05)(43)	&  4.879(13)(24)  &  2.943(09)(15) \\
\hline
B450				&	0.125095(22)	&	1.12972(06)(82)	&  3.993(07)(26)  &  2.620(05)(17) \\
S400				&	0.125252(20)	&	1.11159(13)(88)	&  4.047(08)(31)  &  2.702(06)(21) \\
N451			&	0.125439(15)	&	1.11412(04)(58)	&  4.255(02)(23)  &  2.837(01)(15) \\
D450			&	0.125585(07) 	&	1.10790(04)(26)	&  4.372(01)(12)  &  2.934(01)(08) \\
D452			&	0.125640(06)	&	1.10790(02)(23)	&  4.445(01)(09)  & 2.985(01)(06) \\
\hline
H200 			&	0.127579(16)	&	1.04843(03)(85)	&  3.503(10)(27)  &  2.590(08)(20) \\
N202 			&	0.127579(16)	&	1.04843(03)(85)	&  3.517(10)(27)  &  2.600(08)(20) \\
N203			&	0.127714(11)	&	1.04534(03)(39)  	&  3.623(08)(19)  &  2.686(06)(14) \\
N200  			&	0.127858(07)	&	1.04012(03)(25)	&  3.758(11)(13)  &  2.802(09)(10) \\
D200 			&	0.127986(06)	&	1.03587(04)(21)	&  3.883(17)(11)  &  2.908(13)(09) \\
E250 			&	0.128054(03)	&	1.03310(01)(11)	&  3.961(01)(11)  &  2.977(01)(08) \\
\hline
N300			&	0.130099(18)	&	0.97722(03)(60)	&  3.030(08)(36)  &  2.513(07)(30) \\
N302    			&	0.130247(09)	&	0.97241(03)(30) 	&  3.218(07)(19)  &  2.681(06)(16) \\
J303				&	0.130362(09)	&	0.96037(10)(38) 	&  3.306(12)(18)  &  2.790(11)(16) \\ 
E300				&	0.130432(06)	&	0.96639(02)(26)  	&  3.447(02)(25)  &  2.891(02)(21) \\
\hline
J500				&	0.131663(16)	&	0.93412(02)(51)	& 2.816(11)(40)  &  2.503(11)(35)	\\
\hline
\end{tabular} 
\label{tab:charm}
\end{table}

\medskip
\bibliographystyle{jhep_collab}
\bibliography{biblio} 

\end{document}